\documentclass[12pt,twoside, a4paper]{article}
\pdfoutput=1
\def\pd{\partial}
\def\mc{\mathcal}

\def\ul{\underline}
\def\bb{\mathbb}

\usepackage[dvips]{graphicx}
\usepackage{amssymb}
\usepackage{amssymb,amsmath}
\usepackage{graphicx}
\usepackage{dsfont}
\usepackage{caption}
\usepackage{subcaption}
\usepackage{mathtools}
\usepackage{verbatim}
\usepackage{graphicx}
\usepackage{multirow}
\usepackage[outline]{contour}
\usepackage{xcolor,colortbl}
\usepackage{pdflscape}
\input{epsf.sty}  
\begin{document}
\begin{center}
\LARGE{\textbf{Supersymmetric Janus solutions in $\omega$-deformed $N=8$ gauged supergravity}}
\end{center}
\vspace{1 cm}
\begin{center}
\large{\textbf{Parinya Karndumri}$^a$ and \textbf{Chawakorn Maneerat}$^b$}
\end{center}
\begin{center}
String Theory and Supergravity Group, Department
of Physics, Faculty of Science, Chulalongkorn University, 254 Phayathai Road, Pathumwan, Bangkok 10330, Thailand
\end{center}
E-mail: $^a$parinya.ka@hotmail.com \\
E-mail: $^b$chawakorn.manee@gmail.com
\vspace{1 cm}\\
\begin{abstract}
We give a large class of supersymmetric Janus solutions in $\omega$-deformed (dyonic) $SO(8)$ maximal gauged supergravity with $\omega=\frac{\pi}{8}$. Unlike the purely electric counterpart, the dyonic $SO(8)$ gauged supergravity exhibits a richer structure of $AdS_4$ vacua with $N=8,2,1,1$ supersymmetries and $SO(8)$, $U(3)$, $G_2$ and $SU(3)$ symmetries, respectively. Similarly, domain walls interpolating among these critical points show a very rich structure as well. In this paper, we show that this gauged supergravity also accommodates a number of interesting supersymmetric Janus solutions in the form of $AdS_3$-sliced domain walls asymptotically interpolating between the aforementioned $AdS_4$ geometries. These solutions could be holographically interpreted as two-dimensional conformal defects within the superconformal field theories (SCFTs) of ABJM type dual to the $AdS_4$ vacua. We also give a class of solutions interpolating among the $SO(8)$, $G_2$ and $U(3)$ $AdS_4$ vacua in the case of $\omega=0$ which have not previously appeared in the presently known Janus solutions of electric $SO(8)$ gauged supergravity.  
\end{abstract}
\newpage
%%%%%%%%%%%%%%%%%%%%%%%%%%%%%%%%%%%%%%%%%%%%%%%%%%%%%%%%%%%%%%%%%%%%%%%%%%%%%%%%%%%%%%%%%%%%%%%%%%%%%%%%%%%%%%%%%%%%%%%%%%%%%%%%%%%%%%%%%
\section{Introduction}
Janus solutions of $D$-dimensional gauged supergravity take the form of $AdS_{D-1}$-sliced domain walls. Regular solutions of this type are asymptotic to $AdS_D$ geometries on both sides of the $AdS_{D-1}$ slice. According to the AdS/CFT correspondence \cite{maldacena,Gubser_AdS_CFT,Witten_AdS_CFT}, these configurations are dual to $(D-2)$-dimensional conformal interfaces or defects in the $(D-1)$-dimensional CFT dual to the $AdS_D$ vacuum. Since the original Janus solution found in \cite{Bak_Janus} by considering a deformation of the $AdS_5\times S^5$ geometry in type IIB theory, a number of works have studied this type of solutions both in type IIB theory and five-dimensional gauged supergravity along with the corresponding conformal interfaces in the dual $N=4$ super Yang-Mills theory, see for example \cite{Freedman_Janus,DHoker_Janus,Witten_Janus,Freedman_Holographic_dCFT,5D_Janus_CK,5D_Janus_DHoker1,5D_Janus_DHoker2,5D_Janus_Suh} and \cite{Bobev_5D_Janus1,Bobev_5D_Janus2} for more recent results. 
\\
\indent In this paper, we are interested in supersymmetric Janus solutions of dyonic $SO(8)$ gauged supergravity in four dimensions constructed in \cite{omega_N8_1}, see also \cite{omega_N8_2}. This gauged supergravity is a deformation of the original $SO(8)$ gauged supergravity constructed long ago in \cite{SO8_deWit} by an electromagnetic phase usually called $\omega$. For $\omega=0$, a number of Janus solutions have been given in \cite{warner_Janus}, see \cite{N3_Janus,tri-sasakian-flow,orbifold_flow,Minwoo_4DN8_Janus,Kim_Janus,N5_flow,N6_flow} for Janus solutions in four-dimensional gauged supergravities with different numbers of supersymmetries and \cite{3D_Janus_de_Boer,3D_Janus_Bachas,3D_Janus_Bak,half_BPS_AdS3_S3_ICFT,exact_half_BPS_string,multi_face_Janus,
4D_Janus_from_11D,6D_Janus,3D_Janus} for solutions in other dimensions. 
\\
\indent The $SO(8)$ gauged supergravity with $\omega\neq 0$ exhibits a richer structure of supersymmetric $AdS_4$ vacua \cite{omega_vacua} compared to the $\omega=0$ theory. In particular, there exist new $N=1$ supersymmetric critical points with $G_2$ and $SU(3)$ symmetries in addition to the $SO(8)$, $G_2$ and $U(3)\sim SU(3)\times U(1)$ critical points with $N=8,1,2$ supersymmetries which have $\omega=0$ counterparts. Holographic RG flow solutions interpolating between these critical points have been investigated in \cite{Guariano} and \cite{Varella_N8_flow}, see also \cite{Yi_4D_flow}, and also show a richer structure than the $\omega=0$ analogue. We then expect that Janus solutions will exhibit a much richer structure with many possible solutions as well. We will see that this is indeed the case.
\\
\indent Janus solutions given in \cite{warner_Janus} only involve the $SO(8)$ and $G_2$ critical points resulting in $SO(8)/SO(8)$, $SO(8)/G_2$ and $G_2/G_2$ interfaces. The $N=2$ $U(3)$ critical point is however not present in the two-scalar truncation considered in \cite{warner_Janus}. Since in this paper, we study solutions in the full $SU(3)$ invariant scalar sector, we also consider Janus solutions with $\omega=0$ that involve all $AdS_4$ critical points with $SO(8)$, $G_2$ and $U(3)$ symmetry. The resulting solutions could hopefully provide the missing part in the list of known Janus solutions in electric $SO(8)$ gauged supergravity. To the best of our knowledge, all the solutions with $\omega\neq 0$ have not previously appeared.  
\\
\indent The paper is organized as follows. In section \ref{N8_SUGRA},
we review the construction of four-dimensional $N=8$ gauged supergravity with dyonic $SO(8)$ gauge group and the corresponding $AdS_4$ vacua. BPS equations for Janus solutions in $SU(3)$ invariant sector are also given. In sections \ref{Janus_omega_0} and \ref{Janus_omega_n0}, we give numerical Janus solutions for $\omega=0$ and $\omega=\frac{\pi}{8}$ cases, respectively. Conclusions and comments on the results are given in section \ref{conclusion}.

%%%%%%%%%%%%%%%%%%%%%%%%%%%%%%%%%%%%%%%%%%%%%%%%%%%%%%%%%%%%%%%%%%%%%%%%%%%%%%%%%%%%%%%%%%%%%%%%%%%%%%%%%%%%%%%%%%%%%%%%%%%%%%%%%%%%%%%%%
\section{$N=8$ gauged supergravity with dyonic $SO(8)$ gauge group}\label{N8_SUGRA}
We first give a brief review of $N=8$ gauged supergravity in four dimensions with dyonic $SO(8)$ gauge group constructed in \cite{omega_N8_1,omega_N8_2} to which we refer for more detail. We mostly follow the conventions of \cite{Guariano} with $(-+++)$ signature for the space-time metric. The only supermultiplet in $N=8$ supersymmetry is given by the supergravity multiplet with the field content
\begin{equation}
(e^{\hat{\mu}}_\mu,\psi^I_\mu,A^{AB}_\mu,\chi_{IJK},\Sigma_{IJKL}).
\end{equation}
This multiplet consists of the graviton $e^{\hat{\mu}}_\mu$, $8$ gravitini $\psi^I_\mu$, $28$ vectors $A^{AB}_\mu=-A^{BA}_\mu$, $56$ spin-$\frac{1}{2}$ fields $\chi_{IJK}=\chi_{[IJK]}$ and $70$ scalars $\Sigma_{IJKL}=\Sigma_{[IJKL]}$.
\\
\indent Before moving on, we first state the conventions on various indices used throughout the paper. Space-time and tangent space indices are denoted by $\mu,\nu,\ldots =0,1,2,3$ and $\hat{\mu},\hat{\nu},\ldots =0,1,2,3$, respectively. The $N=8$ supergravity admits global $E_{7(7)}$ and local composite $SU(8)$ symmetries with the corresponding fundamental representations are respectively described by indices $\bb{M},\bb{N},\ldots=1,2,3,\ldots ,56$ and $I,J,K,\ldots =1,2,3,\ldots 8$. Indices $A,B,\ldots =1,2,3,\ldots, 8$ refer to fundamental indices of $SL(8)\subset E_{7(7)}$. The scalars $\Sigma_{IJKL}$ are encoded in the $E_{7(7)}/SU(8)$ coset manifold and can be described by the coset representative ${\mc{V}_{\bb{M}}}^{\ul{\bb{M}}}$. The local $SU(8)$ indices $\ul{\bb{M}},\ul{\bb{N}},\ldots$ will further be decomposed as $_{\ul{\bb{M}}}=(_{[IJ]},^{[IJ]})$. Similarly, the global $E_{7(7)}$ indices $\bb{M},\bb{N},\ldots$ will be decomposed in the $SL(8)$ basis as $_{\bb{M}}=(_{[AB]},^{[AB]})$. The scalars $\Sigma_{IJKL}$ are self-dual 
\begin{equation}
\Sigma_{IJKL}=\frac{1}{4!}\epsilon_{IJKLMNPQ}\Sigma^{MNPQ}
 \end{equation}
with $\Sigma^{IJKL}=(\Sigma_{IJKL})^*$ and $\epsilon_{IJKLMNPQ}$ being the invariant tensor of the $SU(8)$ R-symmetry. 
\\
\indent The action of the global $E_{7(7)}$ symmetry includes electric-magnetic duality with the vector fields $A_\mu^{AB}$ together with the magnetic dual $A_{\mu AB}$ transforming in the fundamental $\mathbf{56}$ representation. In general, the Lagrangian of the ungauged $N=8$ supergravity will exhibit only particular subgroups of $E_{7(7)}$ depending on the electric-magnetic or symplectic frames. On the other hand, the full $E_{7(7)}$ symmetry is realized through the field equations together with the Bianchi identities. The most general gaugings of the $N=8$ supergravity can be described by the so-called embedding tensor ${\Theta_{\bb{M}}}^\alpha$ which introduces a minimal coupling to various fields in the ungauged supergravity via the covariant derivative
\begin{equation}
D_\mu=\nabla_\mu-gA^{\bb{M}}_\mu {\Theta_{\bb{M}}}^\alpha t_\alpha
 \end{equation} 
with $\nabla_\mu$ being the usual space-time covariant derivative including the $SU(8)$ composite connection (if any). $t_\alpha$ are $E_{7(7)}$ generators with $\alpha=1,2,3,\ldots , 133$. Supersymmetry requires the embedding tensor to transform as $\mathbf{912}$ representation of $E_{7(7)}$. In addition, the gauge generators $X_{\bb{M}}={\Theta_{\bb{M}}}^\alpha t_\alpha$ must form a closed subalgebra of $E_{7(7)}$. The latter imposes the quadratic constraint on the embedding tensor of the form 
\begin{equation}
\Omega^{\bb{M}\bb{N}}{\Theta_{\bb{M}}}^\alpha{\Theta_{\bb{N}}}^\beta=0\, .
\end{equation}
$\Omega^{\bb{M}\bb{N}}$ is the symplectic form of the duality group $Sp(56,\bb{R})$ in which $E_{7(7)}$ is embedded. The quadratic constraint can be rewritten in terms of the gauge generators as
\begin{equation}
[X_{\bb{M}},X_{\bb{N}}]=-{X_{\bb{M}\bb{N}}}^{\bb{P}}X_{\bb{P}}
\end{equation}
with ${X_{\bb{M}\bb{N}}}^{\bb{P}}={\Theta_{\bb{M}}}^\alpha {(t_\alpha)_{\bb{N}}}^{\bb{P}}$ and ${(t_\alpha)_{\bb{N}}}^{\bb{P}}$ being the $E_{7(7)}$ generators in the fundamental representation.
\\
\indent In this paper, we are interested mainly in the solutions with only the metric and scalars non-vanishing. We will from now on set all the other fields to zero to simplify the presentation. The bosonic Lagrangian of the $N=8$ gauged supergravity can be written as
\begin{equation}
e^{-1}\mc{L}=\frac{1}{2}R-\frac{1}{12}P^{IJKL}_\mu P_{IJKL}^\mu-V\, .
 \end{equation} 
The scalar potential is given in terms of the fermion-shift matrices as
\begin{equation}
V=-\frac{3}{4}g^2A_{1IJ}A_1^{IJ}+\frac{1}{24}g^2{A_{2I}}^{JKL}{A_2^I}_{JKL}
\end{equation}
with $A_1^{IJ}=(A_{1IJ})^*$ and ${A_{2I}}^{JKL}=({A_2^I}_{JKL})^*$. $A_1$ and $A_2$ matrices can be defined in term of the T-tensor by the following relations
\begin{equation}
A_1^{IJ}=\frac{4}{21}{T^{IKJL}}_{KL}\qquad \textrm{and}\qquad {A_{2I}}^{JKL}=2{T_{MI}}^{MJKL}\, .
\end{equation}
The T-tensor is in turn obtained from the embedding tensor via
\begin{equation}
{T_{\ul{\bb{M}}\ul{\bb{N}}}}^{\ul{\bb{P}}}={\mc{V}^{\bb{M}}}_{\ul{\bb{M}}}{\mc{V}^{\bb{N}}}_{\ul{\bb{N}}}{\mc{V}_{\bb{P}}}^{\ul{\bb{P}}}{X_{\bb{M}\bb{N}}}^{\bb{P}}
\end{equation}
with ${\mc{V}^{\bb{M}}}_{\ul{\bb{M}}}={(\mc{V}^{-1})_{\ul{\bb{M}}}}^{\bb{M}}$.
\\
\indent The supersymmetry transformations of $\psi^I_\mu$ and $\chi_{IJK}$, which are needed in finding supersymmetric solutions, are given by
\begin{eqnarray}
\delta\psi^I_\mu&=&2D_\mu \epsilon^I+\sqrt{2}gA_1^{IJ}\gamma_\mu \epsilon_J,\\
\delta \chi^{IJK}&=&-2\sqrt{2}P_\mu^{IJKL}\gamma^\mu \epsilon_L-2g{A_{2L}}^{IJK}\epsilon^L\, .
\end{eqnarray}
The covariant derivative of $\epsilon^I$ is defined by
\begin{equation}
D_\mu \epsilon^I=\pd_\mu \epsilon^I+\frac{1}{4}{\omega_\mu}^{\hat{\mu}\hat{\nu}}\gamma_{\hat{\mu}\hat{\nu}}\epsilon^I+\frac{1}{2}{{Q_\mu}^I}_J\epsilon^J\, .
\end{equation}
The composite connection ${{Q_\mu}^I}_J=({Q_{\mu I}}^J)^*$ and the vielbein $P_\mu^{IJKL}$ on the $E_{7(7)}/SU(8)$ coset are given by
\begin{eqnarray}
{Q_{\mu I}}^J&=&\frac{i}{3}(\mc{V}_{ABIK}\pd_\mu \mc{V}^{JK}-{\mc{V}^{AB}}_{IK}\pd_\mu {\mc{V}_{AB}}^{JK}),\\
P_{\mu IJKL}&=&\frac{i}{2}(\mc{V}_{ABIJ}\pd_\mu {\mc{V}^{AB}}_{KL}-{\mc{V}^{AB}}_{IJ}\pd_\mu \mc{V}_{ABKL})
\end{eqnarray}
with 
\begin{equation}
P_\mu^{IJKL}=\frac{1}{4!}\epsilon^{IJKLMNPQ}P_{\mu MNPQ}\, .
\end{equation}
\indent We finally note that the kinetic term and the scalar potential can be written in term of the symmetric scalar matrix
\begin{equation}
 \mc{M}_{\bb{M}\bb{N}}={\mc{V}_{\bb{M}}}^{\ul{\bb{M}}}{\mc{V}_{\bb{N}}}^{\ul{\bb{N}}}\eta_{\ul{\bb{M}}\ul{\bb{N}}}\qquad \textrm{with}\qquad \eta_{\ul{\bb{M}}\ul{\bb{N}}}=\left(\begin{array}{cc}
 0&\mathbb{I}_{28}\\
 \mathbb{I}_{28}& 0
 \end{array}\right)
\end{equation} 
as   
\begin{equation}
-\frac{1}{12}P^{IJKL}_\mu P_{IJKL}^\mu=\frac{1}{96}\pd_\mu \mc{M}_{\bb{M}\bb{N}}\pd^\mu \mc{M}^{\bb{M}\bb{N}}
\end{equation}
and
\begin{equation}
V=\frac{g^2}{672}\left({X_{\bb{M}\bb{N}}}^{\bb{R}}{X_{\bb{P}\bb{Q}}}^{\bb{S}}\mc{M}^{\bb{M}\bb{P}}\mc{M}^{\bb{N}\bb{Q}}\mc{M}_{\bb{R}\bb{S}}
+7{X_{\bb{M}\bb{N}}}^{\bb{Q}}{X_{\bb{P}\bb{Q}}}^{\bb{N}}\mc{M}^{\bb{M}\bb{P}}\right).
\end{equation}
$\mc{M}^{\bb{M}\bb{N}}$ is the inverse of $\mc{M}_{\bb{M}\bb{N}}$.

\subsection{Dyonic $SO(8)$ gauging}
In general, both electric and magnetic vector fields can participate in the gauging. We now consider gauging of a subgroup $G\subset SL(8)\subset E_{7(7)}$. The $SL(8)$ generators take a block-diagonal form, and various components of the embedding tensor corresponding to the gauge group $G$ are given by
\begin{eqnarray}
& &{X_{[AB][CD]}}^{[EF]}=-8\delta^{[E}_{[A}\theta_{B][C}\delta^{F]}_{D]},\qquad {{X_{[AB]}}^{[CD]}}_{[EF]}=8\delta^{[C}_{[A}\theta_{B][E}\delta^{D]}_{F]},\nonumber \\
& &
{{X^{[AB]}}_{[CD]}}^{[EF]}=-8\delta^{[A}_{[C}\xi^{B][E}\delta^{F]}_{D]},\qquad {X^{[AB][CD]}}_{[EF]}=8\delta^{[A}_{[E}\xi^{B][C}\delta^{D]}_{F]},
\end{eqnarray}
The quadratic constraint gives rise to the condition
\begin{equation}
\theta\xi=\frac{1}{8}\textrm{Tr}(\theta\xi)\mathbb{I}_8
\end{equation}
which implies
\begin{equation}
\xi=c\theta^{-1}\, .
\end{equation}
The tensors $\theta$ and $\xi$ are symmetric and can be diagonalized to have eigenvalues $0,\pm1$. This leads to $CSO(p,q,r)$ gauge group with $p+q+r=8$ for $p$, $q$ and $r$ being numbers of eigenvalues $1$, $-1$ and $0$, respectively. It is also convenient to define another parameter $\omega$ by the following relation
\begin{equation}
\omega=\textrm{Arg}(1+ic)
\end{equation}
with $c=0$ $(\omega=0)$ and $c=\infty$ $(\omega=\frac{\pi}{2})$ leading to purely electric and purely magnetic gauge groups, respectively. The former is the original $SO(8)$ gauged supergravity of \cite{SO8_deWit}. It has been shown in \cite{omega_N8_1,omega_vacua,omega_Range1,deWit_omega} that the values of $\omega$ are equivalent under the identifications $\omega\rightarrow -\omega$ and $\omega\rightarrow \omega+\frac{\pi}{4}$. This results in inequivalent values of $\omega$ in the range $[0,\frac{\pi}{8}]$. In this paper, we are only interested in the case of $\theta_{AB}=\xi^{AB}=\delta_{AB}$ corresponding to the $SO(8)$ gauge group.

\subsection{$SU(3)$ truncation}
In order to make things more manageable, most results on $N=8$ gauged supergravity are obtained by truncating the $70$-dimensional $E_{7(7)}/SU(8)$ manifold to lower-dimensional submanifolds invariant under certain subgroups of the gauge group. In this work, we are interested in scalar fields that are singlets of $SU(3)\subset SO(8)$ \cite{omega_vacua} following the discussion in \cite{Guariano}. The embedding of $SU(3)$ can be identified by decomposing the $\mathbf{8}_v$ representation of $SO(8)$ to $\mathbf{1}+\mathbf{1}+\mathbf{3}+\bar{\mathbf{3}}$ of $SU(3)$. Accordingly, the fundamental $SU(8)$ index $I$ splits as $I=(1,a,\hat{1},\hat{a})$ for $a=2,3,4$ and $\hat{a}=\hat{2},\hat{3},\hat{4}$.
\\
\indent After the truncation, there are six scalars parametrizing the coset space
\begin{equation}
SL(2)/SO(2)\times SU(2,1)/U(2) \label{truncated_singlets}
\end{equation}
and two gauge fields corresponding to $U(1)\times U(1)$ gauge group. The unbroken supersymmetry in the truncated theory is given by $\epsilon^1$ and $\epsilon^{\hat{1}}$ which are singlets of $SU(3)$. The resulting theory is $N=2$ gauged supergravity coupled to one vector multiplet and one hypermultiplet. 
\\
\indent Under $SO(8)$ gauge group, the $70$ scalars $\Sigma_{IJKL}$ decompose into self- and anti-self-dual parts $\Sigma^\pm_{IJKL}$ in representations $\mathbf{35}_s$ and $\mathbf{35}_c$, respectively. The $\mathbf{35}_s$ and $\mathbf{35}_c$ can be rewritten in terms of real and imaginary parts of $35$ complex scalars which are identified respectively with scalars and pseudoscalars. For the $\omega=0$ case with known eleven-dimensional origin, the former arise from the eleven-dimensional metric while the latter come from the three-form potential. They are respectively dual to boson and fermion bilinear operators in the dual $N=8$ SCFT.  
\\
\indent Two of the six singlets in \eqref{truncated_singlets} can be gauged away by the $U(1)\times U(1)$ gauge symmetry \cite{warner_new_extrema}. The remaining four scalars can be described by the four-form $\Sigma_{IJKL}$ of the form 
\begin{equation}
\Sigma=\xi\cos\phi(\Sigma_++\Sigma_+^*)+i\xi\sin\phi (\Sigma_--\Sigma^*_-)-\zeta\cos\chi J^+\wedge J^+-i\zeta\sin\chi J^-\wedge J^-\label{scalar_4form}
\end{equation}
in which the real and complex two-forms $J^\pm$ and $\Sigma_{\pm}$ are defined by
\begin{eqnarray}
J^\pm &=&\pm\frac{i}{2}dz_1\wedge d\bar{z}_{\bar{1}}+\frac{i}{2}\sum_{a=2}^4dz_a\wedge d\bar{z}_{\bar{a}}\\
\textrm{and}\qquad \Sigma_+&=&dz_1\wedge dz_a\wedge dz_b\wedge dz_c,\qquad \Sigma_-=d\bar{z}_{\bar{1}}\wedge dz_a\wedge dz_b\wedge dz_c\, .
 \end{eqnarray} 
The complex coordinates $z_i=(z_1,z_a)$ and their complex conjugate $\bar{z}_{\bar{i}}=(\bar{z}_{\bar{1}},\bar{z}_
{\bar{a}})$ are given in terms of a real vector $(x_1,\ldots,x_4,x_{\hat{1}},\ldots,x_{\hat{4}})$ of $\mathbf{8}_v$ as
\begin{equation}
z_i=x_i+ix_{\hat{i}}\qquad \textrm{and}\qquad \bar{z}_{\bar{i}}=x_i-ix_{\hat{i}}
\end{equation}
Components of $\Sigma_{IJKL}$ given in \eqref{scalar_4form} can be used to construct the scalar coset representative ${\mc{V}_{\bb{M}}}^{\ul{\bb{M}}}$.
\\
\indent In this $SU(3)$ truncation, the scalar kinetic term can be written as
\begin{equation}
\mc{L}_{\textrm{kin}}=-3\pd_\mu \zeta\pd^\mu \zeta-\frac{3}{4}\sinh^2(2\zeta)\pd_\mu \chi\pd^\mu \chi-4\pd_\mu \xi\pd^\mu \xi-\sinh^2(2\xi)\pd_\mu \phi\pd^\mu \phi\, .
\end{equation}
The resulting components of the $A_1^{IJ}$ tensor split into $(A^{11}_1,A^{aa}_1,A^{\hat{1}\hat{1}}_1,A^{\hat{a}\hat{a}}_1)$ and take the form of
\begin{equation}
A_1^{IJ}=e^{i\omega}A^{IJ}_++e^{-i\omega}A^{IJ}_-\, .
 \end{equation} 
$A^{11}_1$ and $A_1^{\hat{1}\hat{1}}$ correspond to the unbroken $N=2$ supersymmetry giving rise to the superpotentials $\mc{W}_1$ and $\mc{W}_{\hat{1}}$. The scalar potential can be written in term of the real superpotential $W=|\mc{W}_1|=|\mc{W}_{\hat{1}}|$ as
\begin{equation}
V=\frac{1}{3}\frac{\pd W}{\pd \zeta}\frac{\pd W}{\pd \zeta}+\frac{4}{3\sinh^2(2\zeta)}\frac{\pd W}{\pd \chi}\frac{\pd W}{\pd \chi}+\frac{1}{4}\frac{\pd W}{\pd \xi}\frac{\pd W}{\pd \xi}+\frac{1}{\sinh^2(2\xi)}\frac{\pd W}{\pd \phi}\frac{\pd W}{\pd \phi}-3W^2\, .
\end{equation}
In our convention, the superpotentials $\mc{W}_{1,\hat{1}}$ are related to $A^{11}_1$ and $A_1^{\hat{1}\hat{1}}$ by
\begin{equation}
\mc{W}_{1}=\sqrt{2}gA^{11}_1\qquad \textrm{and}\qquad \mc{W}_{\hat{1}}=\sqrt{2}gA^{\hat{1}\hat{1}}_1\, .
 \end{equation} 
It should be noted that this definition is different from that used in \cite{Guariano} in which $\mc{W}_{1,\hat{1}}$ are directly defined by $A^{11}_1$ and $A_1^{\hat{1}\hat{1}}$. This results in different numerical factors for some expressions, but the final result is the same.
\\
\indent The explicit form of $A^{11}_1$ and $A_1^{\hat{1}\hat{1}}$ is obtained from $A^{11}_\pm$ and $A_\pm^{\hat{1}\hat{1}}$ given by
\begin{eqnarray}
A^{11}_+&=&\frac{3}{2}e^{2i(\chi+\phi)}\cosh\zeta\sinh^2\zeta\sinh^2(2\xi)+\cosh^3\zeta(\cosh^4\xi+e^{4i\phi}\sinh^4\xi),\\
A^{11}_-&=&\frac{3}{2}e^{i(\chi+2\phi)}\cosh\zeta\sinh^2\zeta\sinh^2(2\xi)+e^{3i\chi}\sinh^3\zeta(\cosh^4\xi+e^{4i\phi}\sinh^4\xi)\nonumber \\
& &
\end{eqnarray}
with 
\begin{equation}
A^{\hat{1}\hat{1}}_\pm=A^{11}_\pm (\phi\rightarrow-\phi).
\end{equation}
Since the explicit form of the scalar potential is rather complicated, we refrain from giving it here but simply refer to \cite{Guariano}.
\\
\indent A number of supersymmetric $AdS_4$ vacua of the scalar potential have been identified, and some of them do not have counterparts in the $\omega=0$ case. Before considering supersymmetric Janus solutions, for convenience, we collect all of the known supersymmetric $AdS_4$ critical points for $\omega=\frac{\pi}{8}$ in table \ref{table1}.

\begin{table}[h]
\centering
\begin{tabular}{|c | c | c|}
\hline
Supersymmetry & Residual symmetry & $(\zeta,\chi,\xi,\phi)$ \\
\hline
$N=8$ & $SO(8)$ & $(0,0,0,0)$\\ \hline
$N=2$ & $U(3)\sim SU(3)\times U(1)$ & $(0.315,0.171\pi,0.375,\pm\frac{\pi}{2})$ \\
 & & $(0.315,1.329\pi,0.375,0)$ \\
& & $(0.315,1.329\pi,0.375,\pi)$ \\ \hline
$N=1$ $W_{1}$ & $G_2$ & $(0.329,0.373\pi,0.329,\pm0.373\pi)$\\
& & $(0.329,0.373\pi,0.329,\pm1.373\pi)$\\
$\phantom{N=1}$ $W_{\hat{1}}$ & & $(0.329,1.127\pi,0.329,\pm1.127\pi)$\\
& &$(0.329,1.127\pi,0.329,\pm0.127\pi)$\\ \hline
$N=1$ $W_{1}$ & $G_2$ & $(0.242,-\frac{\pi}{4},0.242,\mp\frac{\pi}{4})$\\
$\phantom{N=1}$ $W_{\hat{1}}$& & $(0.242,-\frac{\pi}{4},0.242,\pm\frac{3\pi}{4})$\\ \hline
$N=1$ $W_{1}$ & $SU(3)$ & $(0.275,\frac{3\pi}{4},0.573,\pm\frac{\pi}{4})$\\
$\phantom{N=1}$ $W_{\hat{1}}$ & & $(0.275,\frac{3\pi}{4},0.573,\mp\frac{3\pi}{4})$ \\
\hline
\end{tabular}
\caption{Supersymmetic $AdS_4$ vacua of $SO(8)$ gauged supergravity with $\omega=\frac{\pi}{8}$. The first three critical points have $\omega=0$ counterparts while the last two are genuine critical points of the dyonic $SO(8)$ gauged supergravity without the $\omega=0$ analogues. For $N=1$ critical points, we also note which superpotential among $W_1$ and $W_{\hat{1}}$ corresponds to the unbroken supersymmetry.}\label{table1}
\end{table}      

\subsection{BPS equations for Janus solutions}
We are now in a position to analyze the fermionic supersymmetry transformations and find the corresponding BPS equations. The analysis closely follows that given in \cite{warner_Janus}, see also \cite{N3_Janus}, to which the reader is referred for more detail. 
\\
\indent We begin with the metric ansatz of the form
\begin{equation}
ds^2=e^{2A(r)}\left(e^{\frac{2\rho}{\ell}}dx^2_{1,1}+d\rho^2\right)+dr^2\,
.
\end{equation}
This is the domain wall metric with an $AdS_3$ slice rather than the three-dimensional flat Minkowski space. The latter is recovered in the limit $\ell\rightarrow \infty$. 
\\
\indent With the vielbein components
\begin{equation}
e^{\hat{\alpha}}=e^{A+\frac{\rho}{\ell}}dx^\alpha,\qquad
e^{\hat{\rho}}=e^{A}d\rho,\qquad e^{\hat{r}}=dr,
\end{equation}
non-vanishing components of the spin connection from the above metric are given by 
\begin{equation}
\omega^{\hat{\rho}}_{\phantom{\hat{\rho}}\hat{r}}=A'e^{\hat{\rho}},\qquad
\omega^{\hat{\alpha}}_{\phantom{\hat{\rho}}\hat{\rho}}=\frac{1}{\ell}e^{-A}e^{\hat{\alpha}},\qquad
\omega^{\hat{\alpha}}_{\phantom{\hat{\rho}}\hat{r}}=A'e^{\hat{\alpha}}
\end{equation}
where $'$ denotes the $r$-derivative. Indices $\alpha,\beta$ will take values $0,1$. The other non-vanishing fields are the scalars $(\zeta,\xi,\chi,\phi)$ depending only on $r$. 
\\ 
\indent We will use Majorana representation for gamma matrices with all $\gamma^\mu$ real and $\gamma_5$ purely imaginary. The two chiralities of the supersymmetry parameters $\epsilon^I$ and $\epsilon_I$ are then related to each other by a complex conjugation. We will denote the Killing spinors corresponding to the unbroken supersymmetry by $\epsilon$. In the present case, $\epsilon$ will be $\epsilon^1$ or $\epsilon^{\hat{1}}$, and $\epsilon^*$ is given by $\epsilon_1$ or $\epsilon_{\hat{1}}$. The supersymmetry conditions $\delta\chi^{IJK}=0$ involve only $\gamma_{\hat{r}}$ since the scalars depend only on $r$. Following \cite{warner_Janus}, we impose the projector of the form
\begin{equation}
\gamma_{\hat{r}}\epsilon=e^{i\Lambda}\epsilon^*\label{gamma_r_pro}
\end{equation}
where $\Lambda$ is real. 
\\
\indent We now consider the supersymmetry transformations $\delta\psi^{I}_{\hat{\alpha}}$ which lead to the following conditions
\begin{equation}
A'\gamma_{\hat{r}}\epsilon+\frac{1}{\ell}e^{-A}\gamma_{\hat{\rho}}\epsilon+\mc{W}\epsilon^*=0\,
.\label{dPsi_mu_eq}
\end{equation}
The superpotential $\mc{W}$ is given by $\sqrt{2}gA^{11}_1$ or $\sqrt{2}gA^{\hat{1}\hat{1}}_1$. Taking the complex conjugate and
iterating the above equation, we obtain
\begin{equation}
A'^2=W^2-\frac{1}{\ell^2}e^{-2A}\label{dPsi_BPS_eq}
\end{equation}
for $W=|\mc{W}|$. 
\\
\indent We now move to the equation coming from $\delta \psi^I_{\hat{\rho}}=0$. This takes the form 
\begin{equation}
e^{-A}\pd_\rho\epsilon+\frac{1}{2}A'\gamma_{\hat{\rho}}\gamma_{\hat{r}}\epsilon
+\frac{1}{2}\mc{W}\gamma_{\hat{\rho}}\epsilon^*=0\, .
\end{equation}
Using \eqref{dPsi_mu_eq}, we find
\begin{equation}
\pd_\rho\epsilon=\frac{1}{2\ell}\epsilon
\end{equation}
which gives $\epsilon=e^{\frac{\rho}{2\ell}}\tilde{\epsilon}$ for a $\rho$-independent $\tilde{\epsilon}$.
\\
\indent Finally, using the $\gamma_{\hat{\rho}}$ projection of the form
\begin{equation}
\gamma_{\hat{\rho}}\epsilon=i\kappa e^{i\Lambda}\epsilon^*\label{gamma_rho_pro}
\end{equation}
with $\kappa^2=1$ and $\delta\psi^I_{\hat{r}}=0$, we can determine the explicit form of the Killing spinor to be
\begin{equation}
\epsilon=e^{\frac{A}{2}+\frac{\rho}{2\ell}+i\frac{\Lambda}{2}}\varepsilon^{(0)}
\end{equation}
The spinor $\varepsilon^{(0)}$ might have an $r$-dependent phase and satisfies
\begin{equation}
\gamma_{\hat{r}}\varepsilon^{(0)}=\varepsilon^{(0)*}\qquad
\textrm{and}\qquad
\gamma_{\hat{\rho}}\varepsilon^{(0)}=i\kappa\varepsilon^{(0)*}\,
.
\end{equation}
\indent Using the projector \eqref{gamma_rho_pro} in equation \eqref{dPsi_mu_eq}, we find
\begin{equation}
e^{i\Lambda}=\frac{\mc{W}}{A'+\frac{i\kappa}{\ell}e^{-A}}\,
.\label{complex_W_phase}
\end{equation}
We can then use the projector \eqref{gamma_r_pro} with this phase in the $\delta\chi^{IJK}=0$ conditions and find the BPS equations for the scalar fields. The resulting BPS equations can be written in terms of the superpotential as
\begin{eqnarray}
\zeta'&=&-\frac{1}{3}\frac{A'}{W}\frac{\pd W}{\pd \zeta}-\frac{1}{3}\frac{2}{\sinh(2\zeta)}\frac{\kappa e^{-A}}{W\ell}\frac{\pd W}{\pd \chi},\\
\chi'&=&-\frac{1}{3}\frac{A'}{W}\frac{4}{\sinh^2(2\zeta)}\frac{\pd W}{\pd \chi}+\frac{1}{3}\frac{2}{\sinh(2\zeta)}\frac{\kappa e^{-A}}{W\ell}\frac{\pd W}{\pd \zeta},\\
\xi'&=&-\frac{1}{4}\frac{A'}{W}\frac{\pd W}{\pd \xi}-\frac{1}{4}\frac{2}{\sinh(2\xi)}\frac{\kappa e^{-A}}{W\ell}\frac{\pd W}{\pd \phi},\\
\phi'&=&-\frac{A'}{W}\frac{1}{\sinh^2(2\xi)}\frac{\pd W}{\pd \phi}+\frac{1}{4}\frac{2}{\sinh(2\xi)}\frac{\kappa e^{-A}}{W\ell}\frac{\pd W}{\pd \xi}\, .
\end{eqnarray}
As expected, these equations reduce to those of the RG flows studied in \cite{Guariano} in the limit $\ell\rightarrow \infty$. It can also be shown that these equations satisfy the corresponding field equations. The complete solutions can be obtained by solving these equations together with \eqref{dPsi_BPS_eq}.
\\
\indent Before giving Janus solutions, we note that the constant $\kappa=\pm 1$ corresponds to the chiralities of the Killing spinor on the two-dimensional defects dual to the $AdS_3$ slices. This can be seen by using $\gamma_5\epsilon=\epsilon$ which implies
\begin{equation}
\gamma^0\gamma^1\epsilon=\kappa\epsilon\, .
\end{equation}
In the present case, we will have Janus solutions with only $N=1$ supersymmetry, or $(1,0)$ or $(0,1)$ superconformal symmetry on the defects, since the above BPS equations can be derived from either $\mc{W}_1$ or $\mc{W}_{\hat{1}}$. However, at the $SO(8)$ and $U(3)$ critical points, supersymmetry will enhance to $N=8$ and $N=2$ respectively.  

%%%%%%%%%%%%%%%%%%%%%%%%%%%%%%%%%%%%%%%%%%%%%%%%%%%%%%%%%%%%%%%%%%%%%%%%%%%%%%%%%%%%%%%%%%%%%%%%%%%%%%%%%%%%%%%%%%%%%%%%%%%%%%%%%%%%%%%%%
\section{Supersymmetric Janus solutions with $\omega=0$}\label{Janus_omega_0}
\begin{figure}
%[h!]
  \centering
  \begin{subfigure}[b]{0.4\linewidth}
    \includegraphics[width=\linewidth]{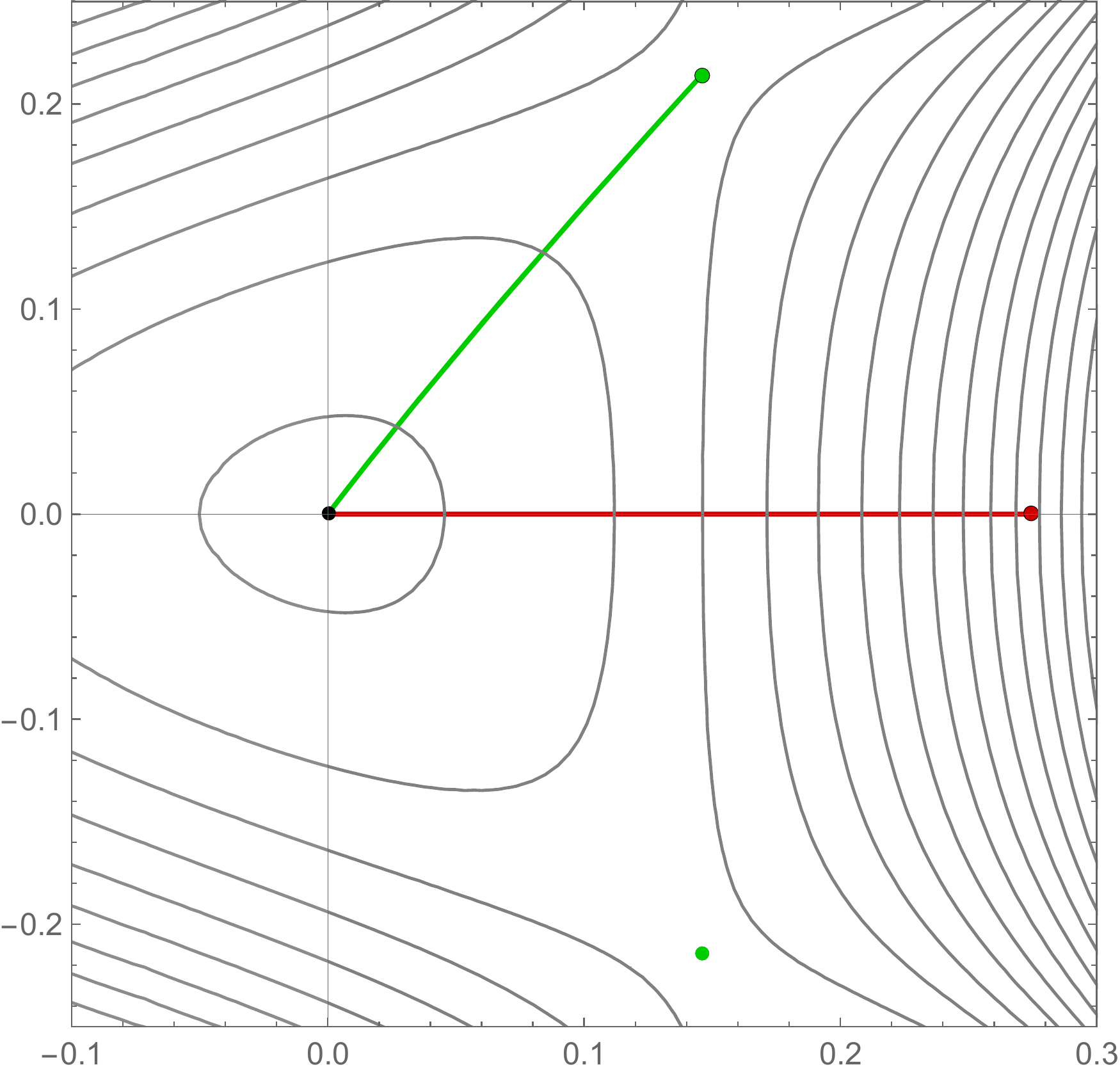}
  \caption{$(\zeta\cos\chi,\zeta\sin\chi)$ plane}
  \end{subfigure}\qquad\quad
  \begin{subfigure}[b]{0.4\linewidth}
    \includegraphics[width=\linewidth]{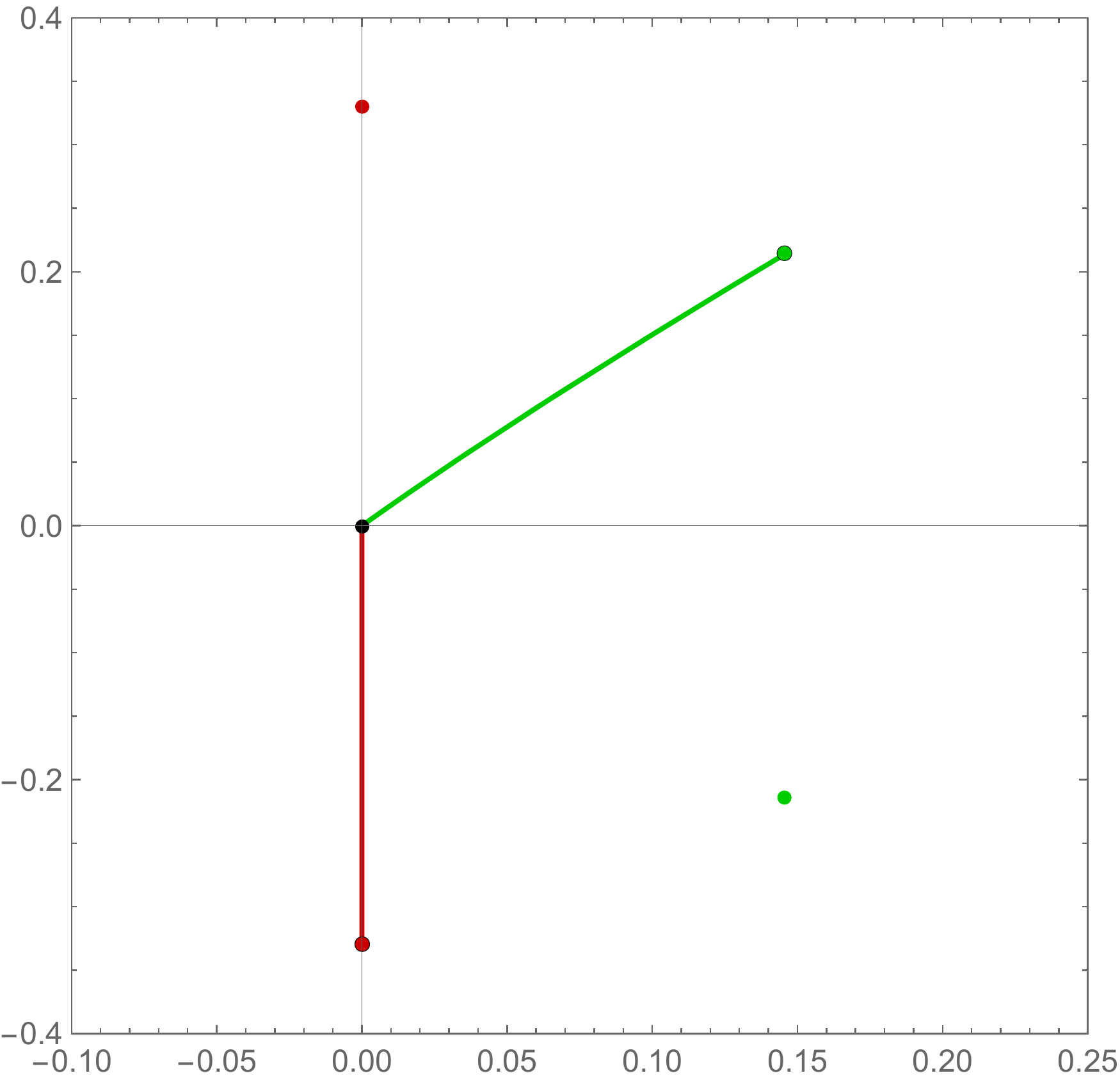}
  \caption{$(\xi\cos\phi,\xi\sin\phi)$ plane}
  \end{subfigure}
  \caption{$SO(8)/SO(8)$ Janus solutions that flow to $G_2$ and $U(3)$ critical points are given respectively by green and red lines on the contour plot of the superpotential. The two green and red dots represent equivalent $G_2$ and $U(3)$ critical points.} 
  \label{direct_O_0}
\end{figure}

\begin{figure}
%[h!]
  \centering
  \begin{subfigure}[b]{0.45\linewidth}
    \includegraphics[width=\linewidth]{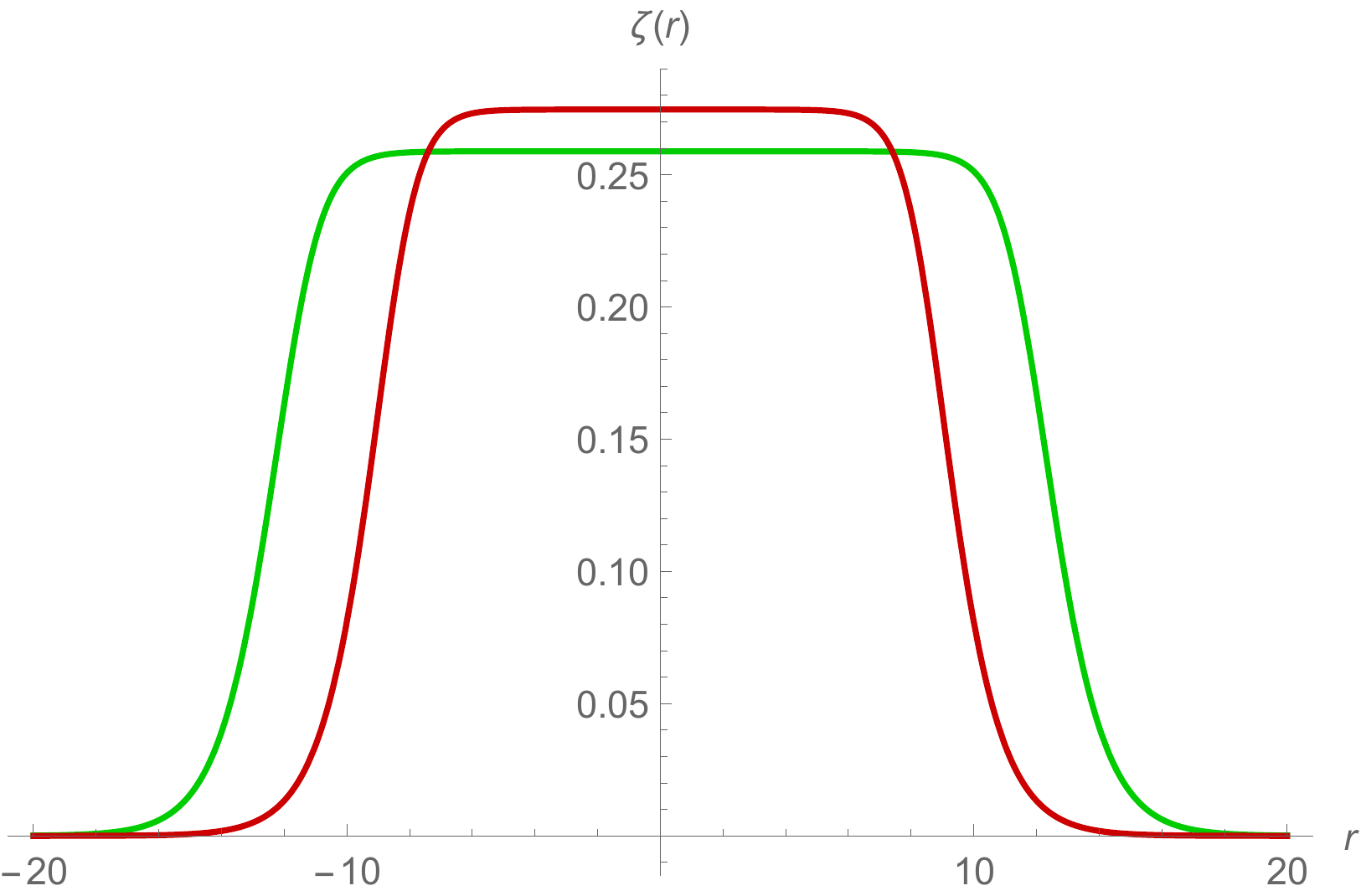}
  \caption{$\zeta(r)$}
  \end{subfigure}
  \begin{subfigure}[b]{0.45\linewidth}
    \includegraphics[width=\linewidth]{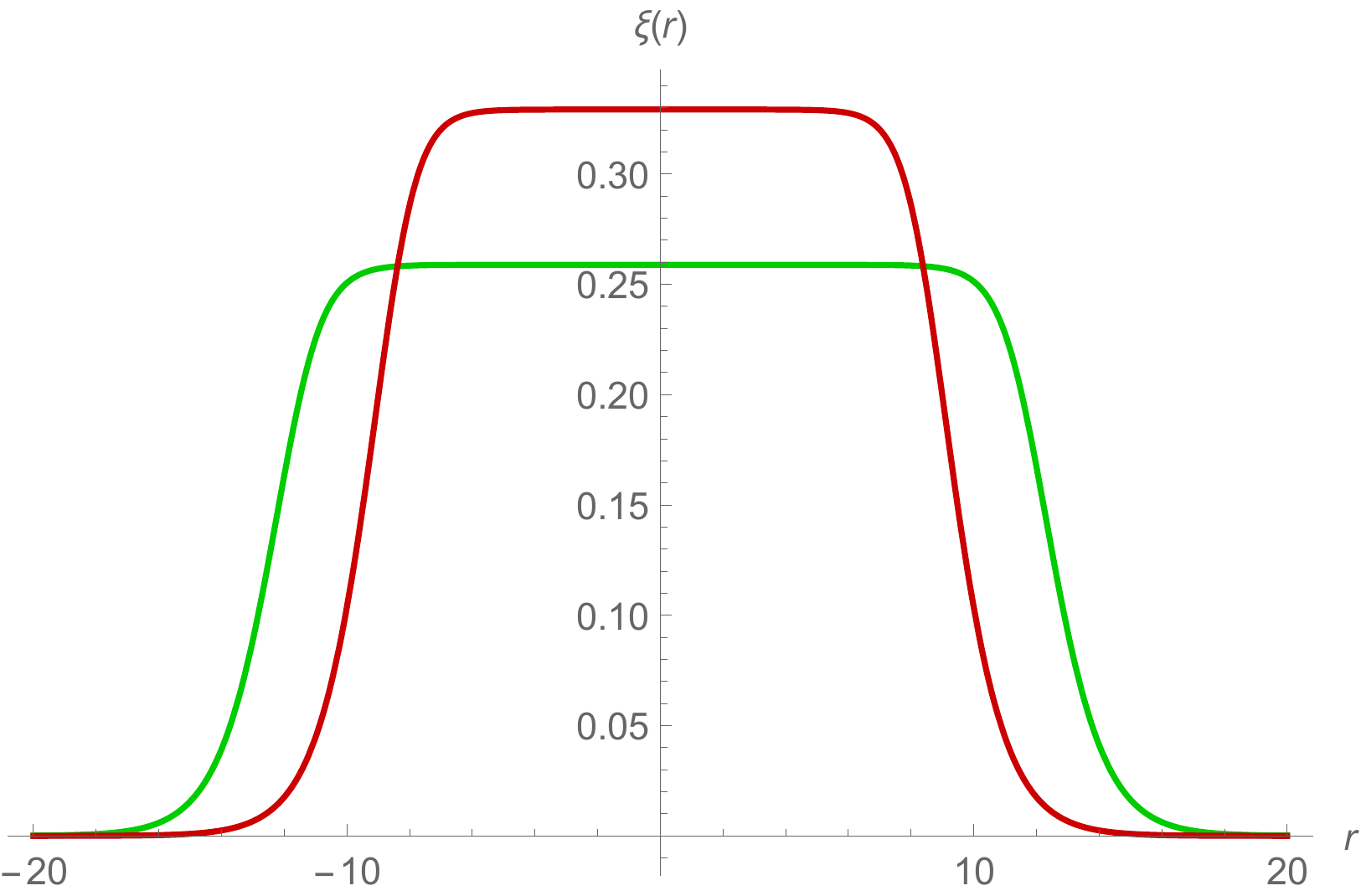}
  \caption{$\xi(r)$}
  \end{subfigure}\\
  \begin{subfigure}[b]{0.45\linewidth}
    \includegraphics[width=\linewidth]{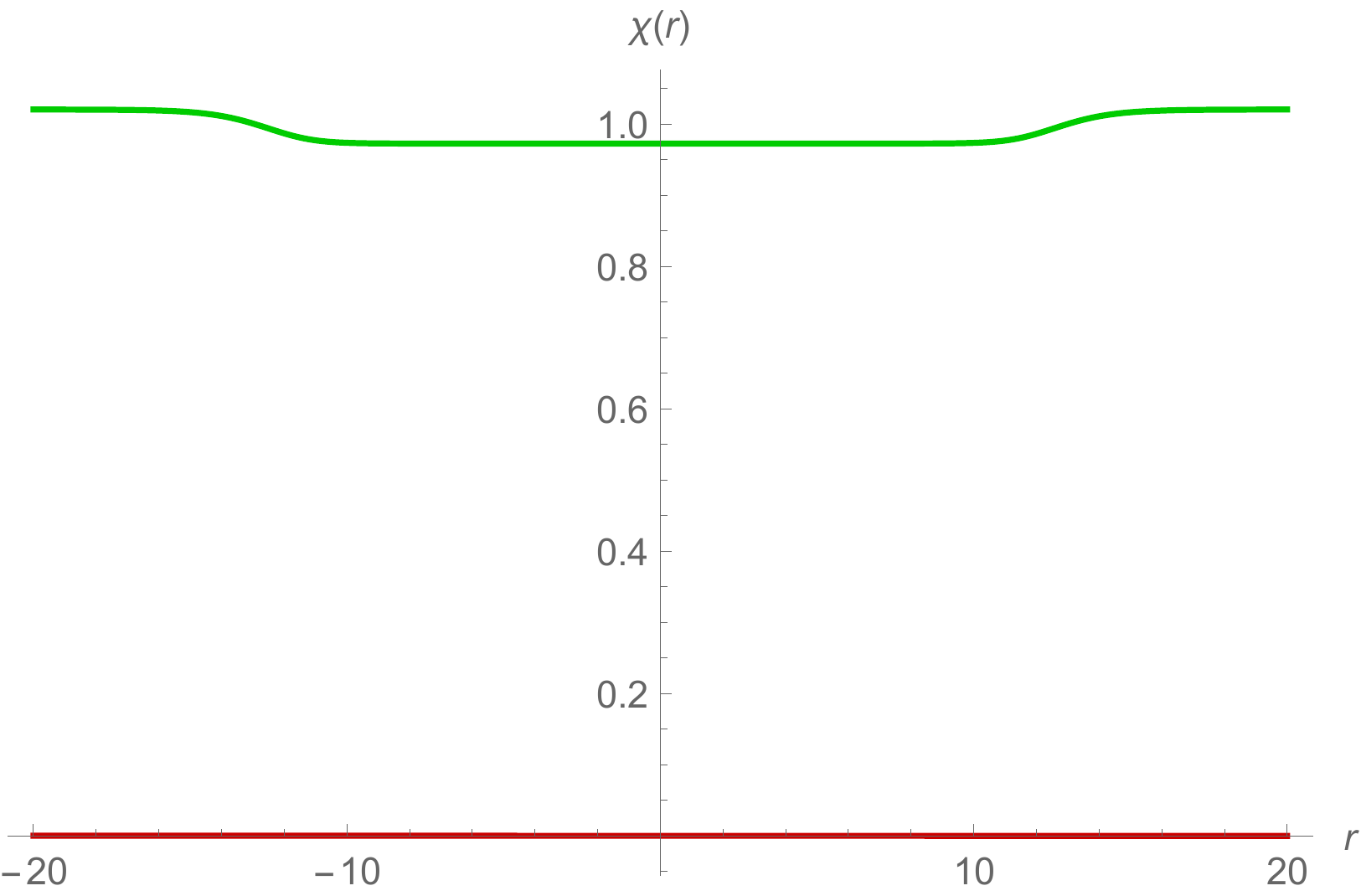}
  \caption{$\chi(r)$}
  \end{subfigure}
  \begin{subfigure}[b]{0.45\linewidth}
    \includegraphics[width=\linewidth]{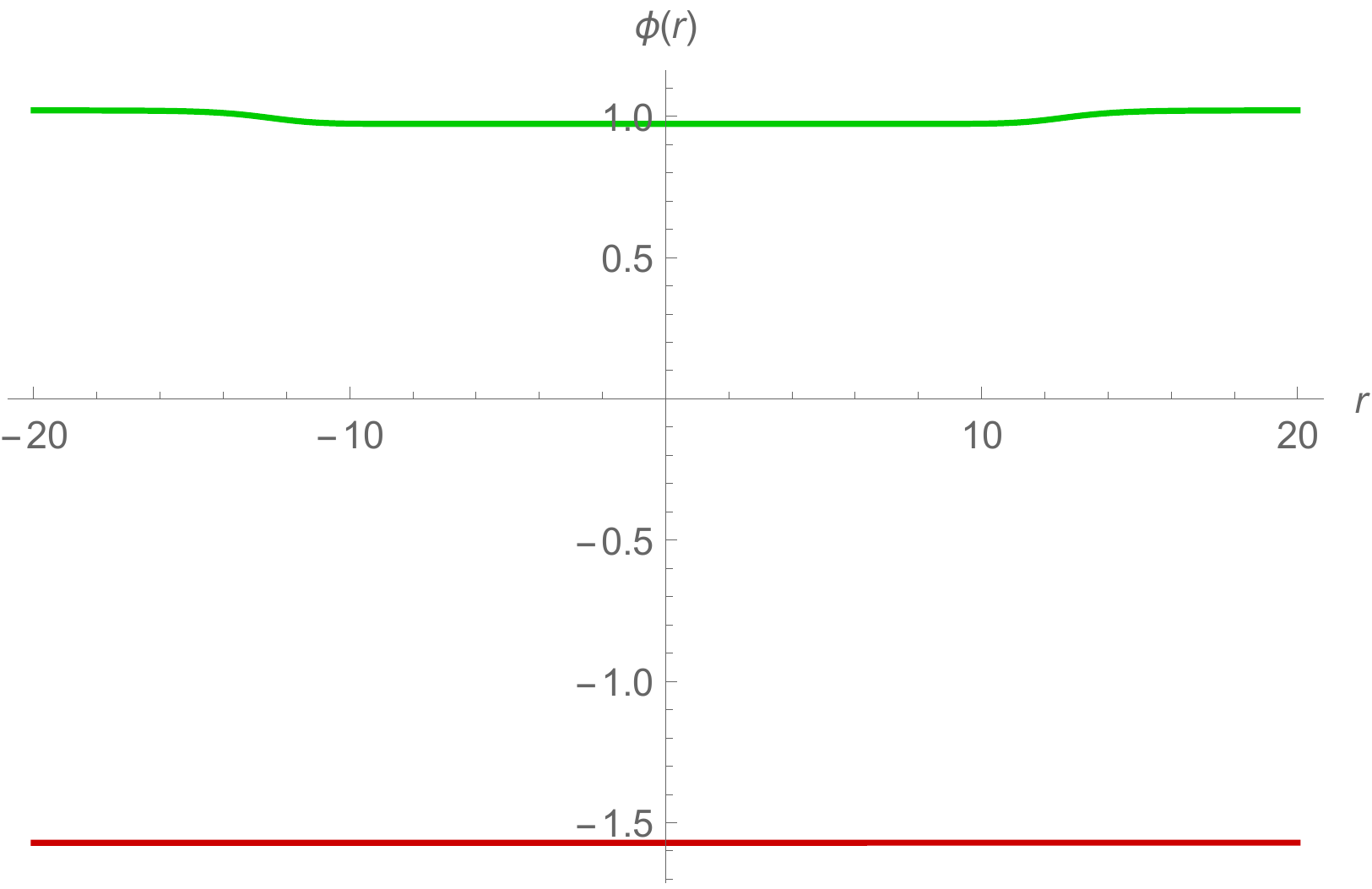}
  \caption{$\phi(r)$}
  \end{subfigure}\\
   \begin{subfigure}[b]{0.45\linewidth}
    \includegraphics[width=\linewidth]{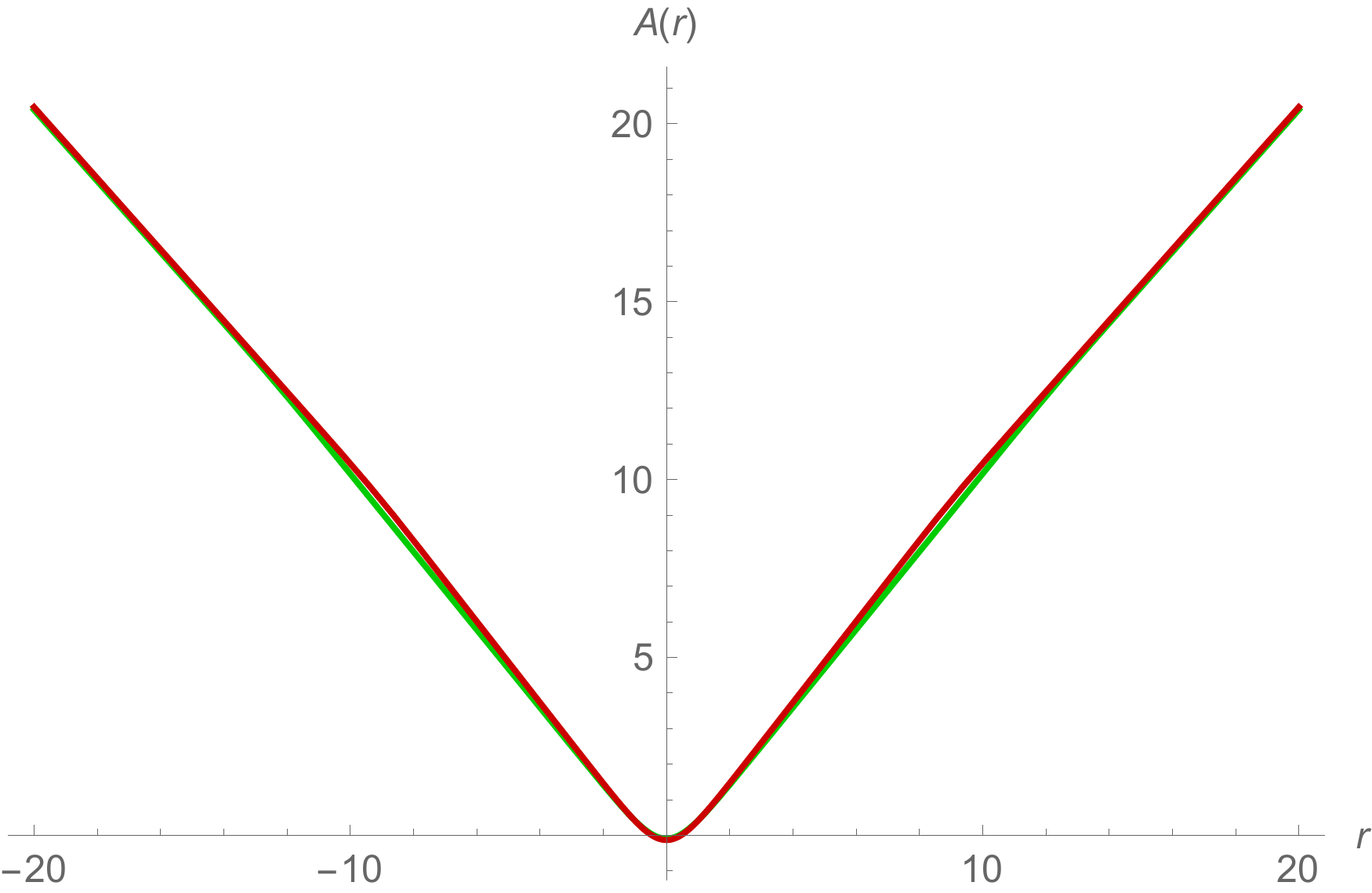}
  \caption{$A(r)$}
   \end{subfigure} 
 \begin{subfigure}[b]{0.45\linewidth}
    \includegraphics[width=\linewidth]{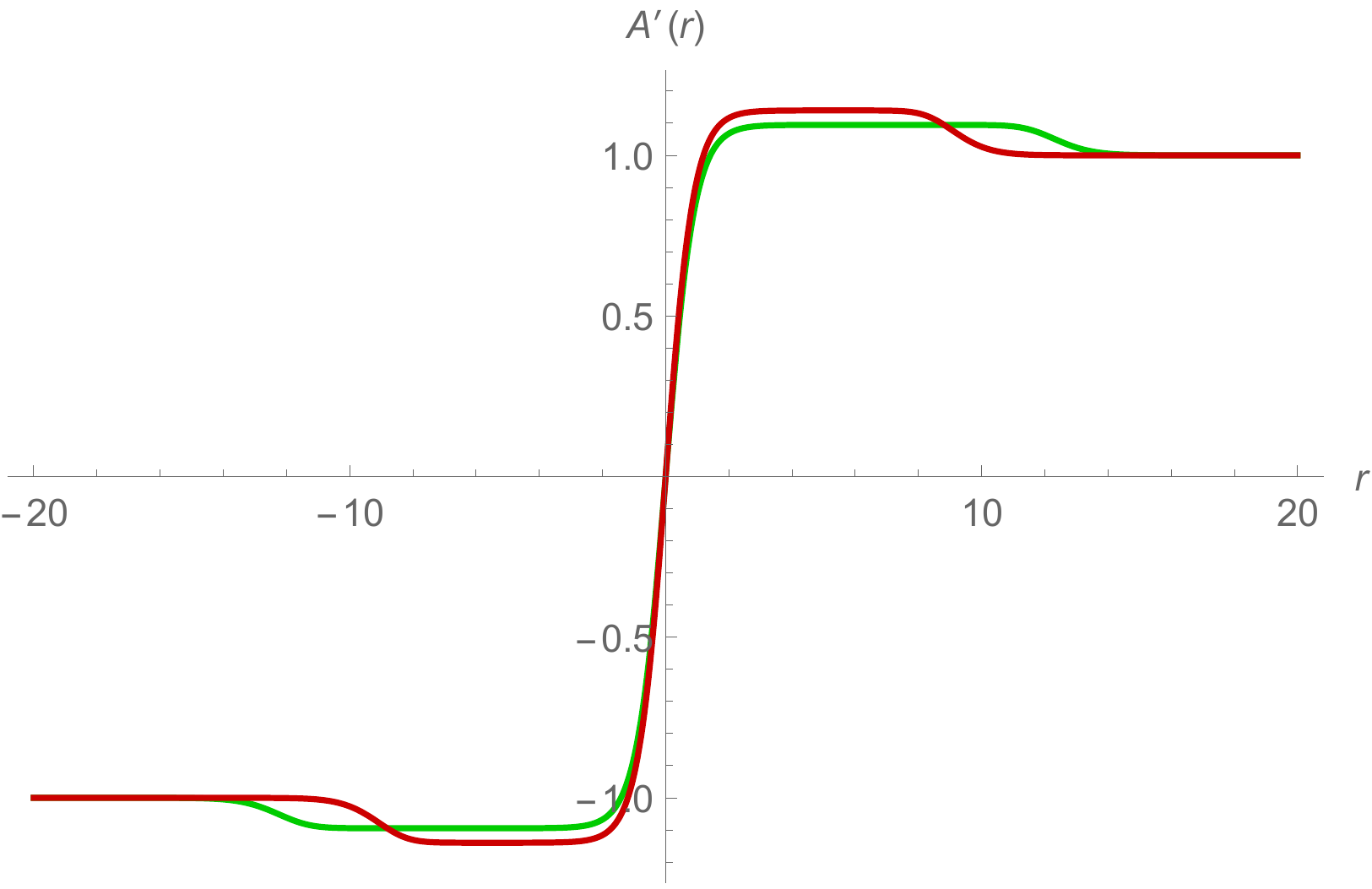}
  \caption{$A'(r)$}
   \end{subfigure} 
  \caption{Profiles of scalar fields $(\zeta,\xi,\chi,\phi)$ and the warped factor $A$ as functions of the radial coordinate $r$ for $SO(8)/SO(8)$ Janus solutions that flow to $G_2$ and $U(3)$ critical points are shown by green and red lines, respectively.}
  \label{derect_Profile}
\end{figure}

In this section, we first consider supersymmetric Janus solutions in the $SO(8)$ gauged supergravity with $\omega=0$. A number of solutions in the truncation with only two scalars non-vanishing have already been given in \cite{warner_Janus}. However, the solutions involving the $N=2$ supersymmetric $AdS_4$ critical point with $SU(3)\times U(1)$ symmetry have not been studied since this vacuum does not arise in that truncation. In this work, we consider the full $SU(3)$ invariant scalar sector, so it is possible to accommodate this type of solutions. Accordingly, we first give solutions with $\omega=0$ for which only $SO(8)$, $G_2$ and $SU(3)\times U(1)$ $AdS_4$ critical points with $N=8,1,2$ supersymmetries exist.
\\
\indent The resulting BPS equations are highly complicated to look for any analytic solutions. Therefore, we will perform a numerical analysis in finding supersymmetric Janus solutions. In subsequent analysis, we will choose the following numerical values
\begin{equation}
g=\frac{1}{\sqrt{2}},\qquad \ell=1,\qquad \kappa=1\, .
\end{equation}
The $A'(r)$ equation involves taking a square root giving rise to a branch cut. To avoid this and work with a smooth numerical analysis, we follow the procedure carried out in \cite{warner_Janus} and instead solve the second order field equations. This process begins with fixing a turning point of $A$ such that $A'(r_0)=0$ for particular values of $\zeta_0$, $\xi_0$, $\chi_0$ and $\phi_0$. We will conveniently choose $r_0=0$ as in \cite{warner_Janus}. We then determine the values of $A(0)$, $\zeta'(0)$, $\xi'(0)$, $\chi'(0)$ and $\phi'(0)$ using the previously obtained BPS equations. This provides a full set of initial conditions to solve the second order field equations. After numerically integrating to find the solution, we check whether the resulting solution satisfies the BPS equations. For convenience, we will denote a solution interpolating between $AdS_4$ critical points with $G$ and $G'$ symmetries on the two sides of the interface by $G/G'$ Janus. 
\\
\indent As in \cite{warner_Janus}, there are solutions describing conformal interfaces between $SO(8)$ symmetric phases. These solutions have been studied extensively in \cite{warner_Janus}, so will not repeat them here but mainly focus on solutions involving $G_2$ and $U(3)$ critical points. We first remark that there exist $SO(8)/SO(8)$ Janus solutions that flow to $G_2$ and $U(3)$ critical points. These solutions are shown in the contour plot of the superpotential in figure \ref{direct_O_0}. In all the contour plots, as in \cite{Varella_N8_flow}, we denote $SO(8)$, $G_2$ and $U(3)$ critical points respectively by black, green and red dots while open dots represent turning points. Profiles of scalars and the warped factor $A$ as functions of the radial coordinate $r$ are shown in figure \ref{derect_Profile}. We also give the profile of $A'$ in order to make the $AdS_4$ critical points involving in the solution more transparent. These solutions have a very similar structure to the solutions given in \cite{Minwoo_4DN8_Janus}. The only difference is that our solutions asymptotically interpolate between $SO(8)$ conformal phases while those in \cite{Minwoo_4DN8_Janus} describe solutions between non-conformal or super Yang-Mills phases. We also note that for this type of solutions the turning points are very close to the critical point to which the solutions flow. 
        
\begin{figure}
%[h!]
  \centering
  \begin{subfigure}[b]{0.4\linewidth}
    \includegraphics[width=\linewidth]{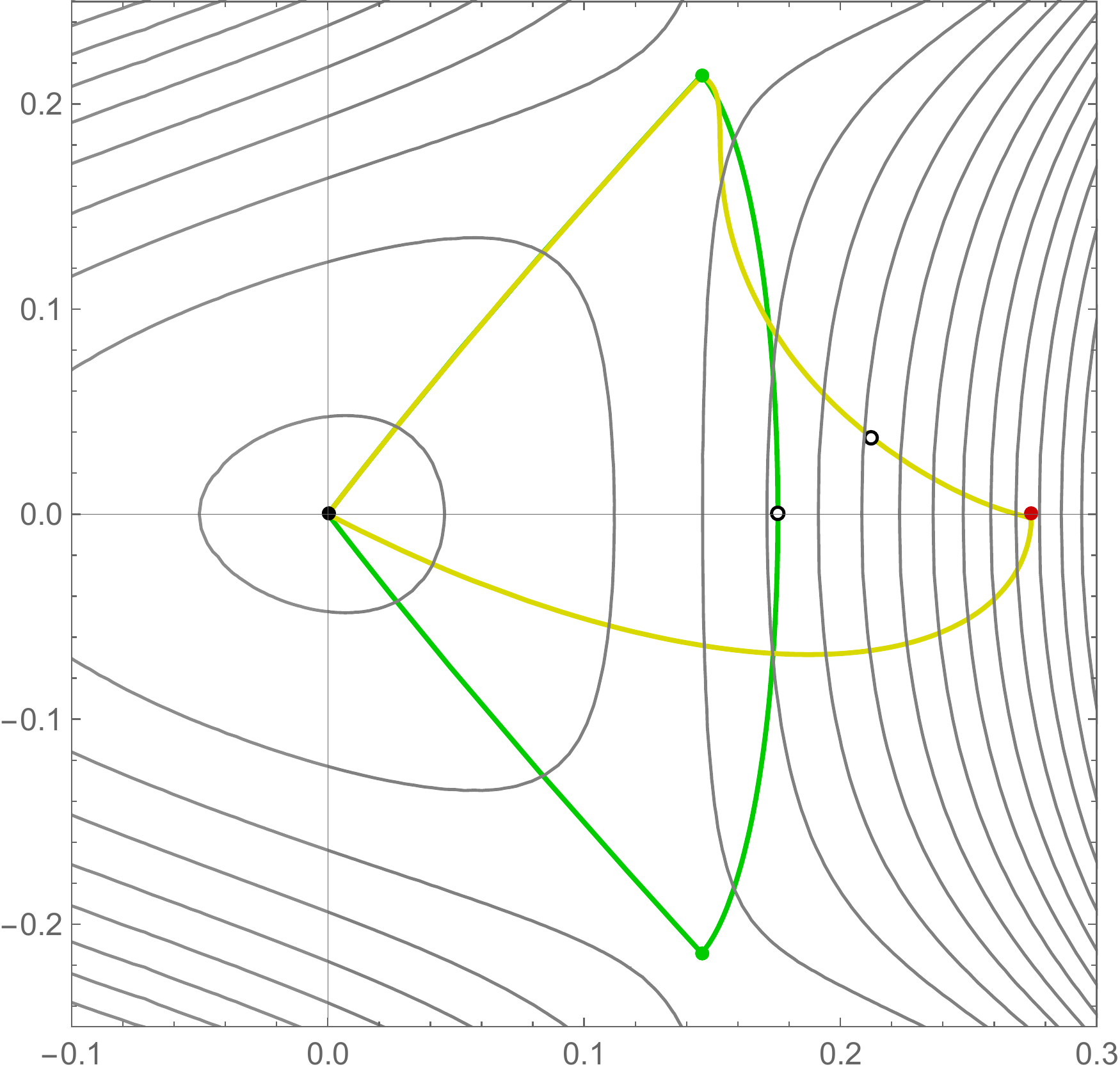}
  \caption{$(\zeta\cos\chi,\zeta\sin\chi)$ plane}
  \end{subfigure}\qquad\quad
  \begin{subfigure}[b]{0.4\linewidth}
    \includegraphics[width=\linewidth]{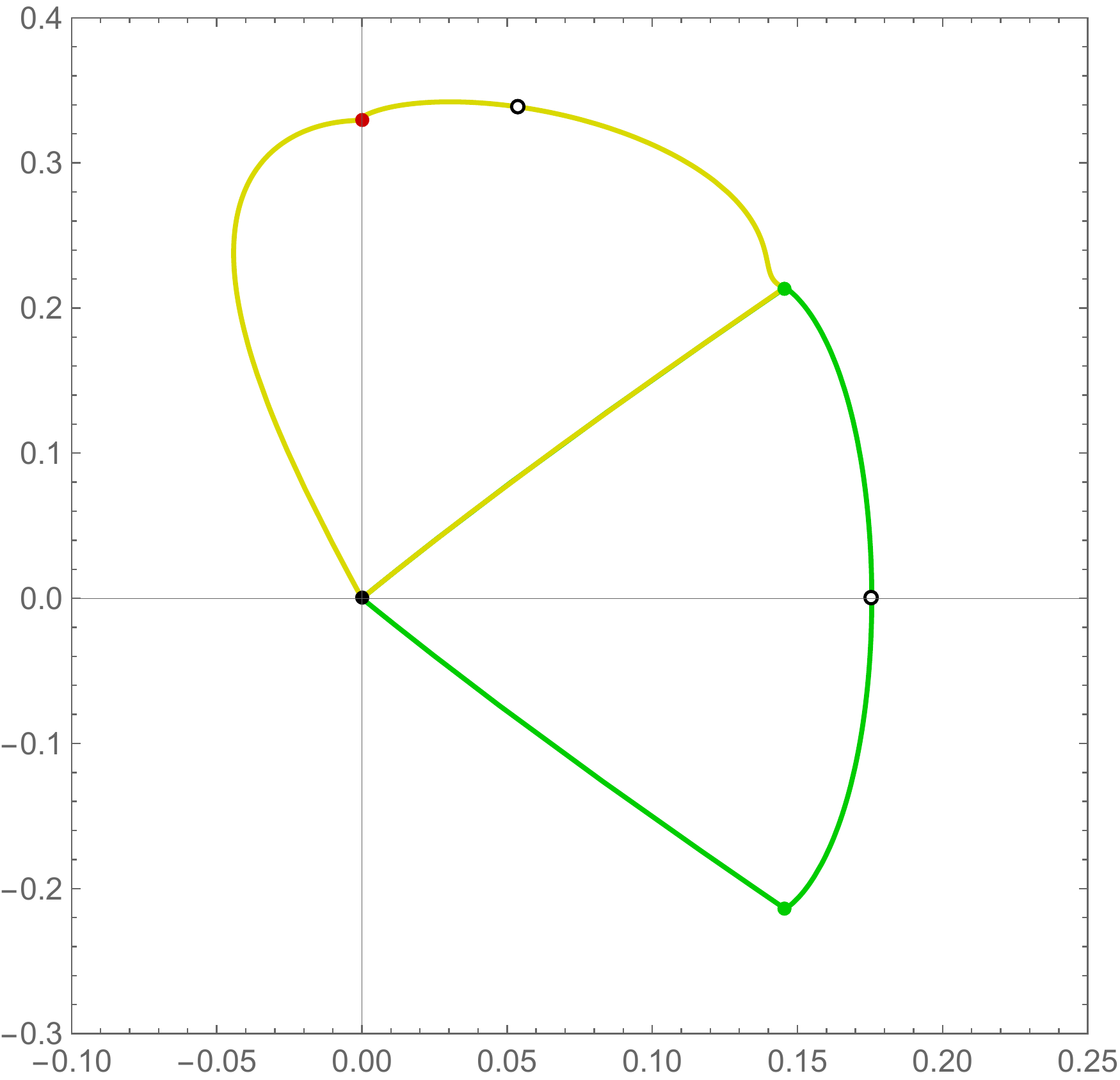}
  \caption{$(\xi\cos\phi,\xi\sin\phi)$ plane}
  \end{subfigure}
  \caption{Janus solutions describing $G_2/U(3)$ and $G_2/G_2$ interfaces are respectively shown by yellow and green lines on the contour plot of the superpotential.} 
  \label{inter_O_0}
\end{figure}

\begin{figure}
%[h!]
  \centering
  \begin{subfigure}[b]{0.45\linewidth}
    \includegraphics[width=\linewidth]{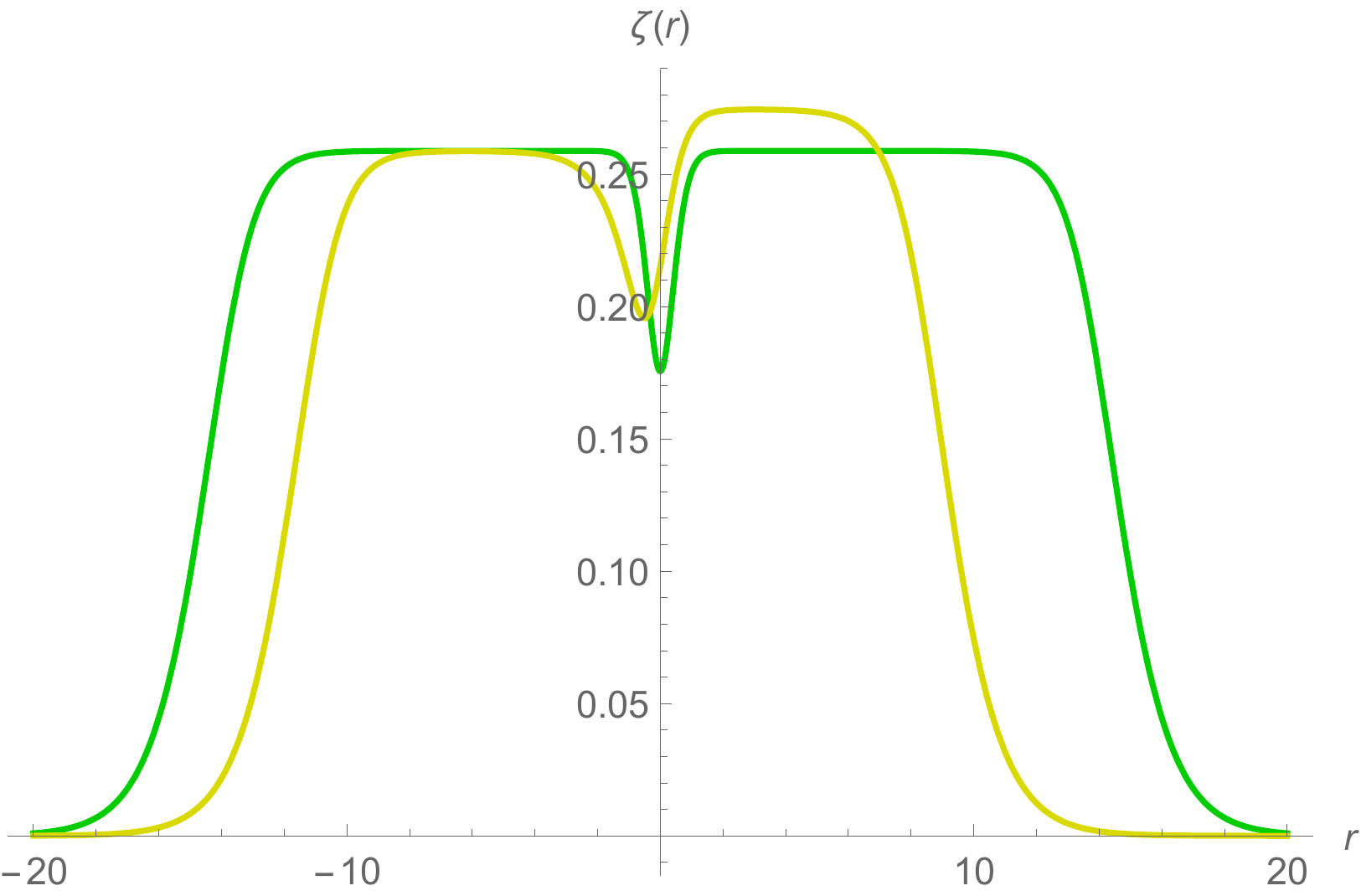}
  \caption{$\zeta(r)$}
  \end{subfigure}
  \begin{subfigure}[b]{0.45\linewidth}
    \includegraphics[width=\linewidth]{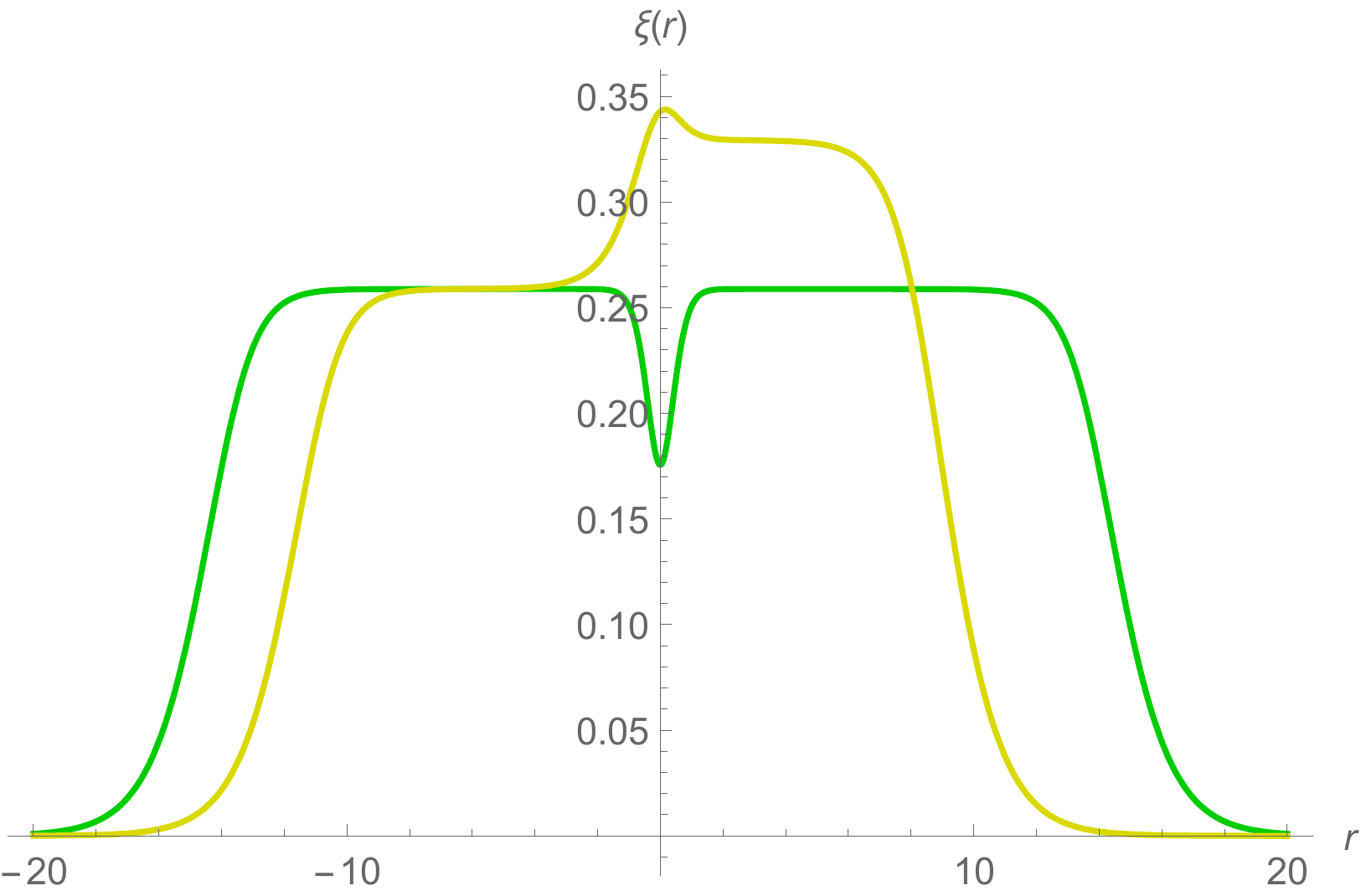}
  \caption{$\xi(r)$}
  \end{subfigure}\\
  \begin{subfigure}[b]{0.45\linewidth}
    \includegraphics[width=\linewidth]{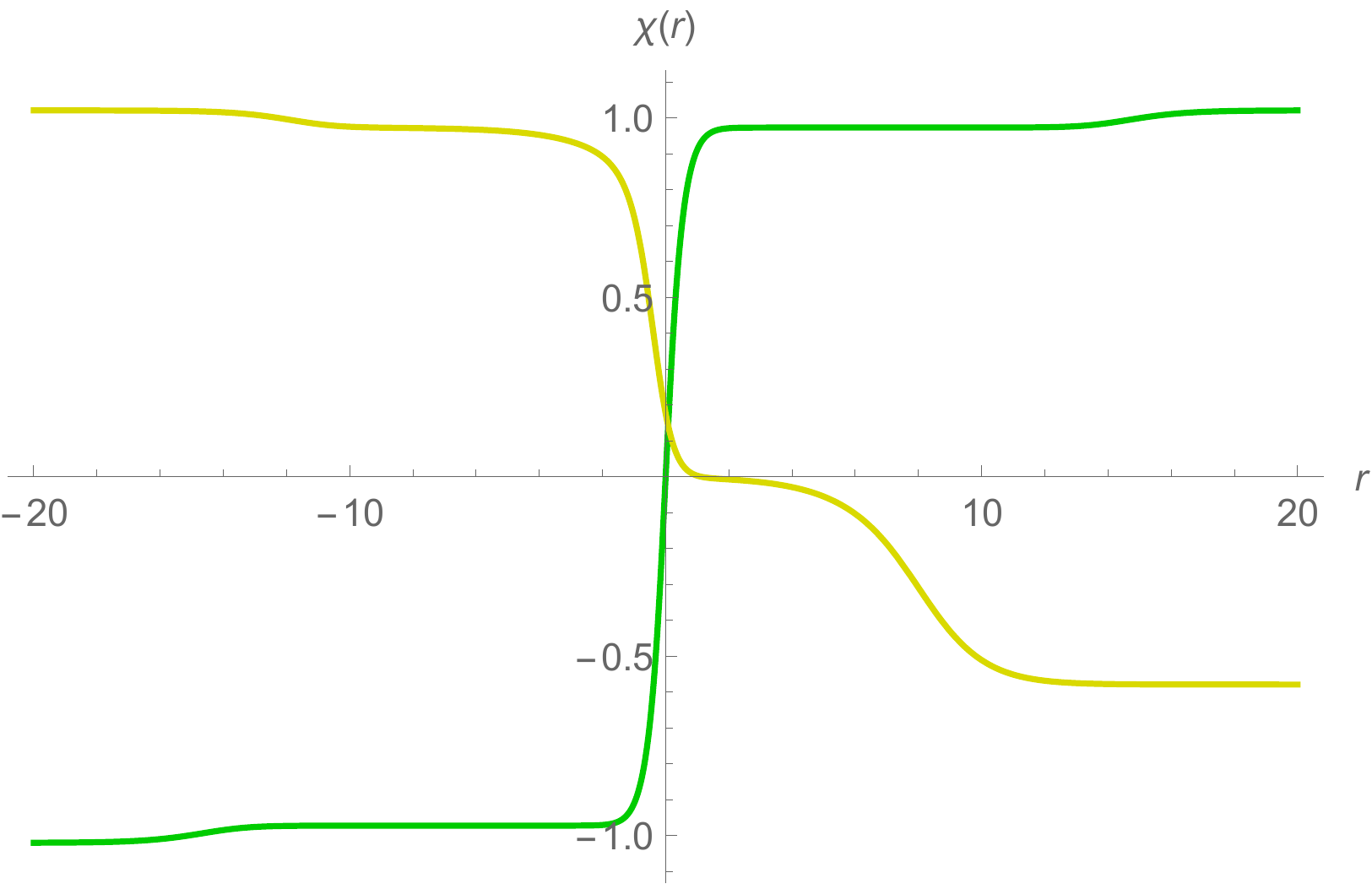}
  \caption{$\chi(r)$}
  \end{subfigure}
  \begin{subfigure}[b]{0.45\linewidth}
    \includegraphics[width=\linewidth]{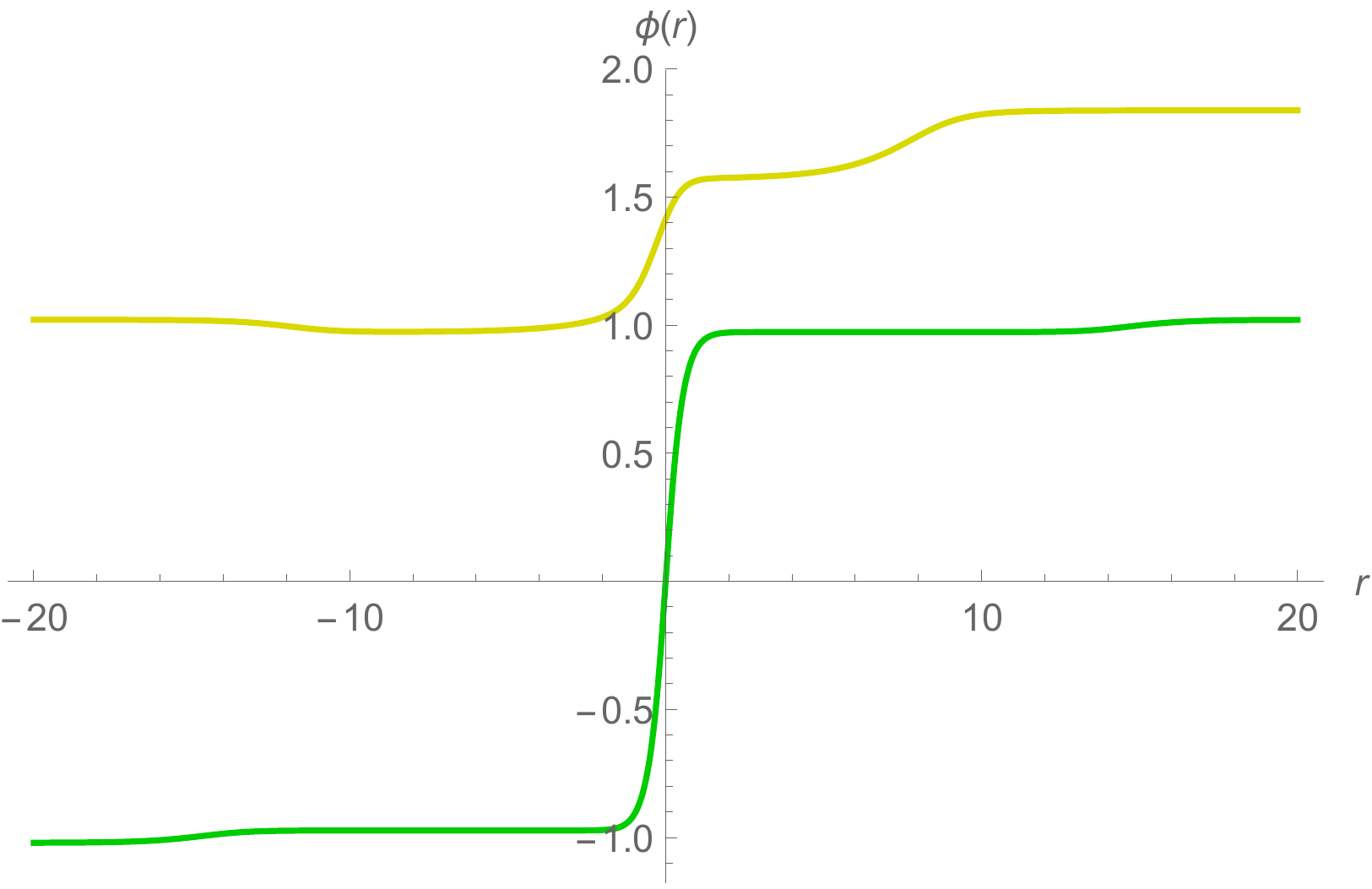}
  \caption{$\phi(r)$}
  \end{subfigure}\\
   \begin{subfigure}[b]{0.45\linewidth}
    \includegraphics[width=\linewidth]{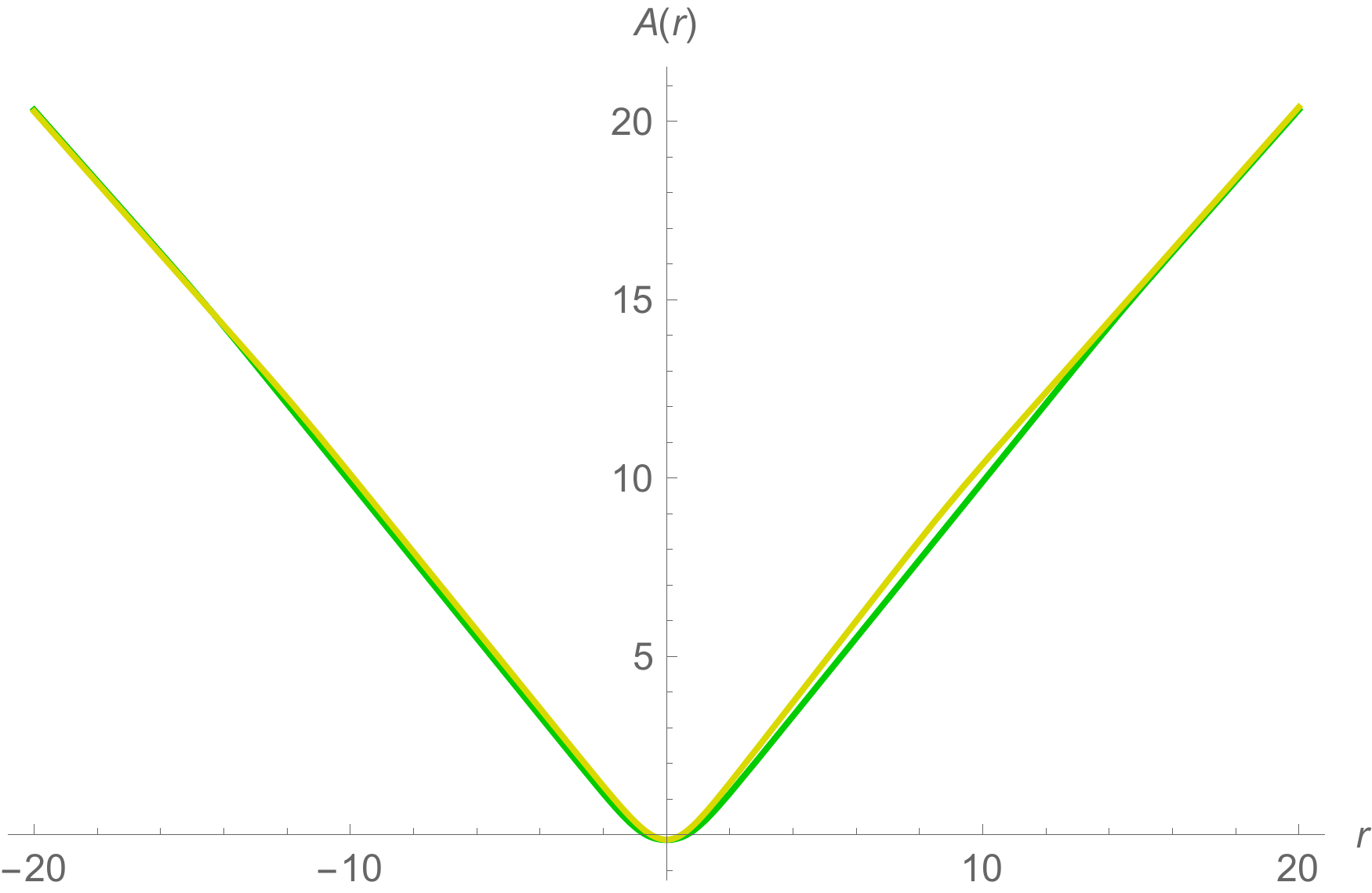}
  \caption{$A(r)$}
   \end{subfigure} 
 \begin{subfigure}[b]{0.45\linewidth}
    \includegraphics[width=\linewidth]{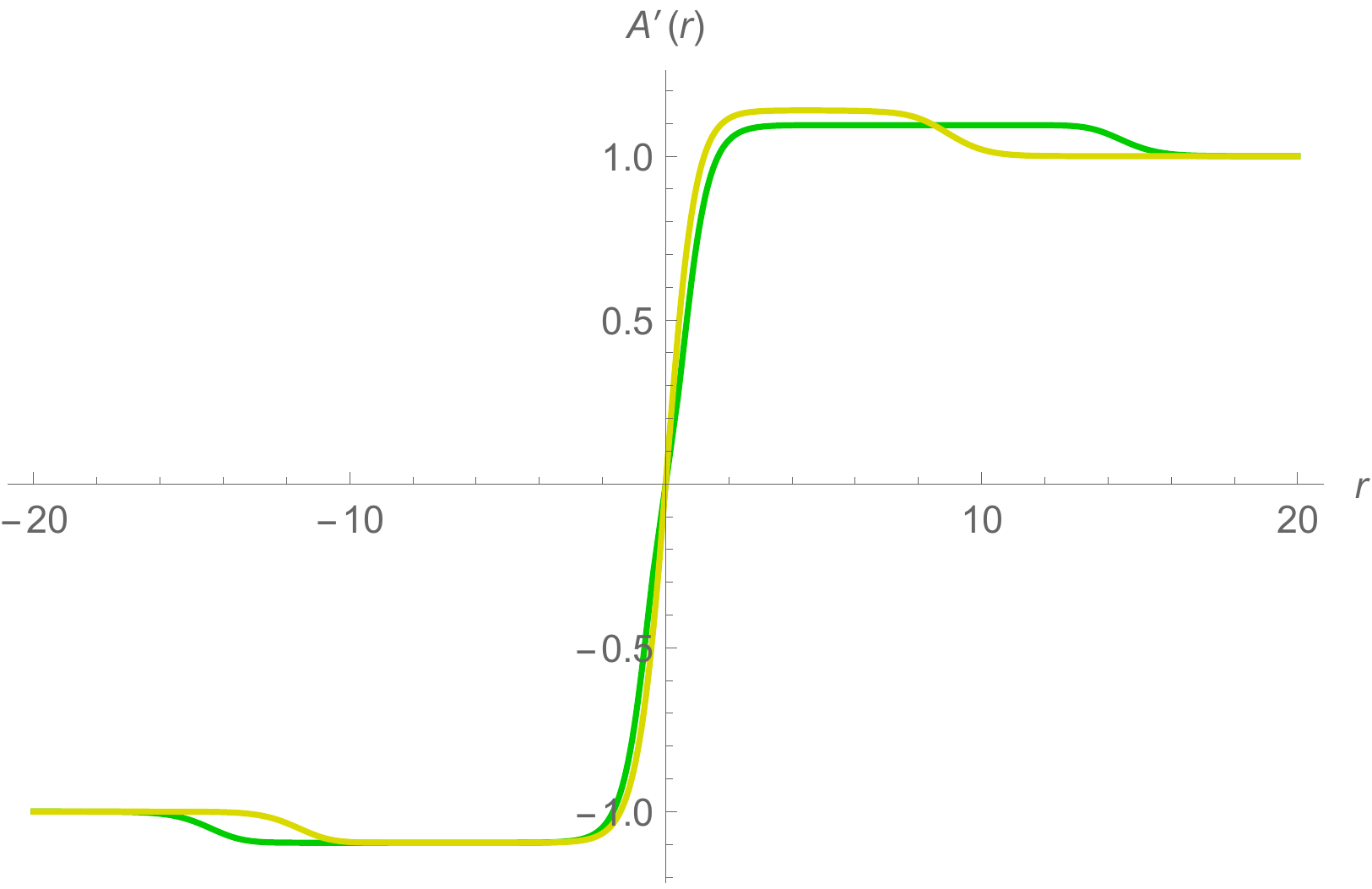}
  \caption{$A'(r)$}
   \end{subfigure} 
  \caption{Profiles of scalar fields $(\zeta,\xi,\chi,\phi)$ and the warped factor $A$ for $G_2/G_2$ and $G_2/U(3)$ Janus solutions as functions of the radial coordinate $r$ are shown by green and yellow lines, respectively.}
  \label{inter_Profile}
\end{figure}

\begin{figure}
%[h!]
  \centering
  \begin{subfigure}[b]{0.4\linewidth}
    \includegraphics[width=\linewidth]{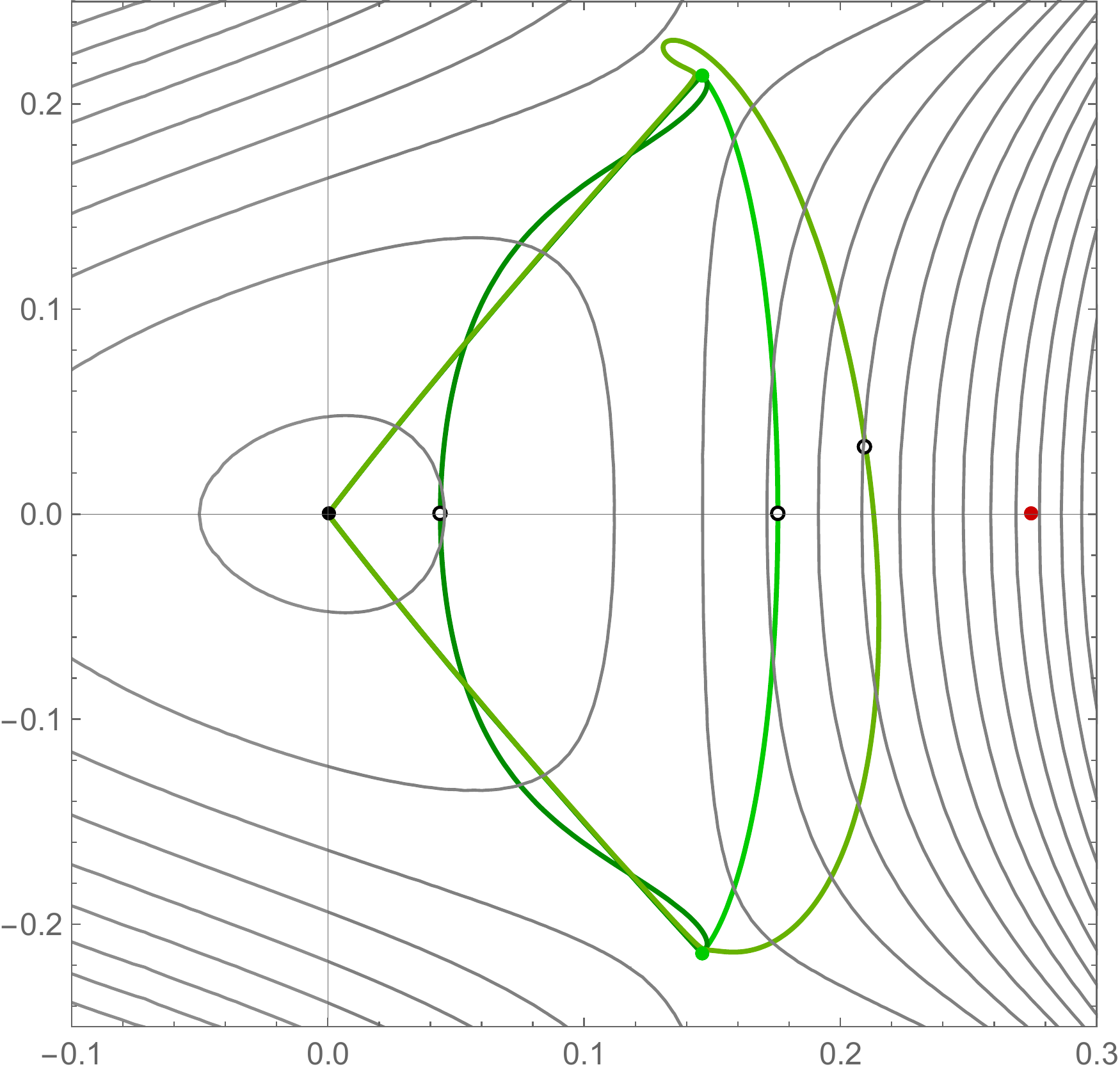}
  \caption{$(\zeta\cos\chi,\zeta\sin\chi)$ plane}
  \end{subfigure}\qquad\quad
  \begin{subfigure}[b]{0.4\linewidth}
    \includegraphics[width=\linewidth]{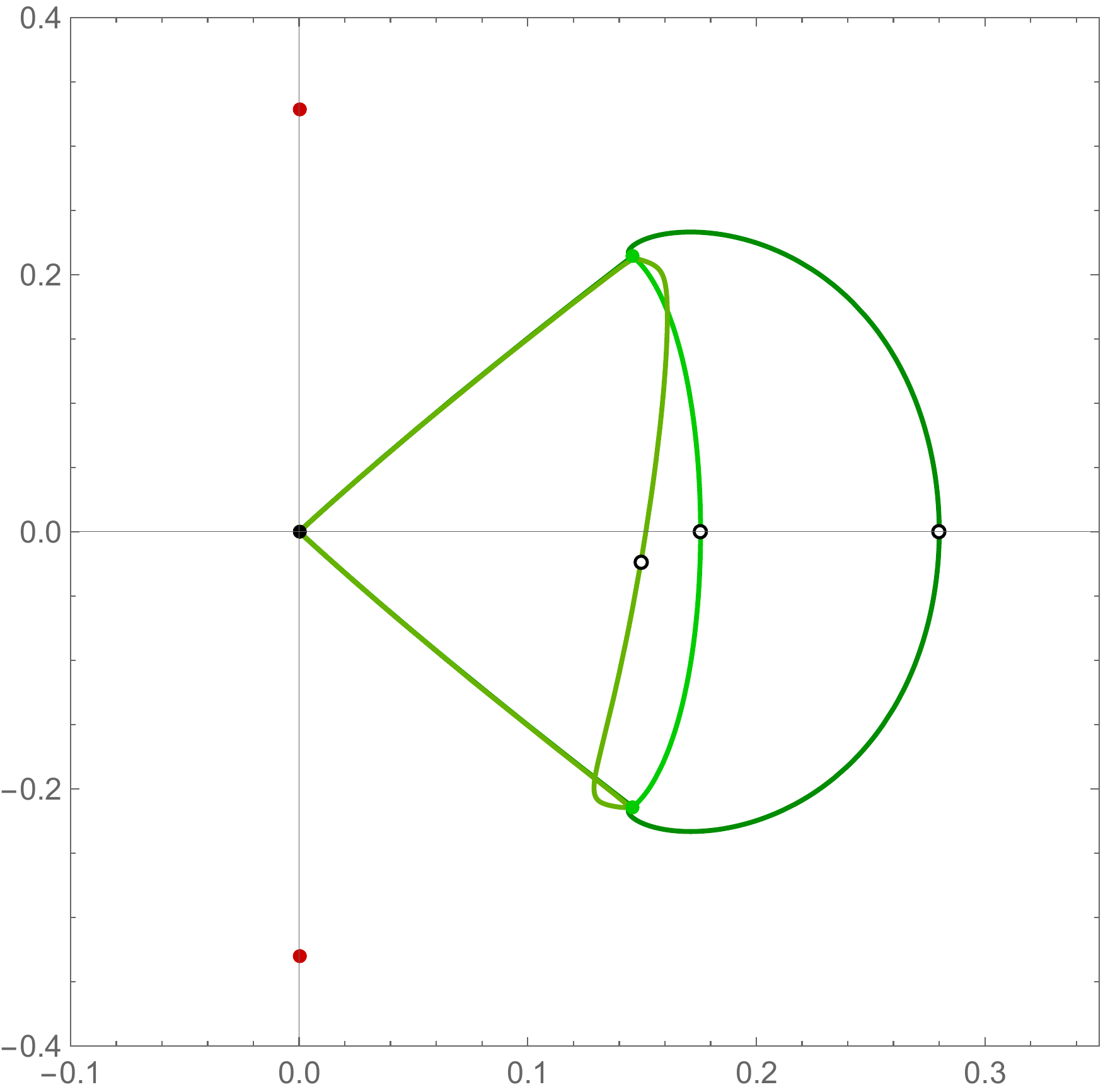}
  \caption{$(\xi\cos\phi,\xi\sin\phi)$ plane}
  \end{subfigure}
  \caption{Examples of a family of $G_2/G_2$ Janus solutions on the contour plot of the superpotential.} 
  \label{G2G2_O_0}
\end{figure}

\begin{figure}
%[h!]
  \centering
  \begin{subfigure}[b]{0.45\linewidth}
    \includegraphics[width=\linewidth]{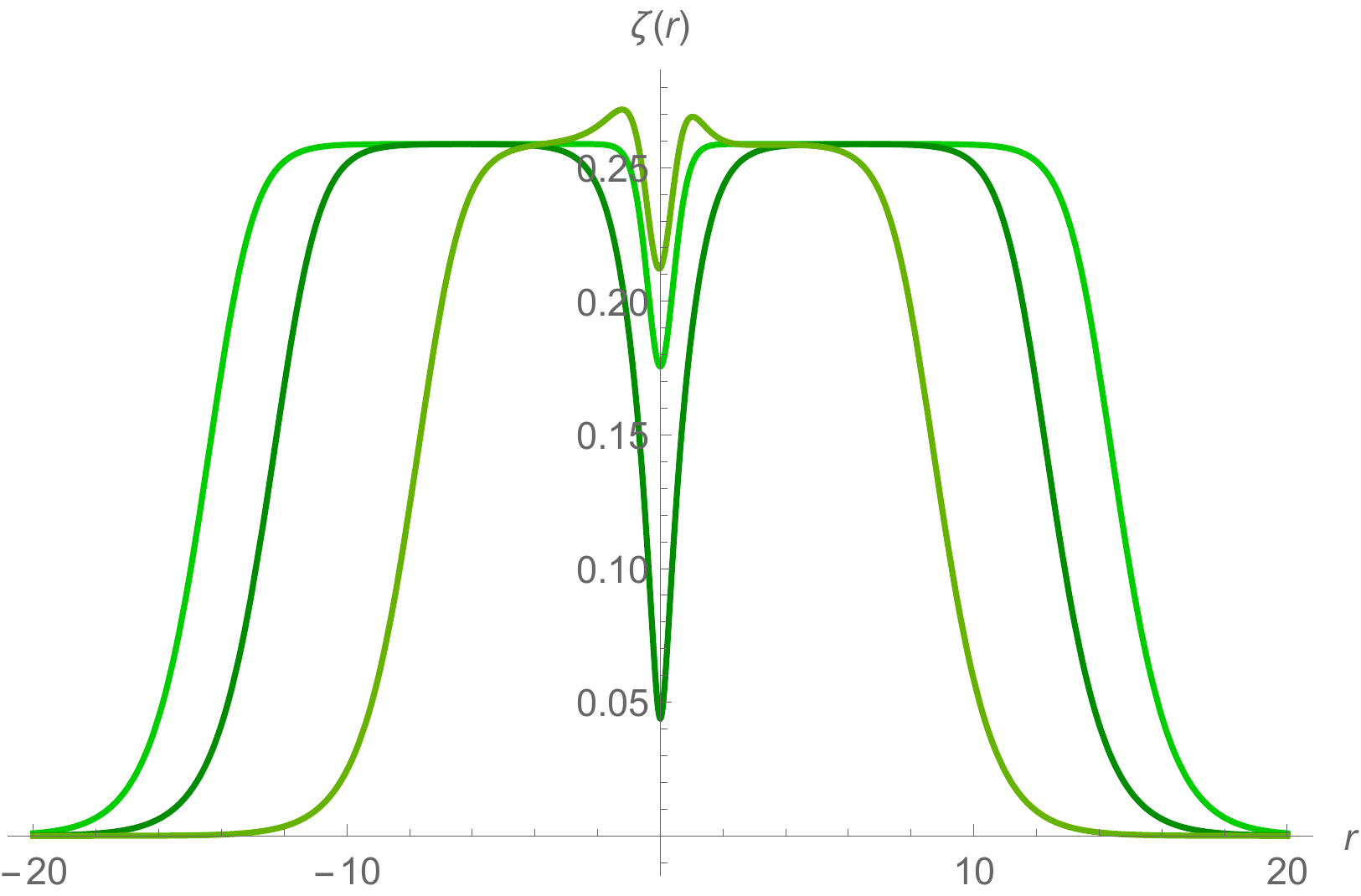}
  \caption{$\zeta(r)$}
  \end{subfigure}
  \begin{subfigure}[b]{0.45\linewidth}
    \includegraphics[width=\linewidth]{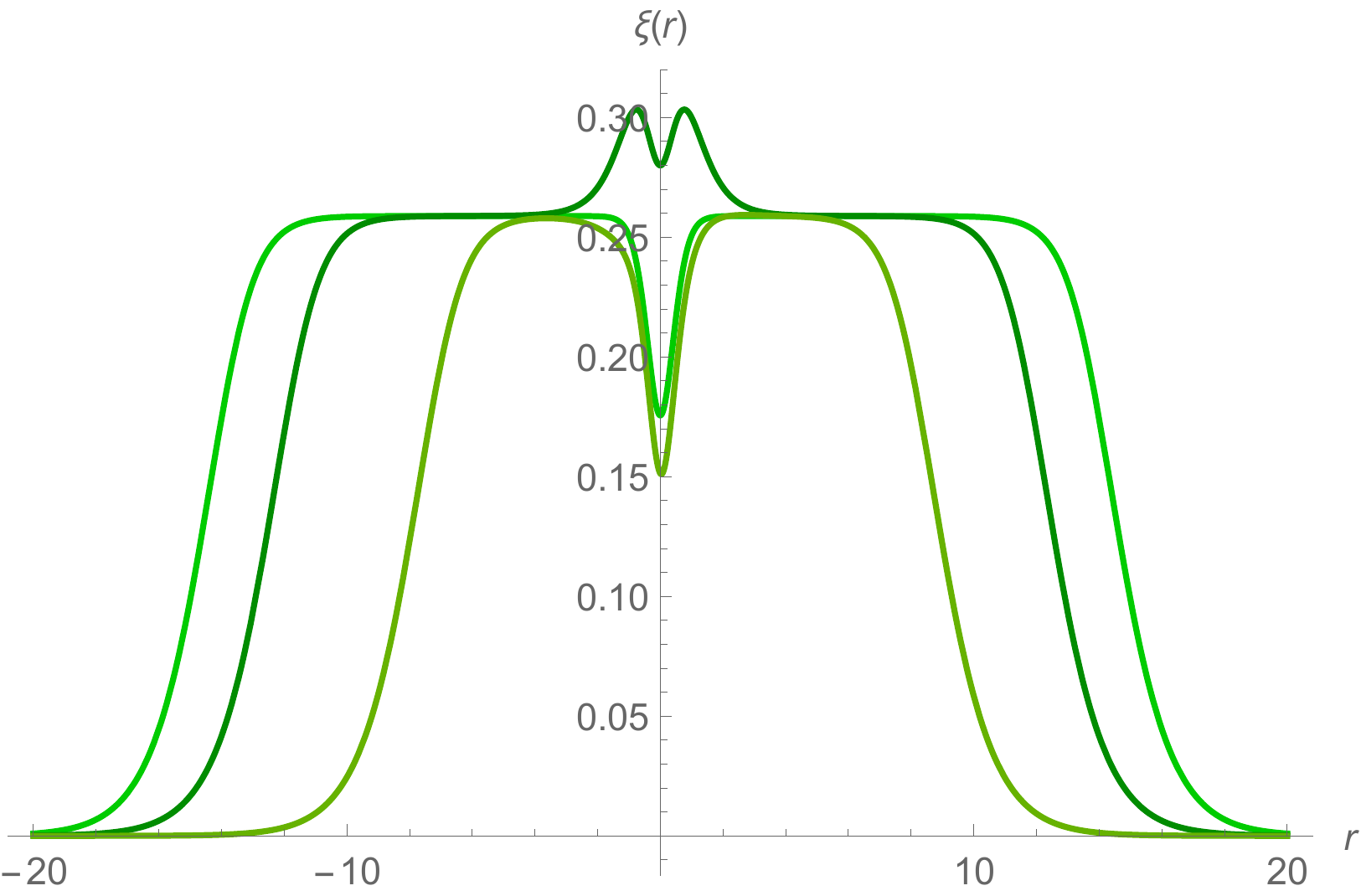}
  \caption{$\xi(r)$}
  \end{subfigure}\\
  \begin{subfigure}[b]{0.45\linewidth}
    \includegraphics[width=\linewidth]{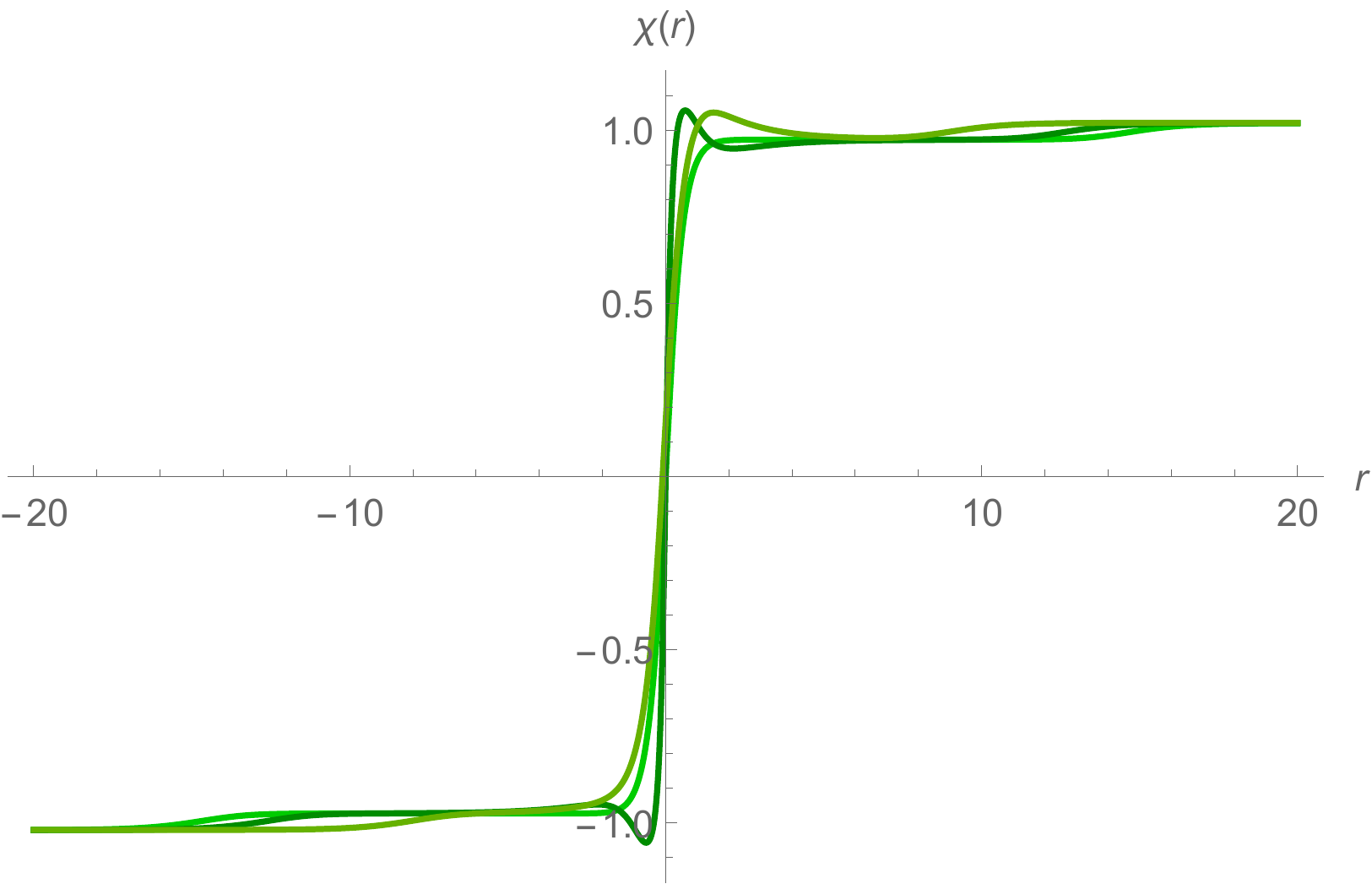}
  \caption{$\chi(r)$}
  \end{subfigure}
  \begin{subfigure}[b]{0.45\linewidth}
    \includegraphics[width=\linewidth]{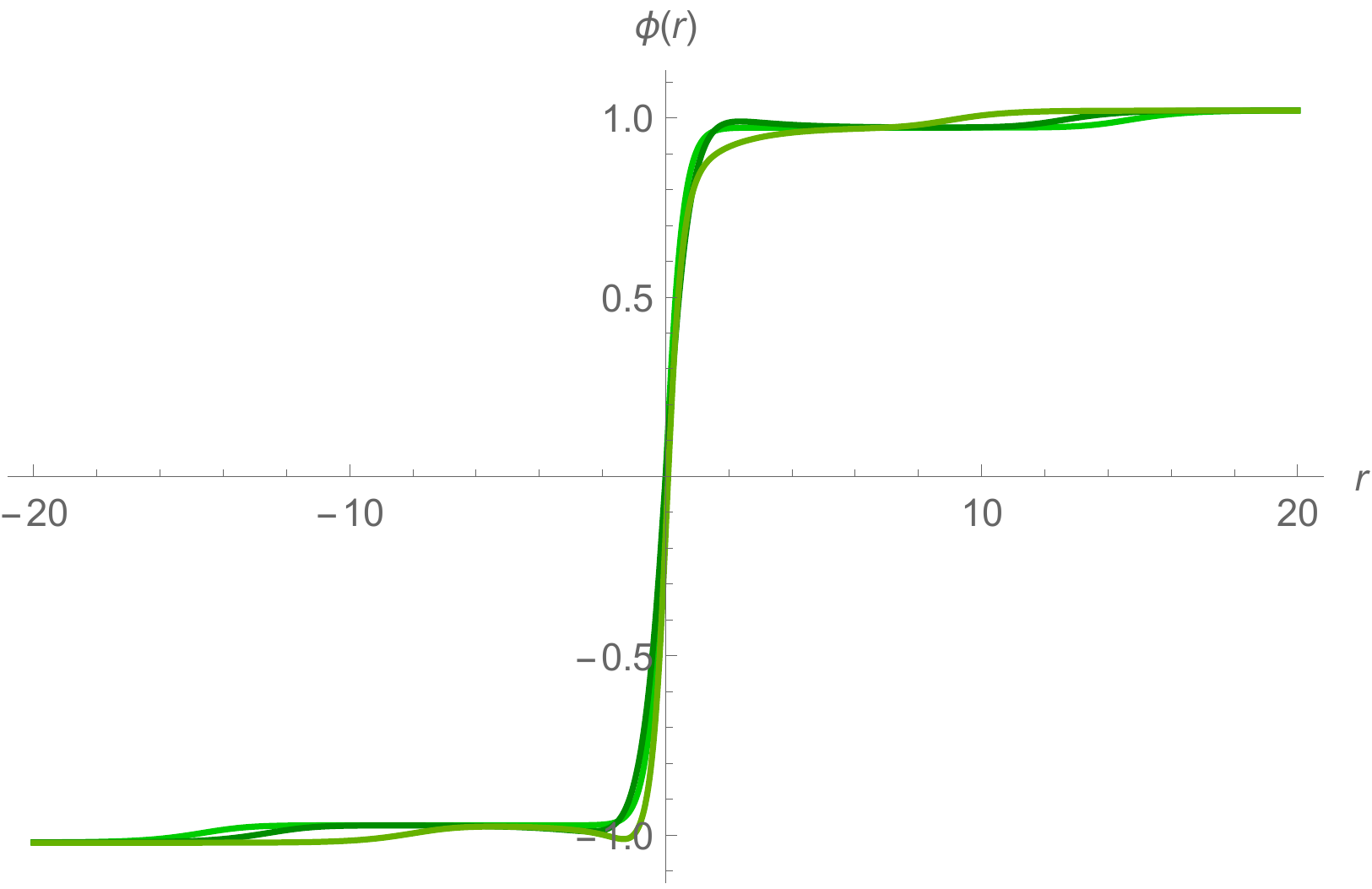}
  \caption{$\phi(r)$}
  \end{subfigure}\\
   \begin{subfigure}[b]{0.45\linewidth}
    \includegraphics[width=\linewidth]{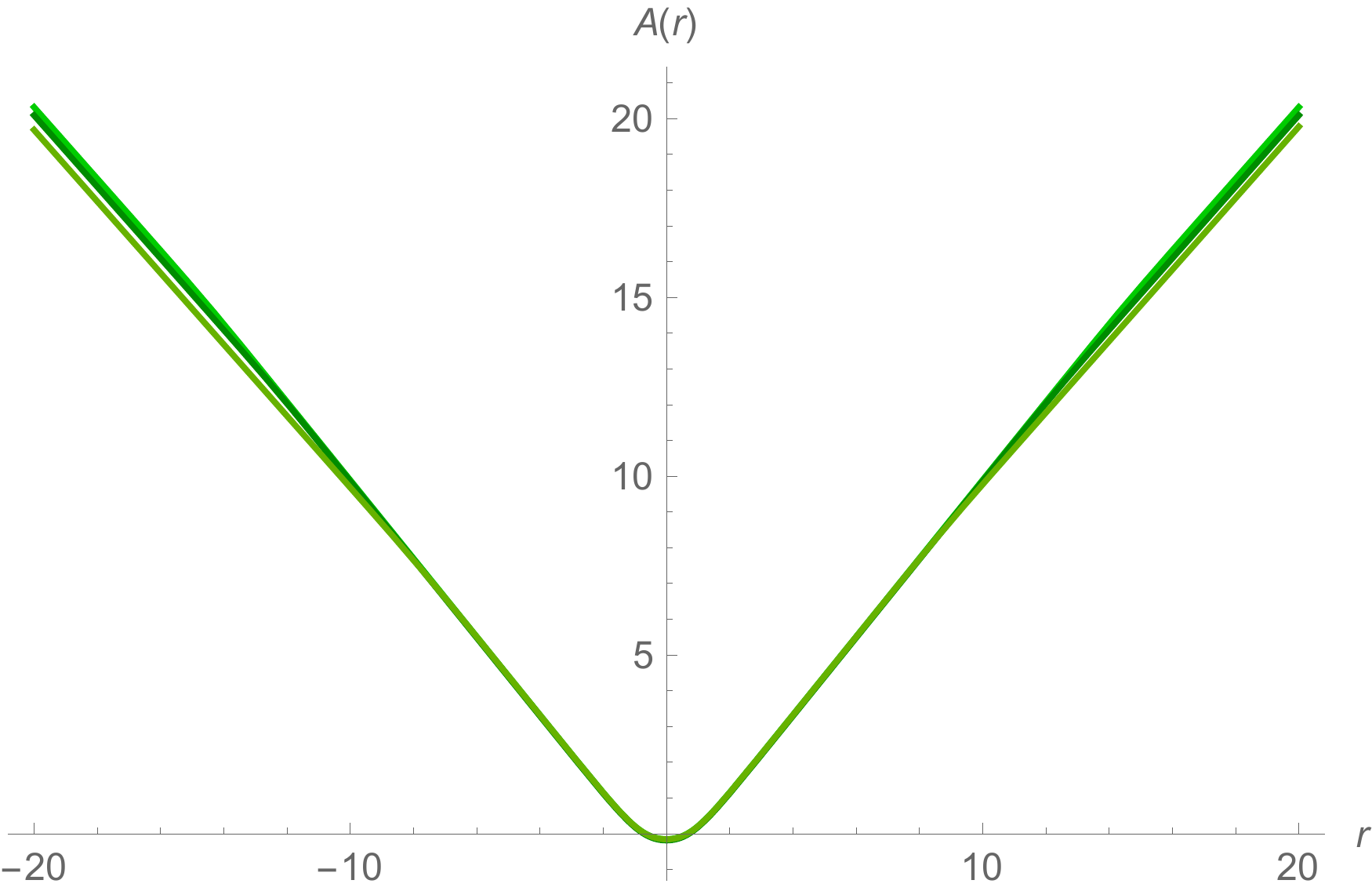}
  \caption{$A(r)$}
   \end{subfigure} 
 \begin{subfigure}[b]{0.45\linewidth}
    \includegraphics[width=\linewidth]{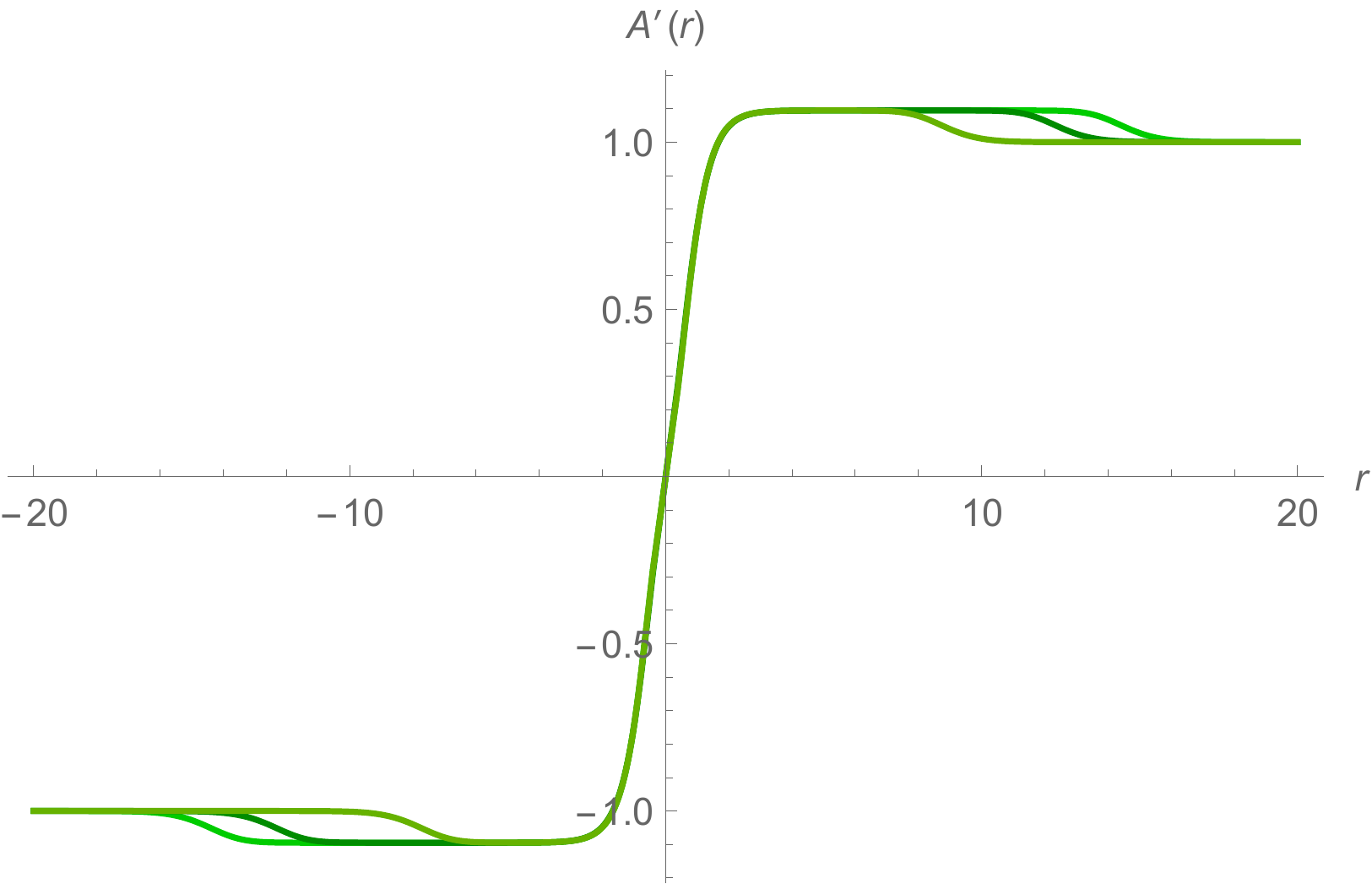}
  \caption{$A'(r)$}
   \end{subfigure} 
  \caption{Profiles of scalar fields $(\zeta,\xi,\chi,\phi)$ and the warped factor $A$ for examples of $G_2/G_2$ Janus solutions as functions of the radial coordinate $r$.}
  \label{G2G2_Profile}
\end{figure}

There exist solutions interpolating among the three critical points. Examples of these are shown in figures \ref{inter_O_0} and \ref{inter_Profile}. The yellow line represents a solution from the $SO(8)$ critical point that first proceeds to the $G_2$ and the $U(3)$ critical points and eventually back to the $SO(8)$ critical point. For comparison, we also show a solution from the $SO(8)$ critical point to the two equivalent $G_2$ critical points and back to the $SO(8)$ critical point by the green line. The latter has already been given in \cite{warner_Janus} in which it has been argued that the solution describes a $G_2/G_2$ interface. In this case, the $G_2$ phase on each side of the interface is generated from the $SO(8)$ phase by a rapid transition via usual RG flows. Similarly, we interpret our new solution represented by the yellow line as an interface between the $G_2$ and $U(3)$ conformal phases since on the two sides, the $SO(8)$ phase undergoes an RG flow to the $G_2$ phase on one side and to the $U(3)$ phase on the other. We also point out that within a larger truncation considered here, there exists a family of $G_2/G_2$ solutions. Examples of these solutions are shown in figure \ref{G2G2_O_0} and \ref{G2G2_Profile}.
\\
\indent We also note here that the two $G_2$ critical points have the same mass spectra and cosmological constants. Therefore, they are equivalent within the $SO(8)$ gauged supergravity. However, upon uplifted to eleven dimensions, the two critical points give inequivalent eleven-dimensional geometries with opposite magnetic charges for the three-form potential due to the opposite signs of the pseudoscalars. Since in this paper, we will not consider the eleven-dimensional uplift, we simply consider the two critical points equivalent.  

\begin{figure}
%[h!]
  \centering
  \begin{subfigure}[b]{0.4\linewidth}
    \includegraphics[width=\linewidth]{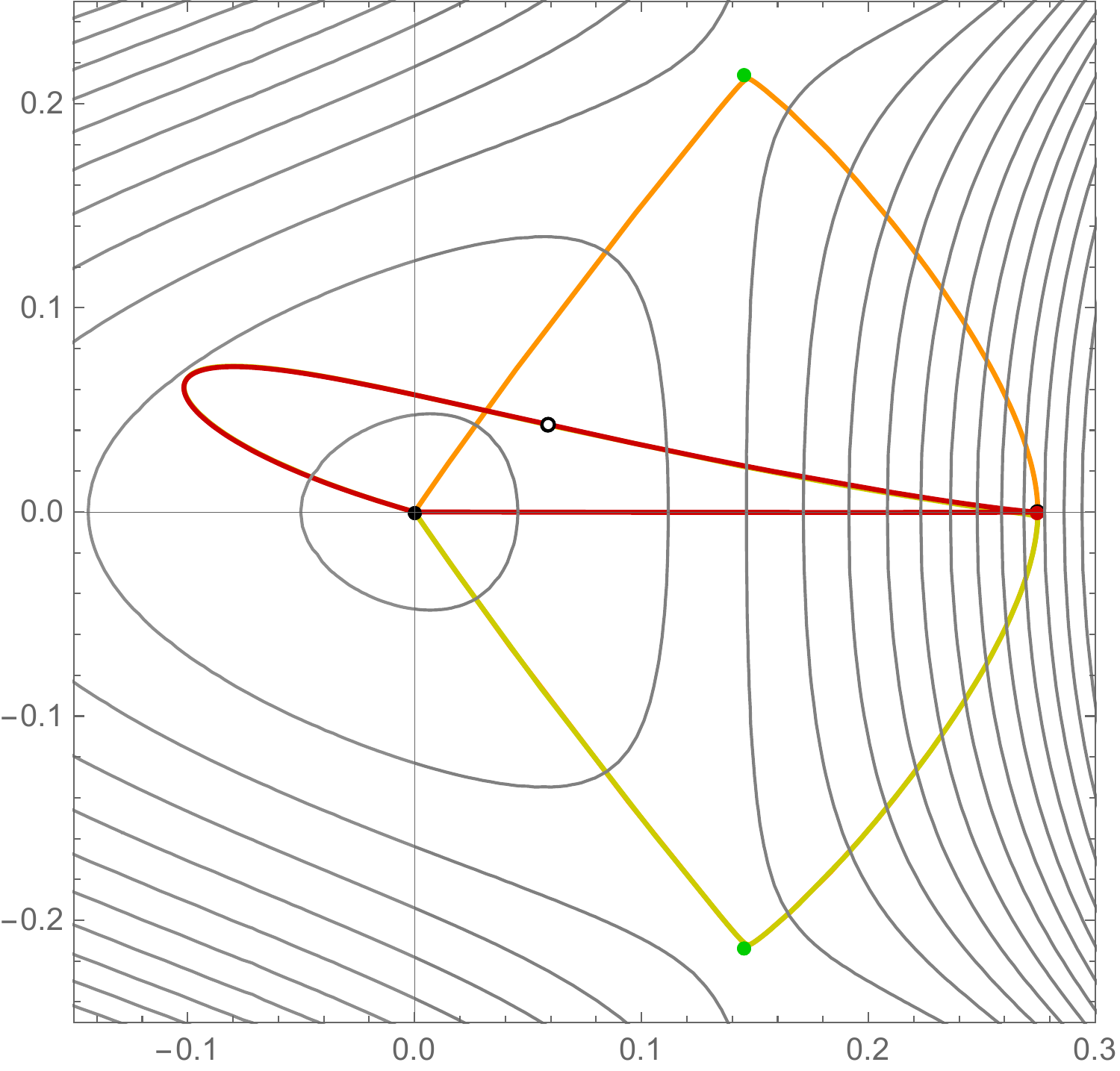}
  \caption{$(\zeta\cos\chi,\zeta\sin\chi)$ plane}
  \end{subfigure}\qquad\quad
  \begin{subfigure}[b]{0.4\linewidth}
    \includegraphics[width=\linewidth]{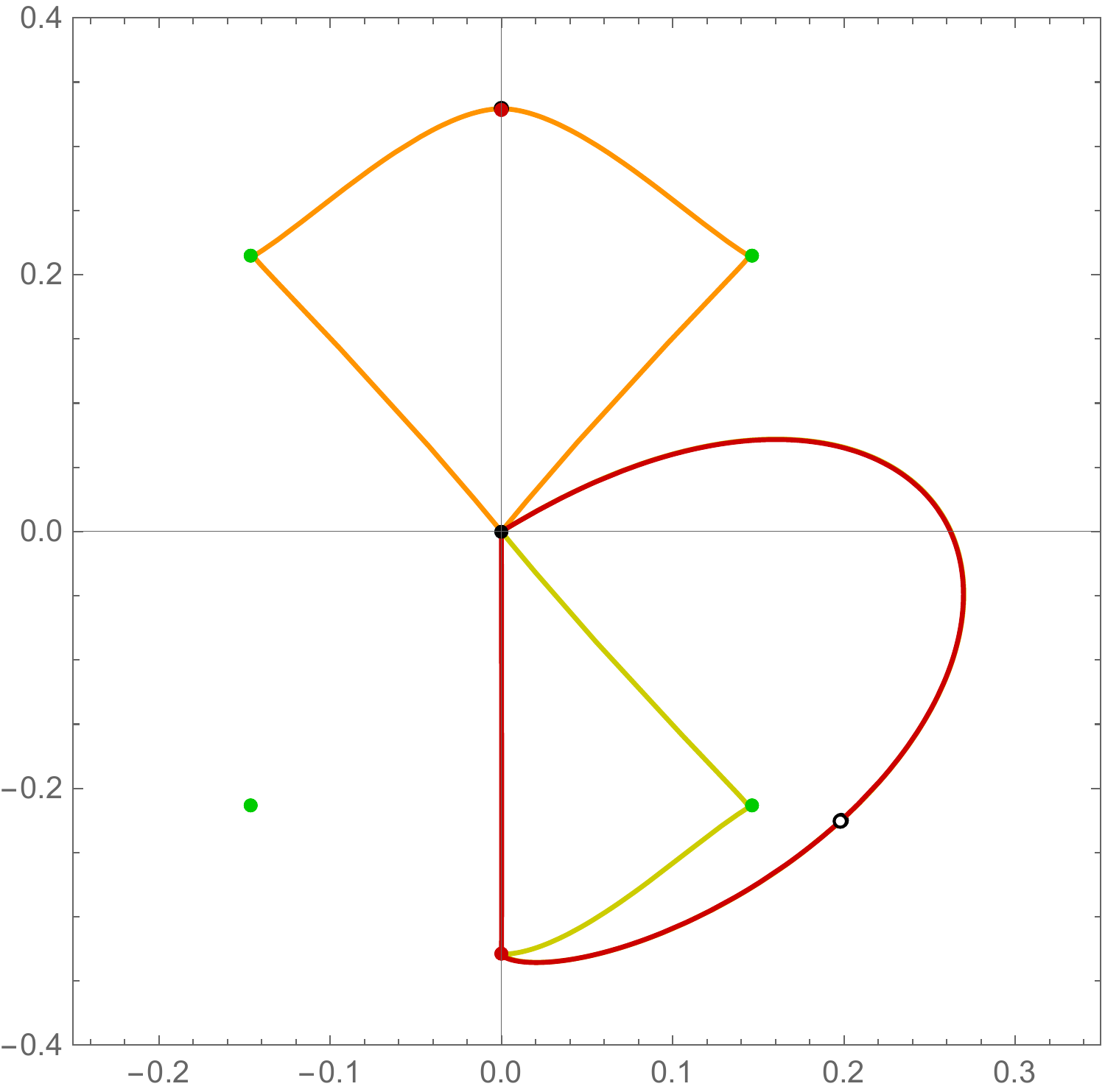}
  \caption{$(\xi\cos\phi,\xi\sin\phi)$ plane}
  \end{subfigure}
  \caption{An $SO(8)/U(3)$ Janus solution is shown on the contour plot of the superpotential by the red line. The yellow and orange lines represent $SO(8)/G_2$ and $G_2/G_2$ solutions that flow to the $U(3)$ critical point.} 
  \label{SO8U3_O_0}
\end{figure}

\begin{figure}
%[h!]
  \centering
  \begin{subfigure}[b]{0.45\linewidth}
    \includegraphics[width=\linewidth]{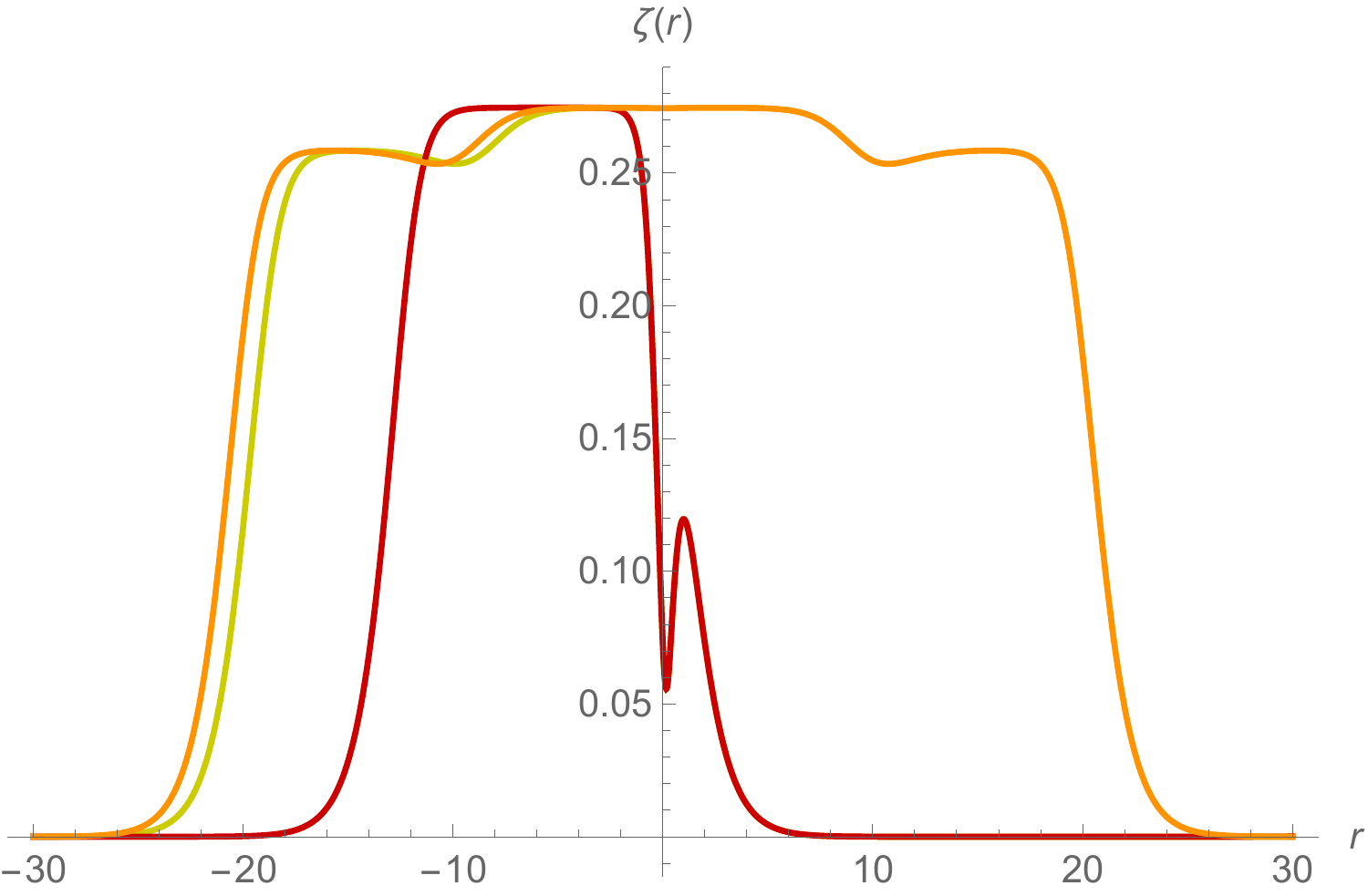}
  \caption{$\zeta(r)$}
  \end{subfigure}
  \begin{subfigure}[b]{0.45\linewidth}
    \includegraphics[width=\linewidth]{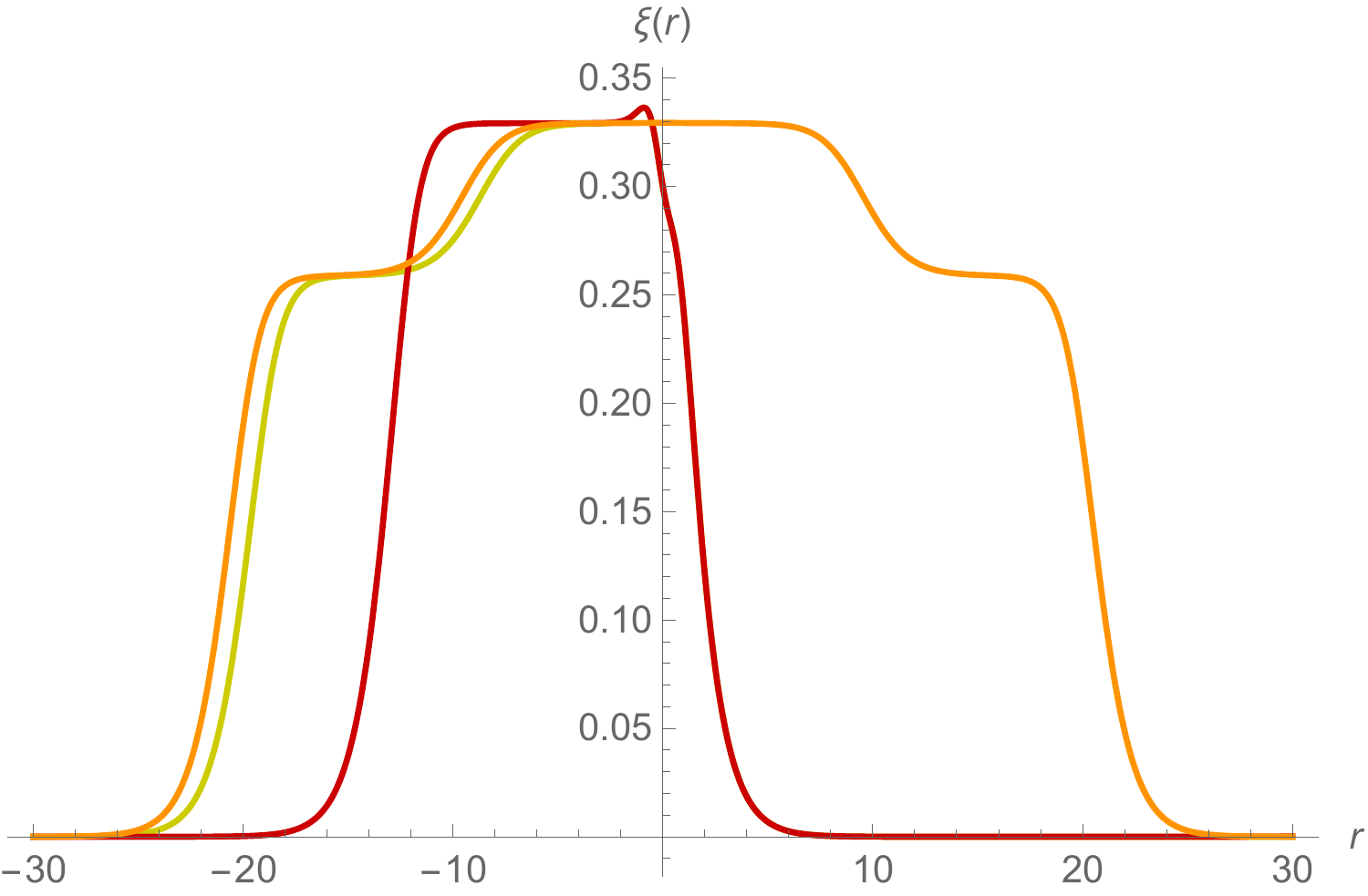}
  \caption{$\xi(r)$}
  \end{subfigure}\\
  \begin{subfigure}[b]{0.45\linewidth}
    \includegraphics[width=\linewidth]{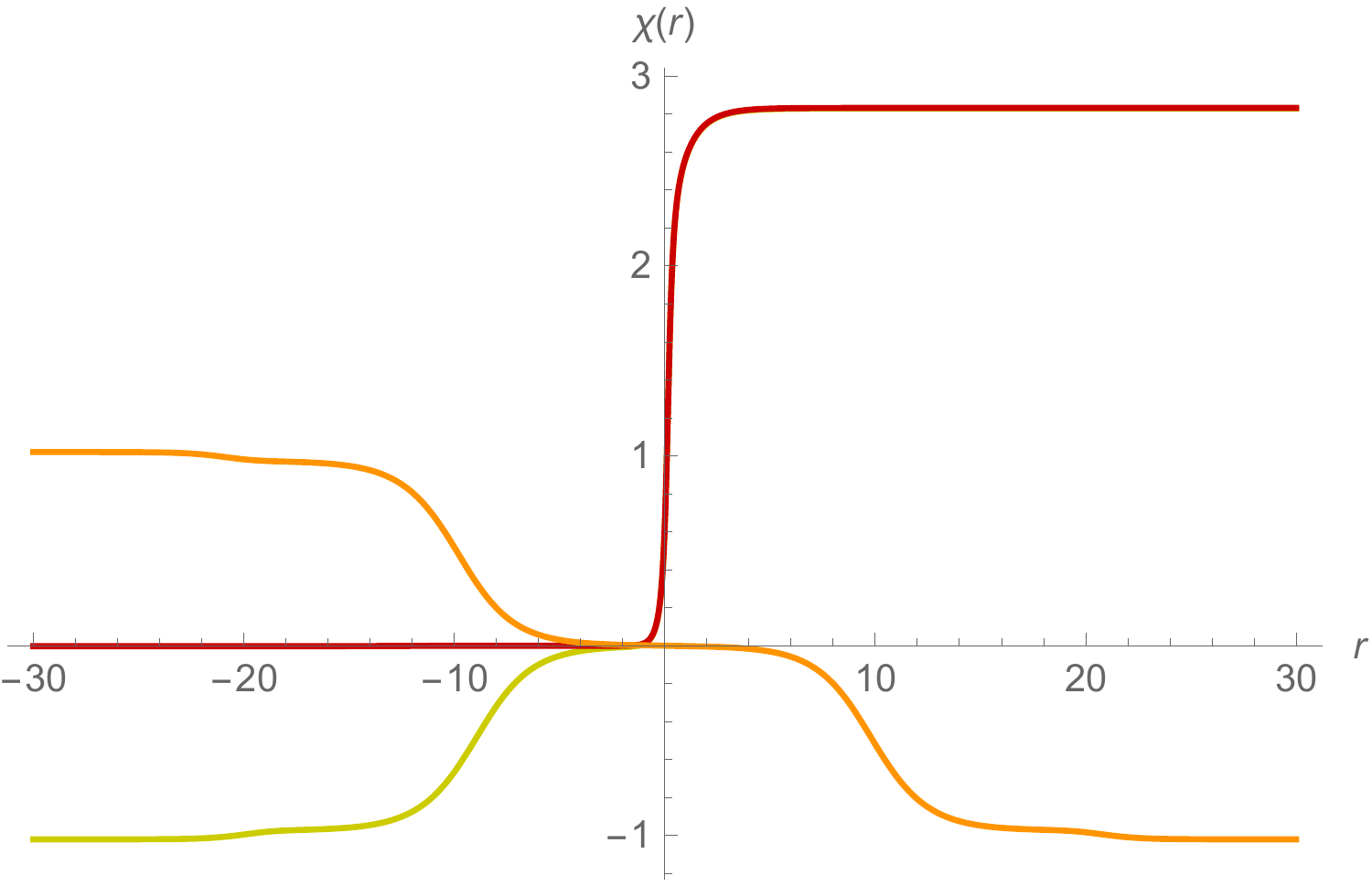}
  \caption{$\chi(r)$}
  \end{subfigure}
  \begin{subfigure}[b]{0.45\linewidth}
    \includegraphics[width=\linewidth]{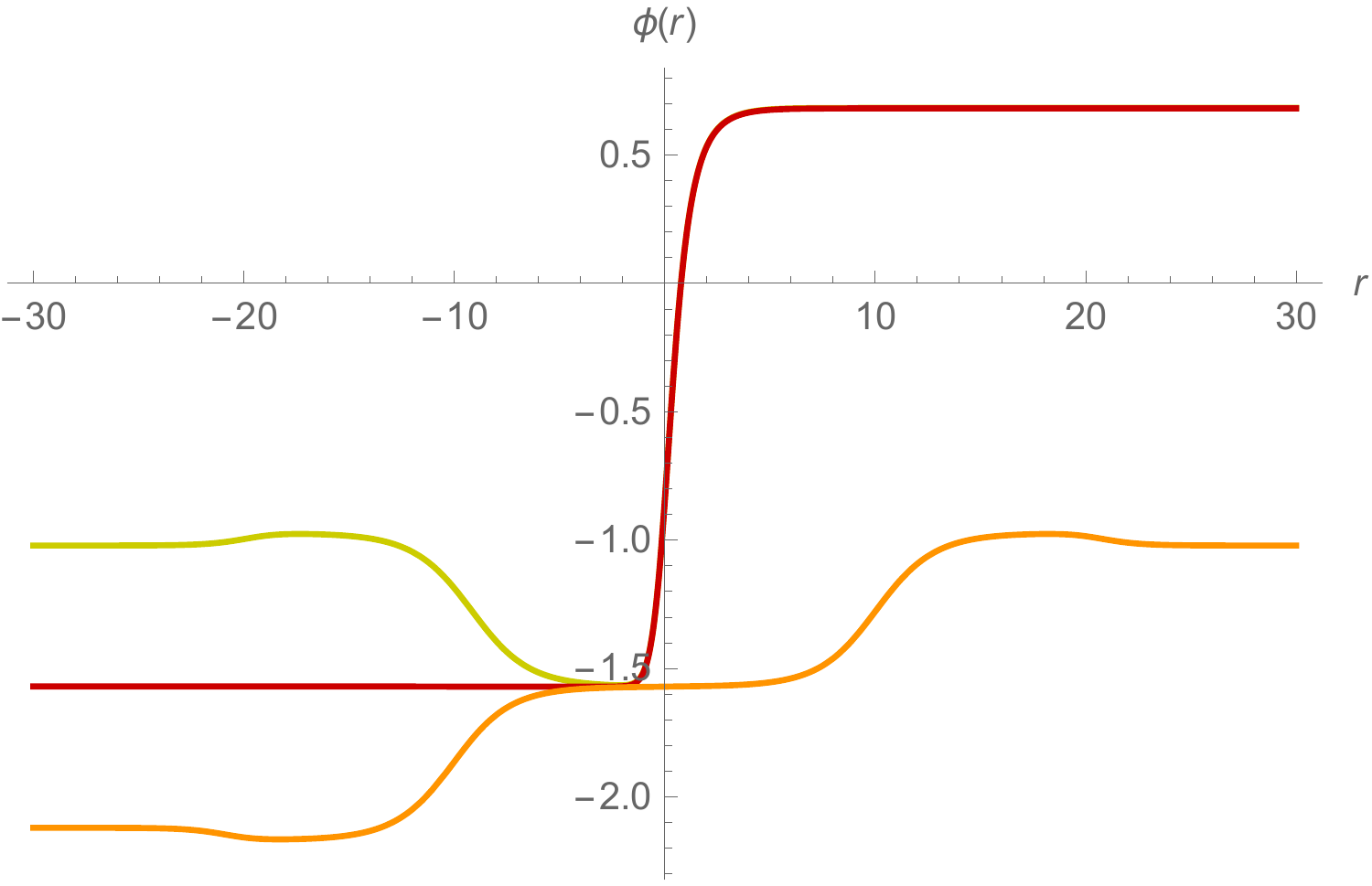}
  \caption{$\phi(r)$}
  \end{subfigure}\\
   \begin{subfigure}[b]{0.45\linewidth}
    \includegraphics[width=\linewidth]{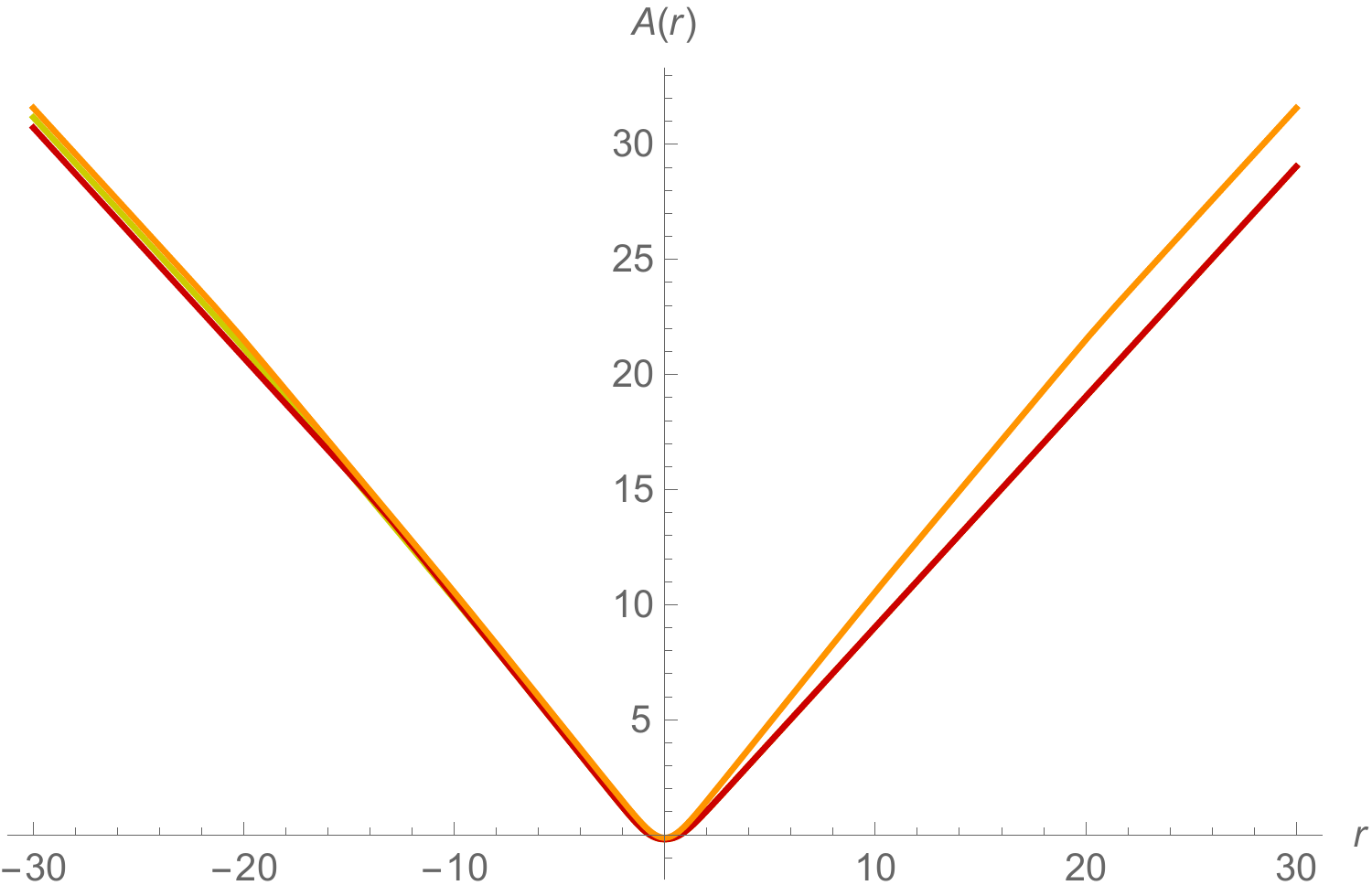}
  \caption{$A(r)$}
   \end{subfigure} 
 \begin{subfigure}[b]{0.45\linewidth}
    \includegraphics[width=\linewidth]{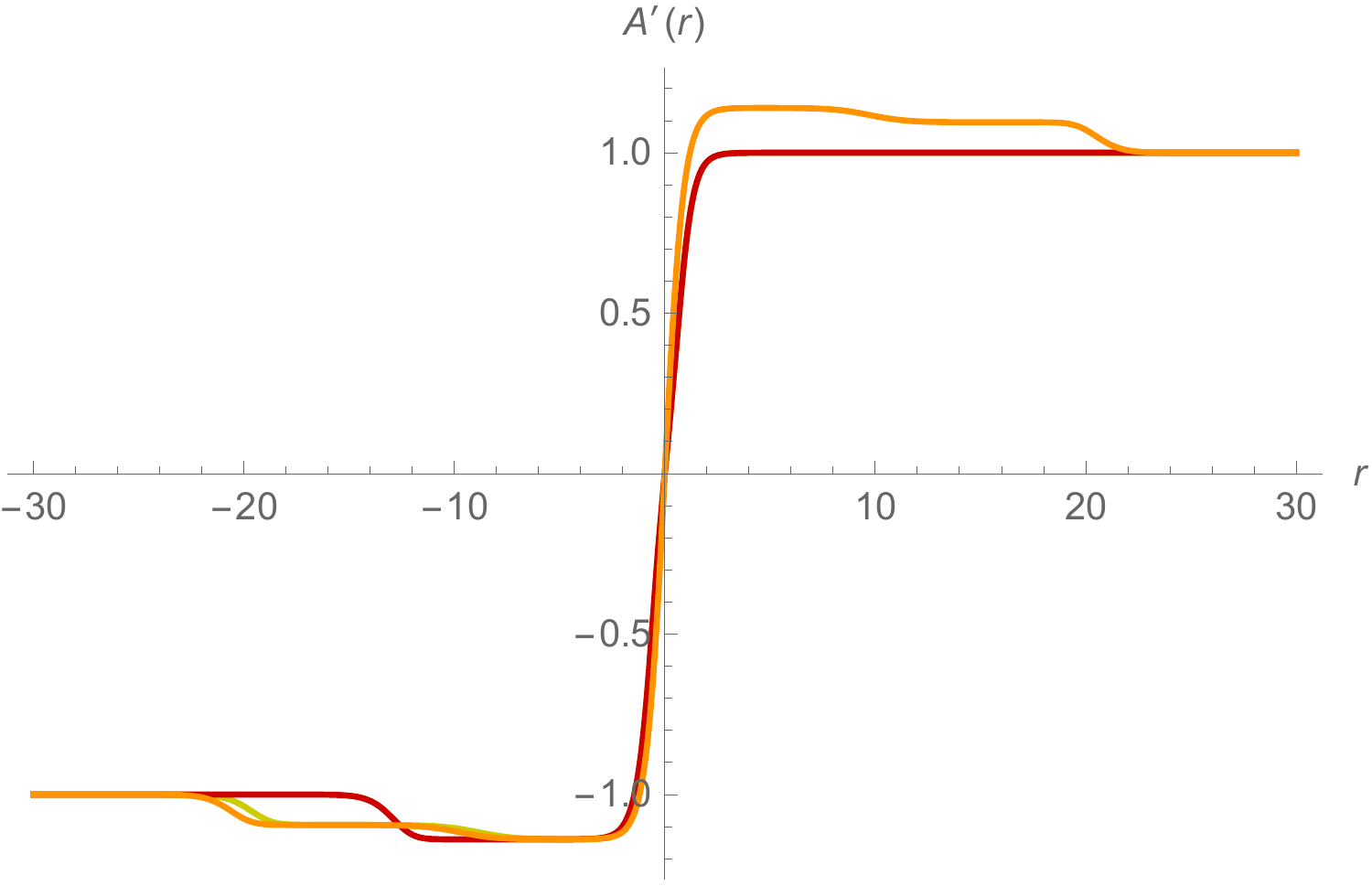}
  \caption{$A'(r)$}
   \end{subfigure} 
  \caption{Profiles of scalar fields $(\zeta,\xi,\chi,\phi)$ and the warped factor $A$ for $SO(8)/U(3)$ Janus together with $SO(8)/G_2$ and $G_2/G_2$ solutions that flow to the $U(3)$ critical point as functions of the radial coordinate $r$.}
  \label{SO8U3_Profile}
\end{figure}

For other solutions involving the $U(3)$ critical point, there exists a $G_2/G_2$ solution that flows to $U(3)$ critical point as shown by the orange line in figures \ref{SO8U3_O_0} and \ref{SO8U3_Profile}. The red line in these figures represents an $SO(8)/U(3)$ interface in which the $SO(8)$ phase undergoes an RG flow to the $U(3)$ phase on one side. The yellow line in figures \ref{SO8U3_O_0} and \ref{SO8U3_Profile} can also be considered as an interface between $SO(8)$ and $G_2$ conformal phases. Unlike the $SO(8)/G_2$ solution found in \cite{warner_Janus}, this $SO(8)/G_2$ solution flows to the $U(3)$ critical point near the interface.

%%%%%%%%%%%%%%%%%%%%%%%%%%%%%%%%%%%%%%%%%%%%%%%%%%%%%%%%%%%%%%%%%%%%%%%%%%%%%%%%%%%%%%%%%%%%%%%%%%%%%%%%%%%%%%%%%%%%%%%%%%%%%%%%%%%%%%%%%
\section{Supersymmetric Janus solutions with $\omega\neq0$}\label{Janus_omega_n0}
\begin{figure}
%[h!]
  \centering
  \begin{subfigure}[b]{0.4\linewidth}
    \includegraphics[width=\linewidth]{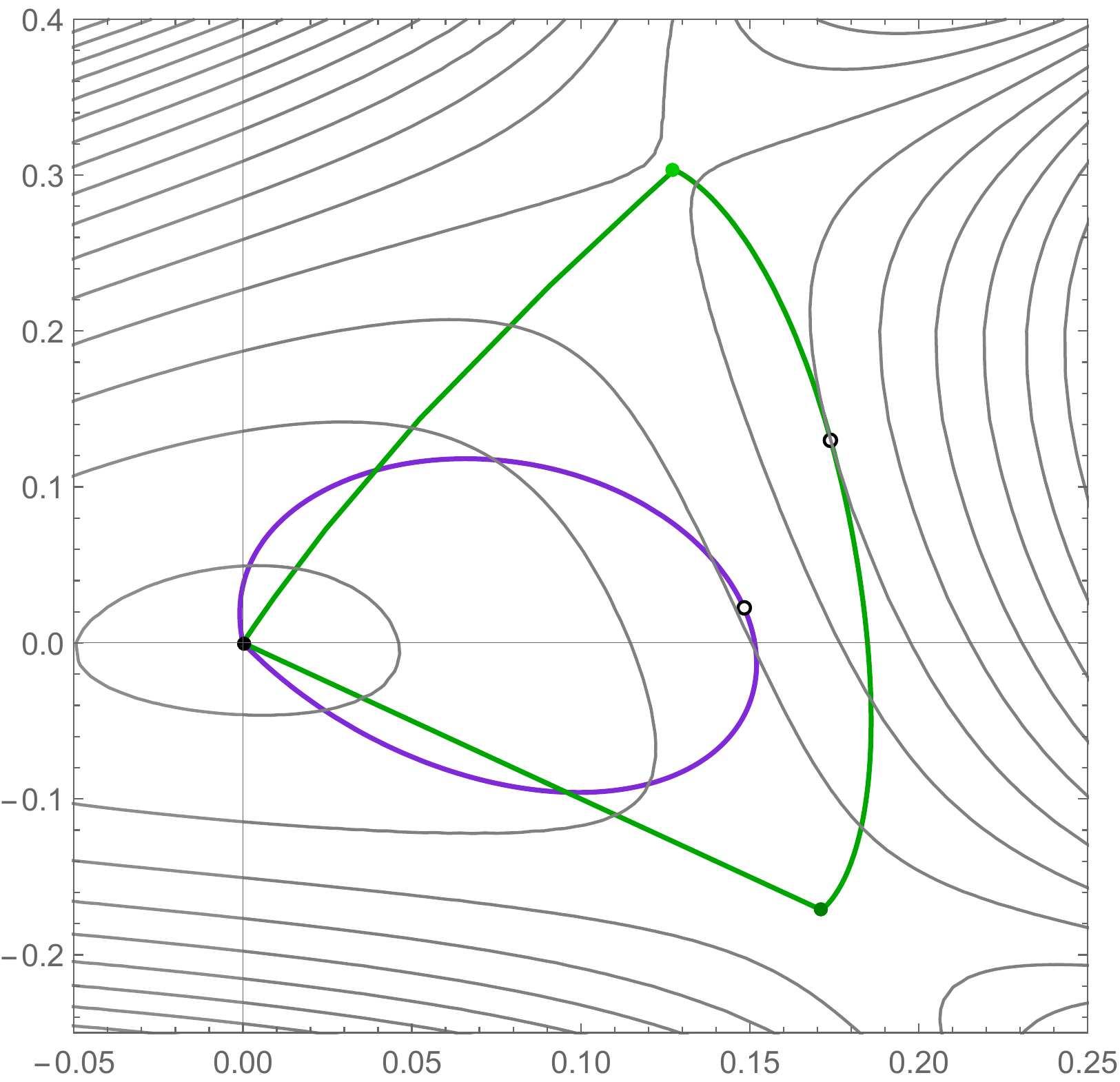}
  \caption{$(\zeta\cos\chi,\zeta\sin\chi)$ plane}
  \end{subfigure}\qquad\quad
  \begin{subfigure}[b]{0.4\linewidth}
    \includegraphics[width=\linewidth]{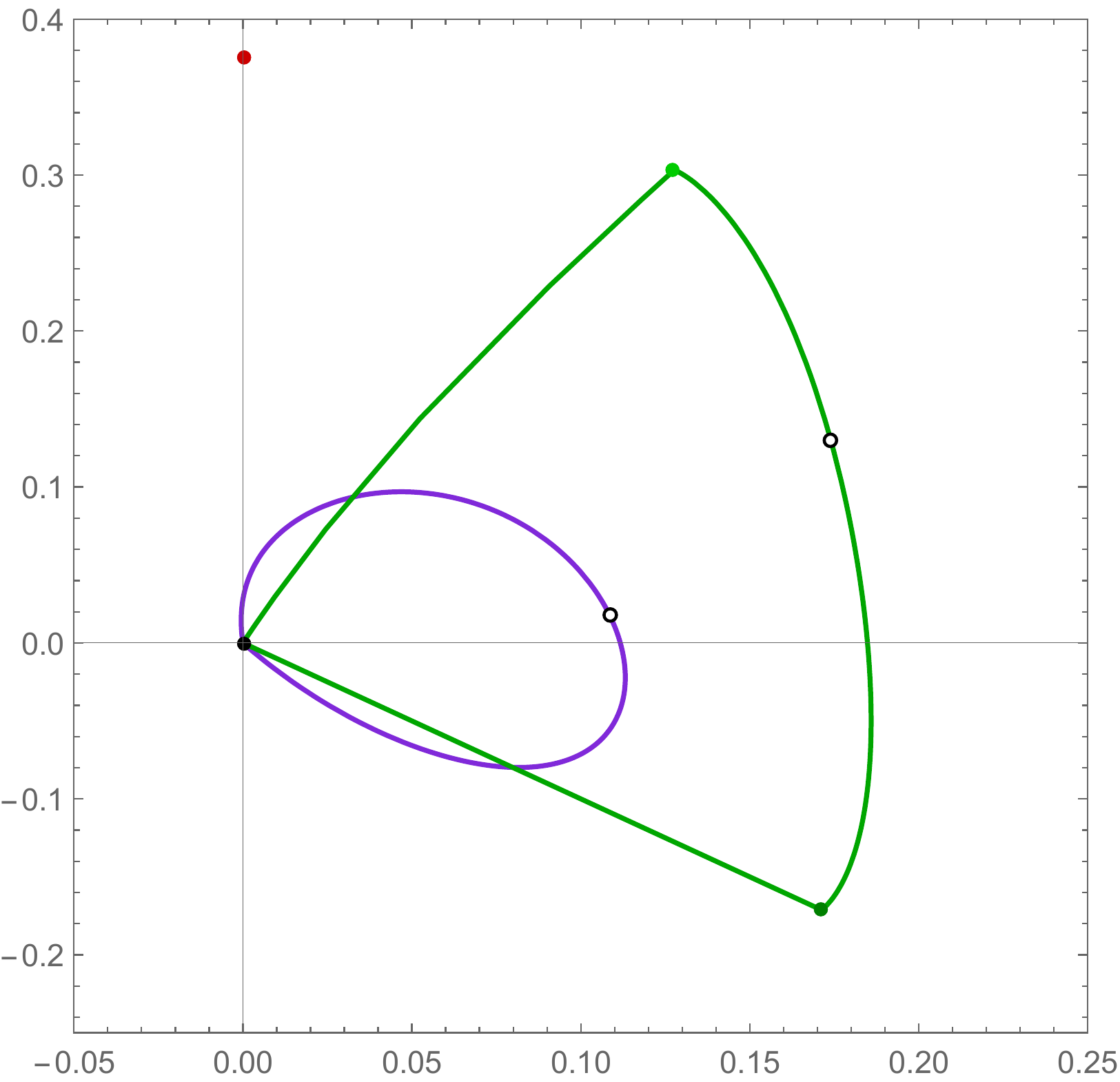}
  \caption{$(\xi\cos\phi,\xi\sin\phi)$ plane}
  \end{subfigure}
  \caption{$SO(8)/SO(8)$ and $G_2/\overline{G}_2$ Janus solutions on the contour plot of the superpotential with $\omega=\frac{\pi}{8}$ are shown by purple and green lines, respectively.} 
  \label{G2G2_O}
\end{figure}

\begin{figure}
%[h!]
  \centering
  \begin{subfigure}[b]{0.45\linewidth}
    \includegraphics[width=\linewidth]{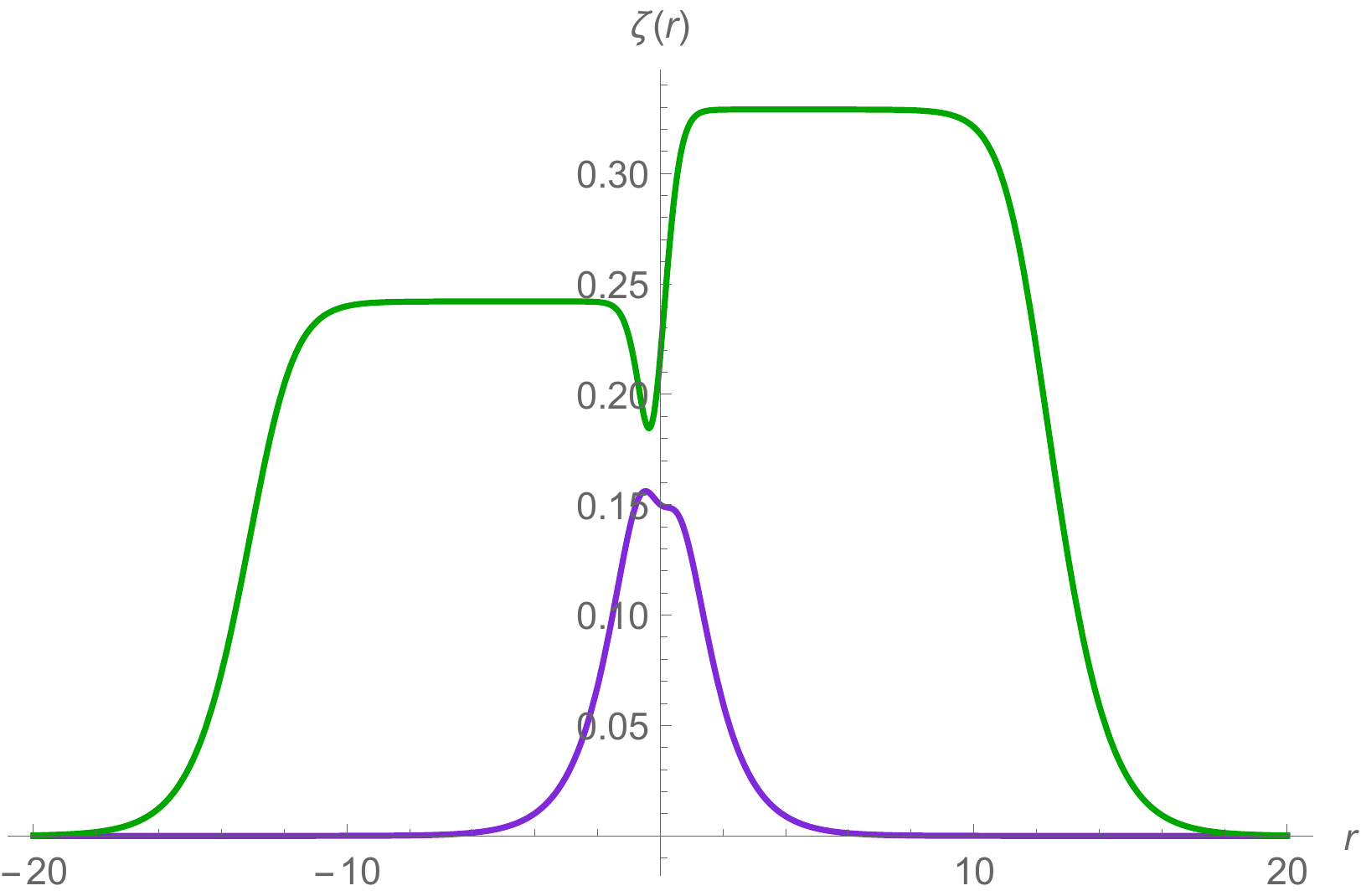}
  \caption{$\zeta(r)$}
  \end{subfigure}
  \begin{subfigure}[b]{0.45\linewidth}
    \includegraphics[width=\linewidth]{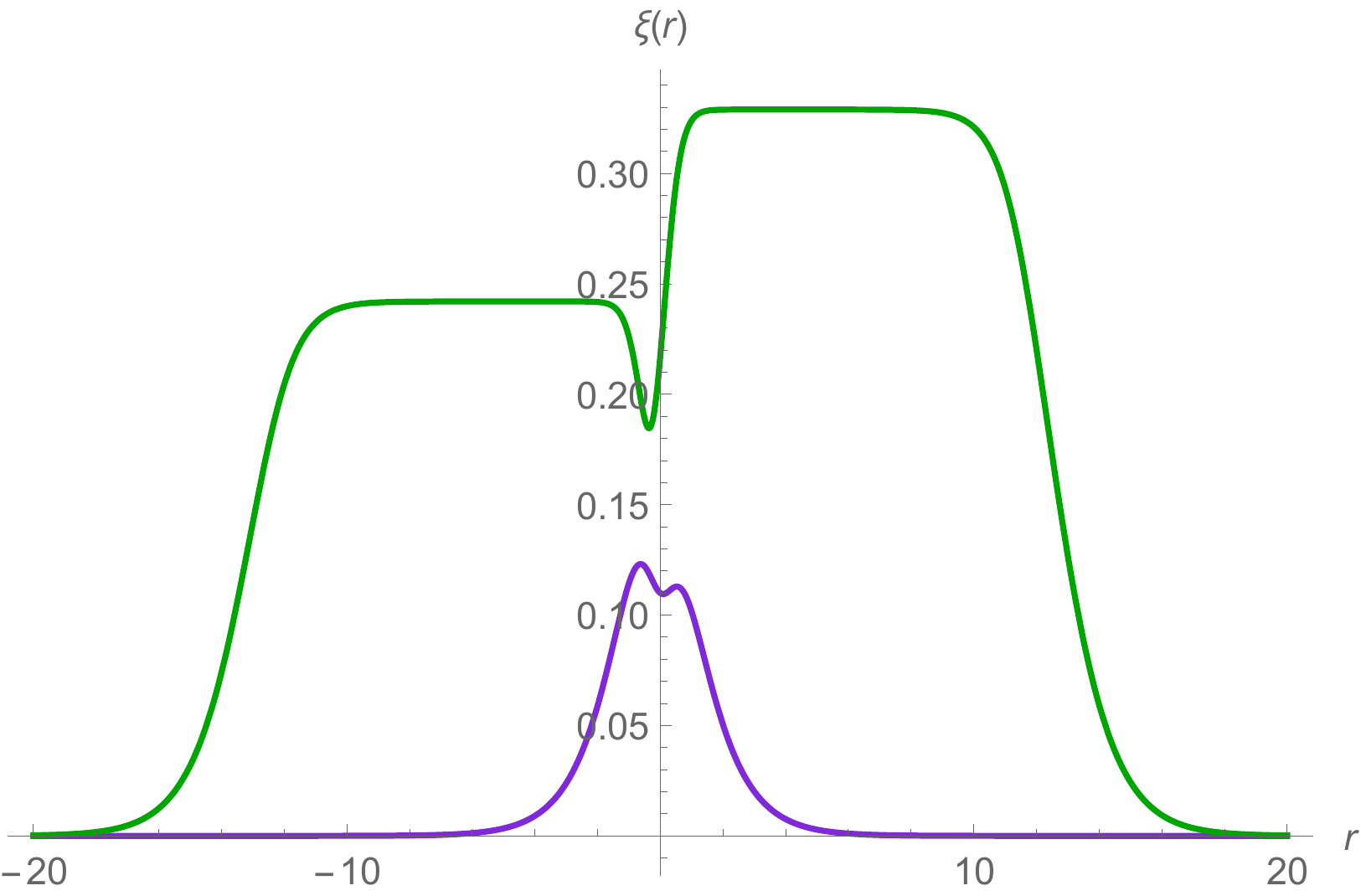}
  \caption{$\xi(r)$}
  \end{subfigure}\\
  \begin{subfigure}[b]{0.45\linewidth}
    \includegraphics[width=\linewidth]{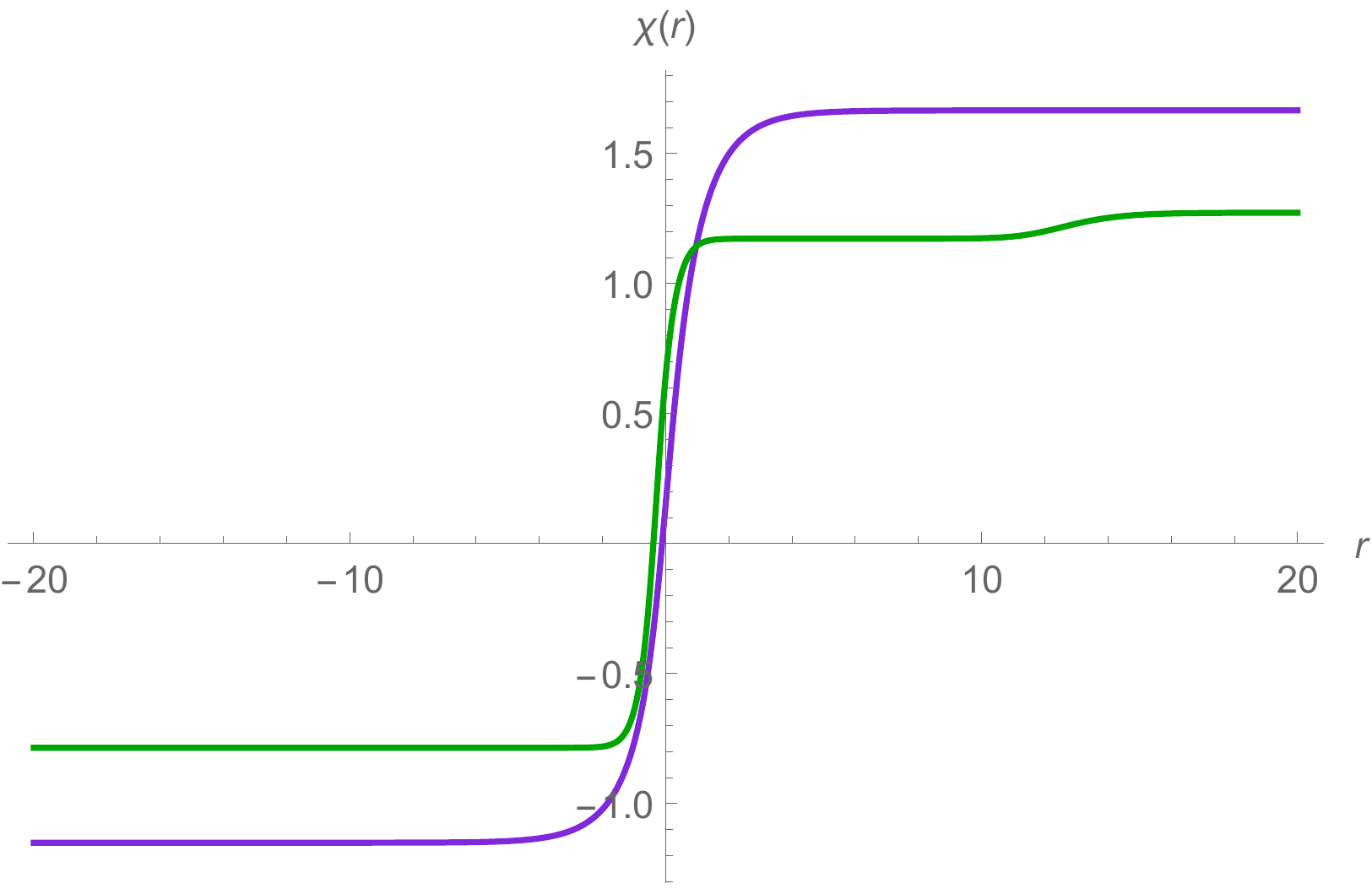}
  \caption{$\chi(r)$}
  \end{subfigure}
  \begin{subfigure}[b]{0.45\linewidth}
    \includegraphics[width=\linewidth]{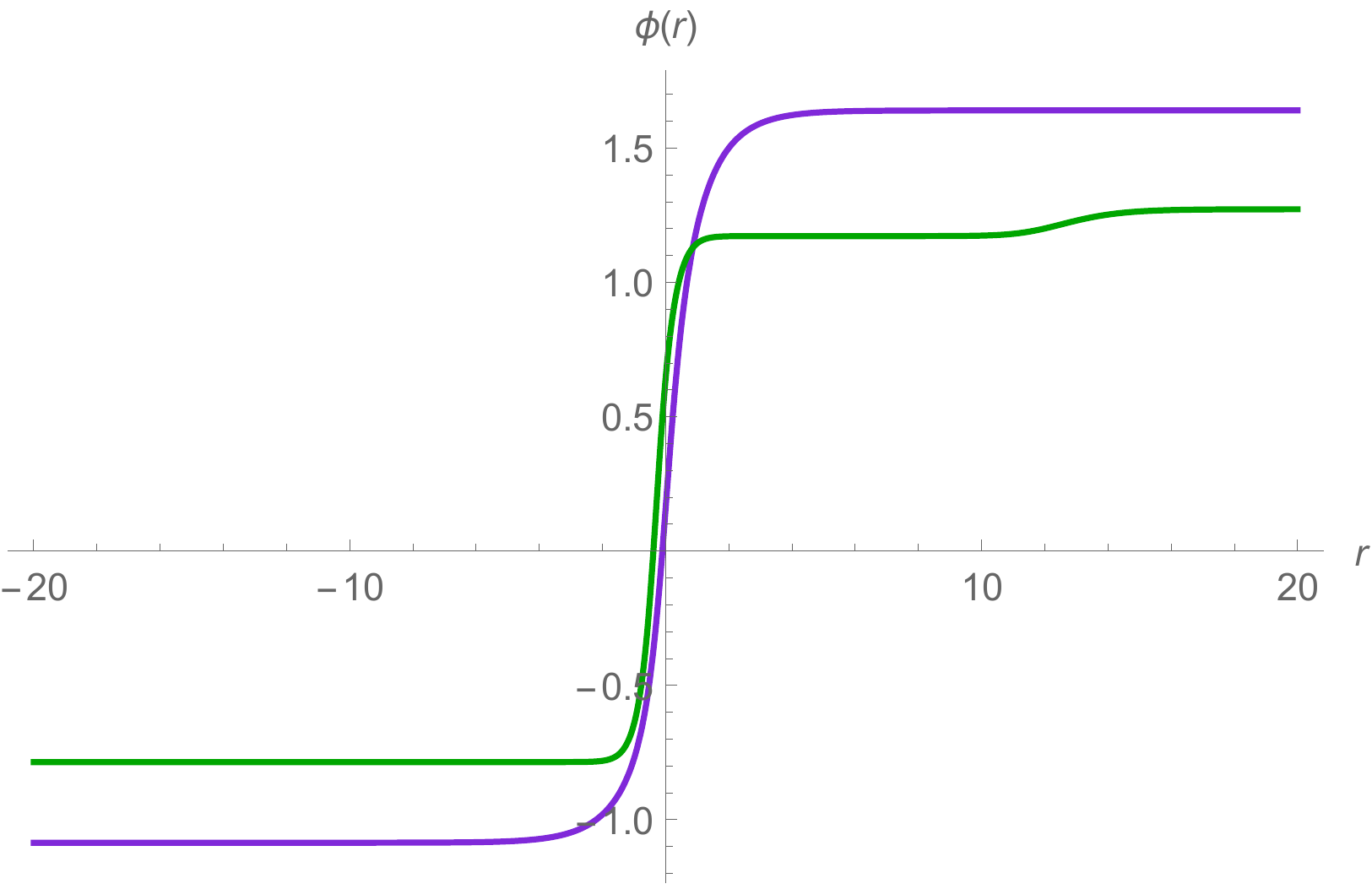}
  \caption{$\phi(r)$}
  \end{subfigure}\\
   \begin{subfigure}[b]{0.45\linewidth}
    \includegraphics[width=\linewidth]{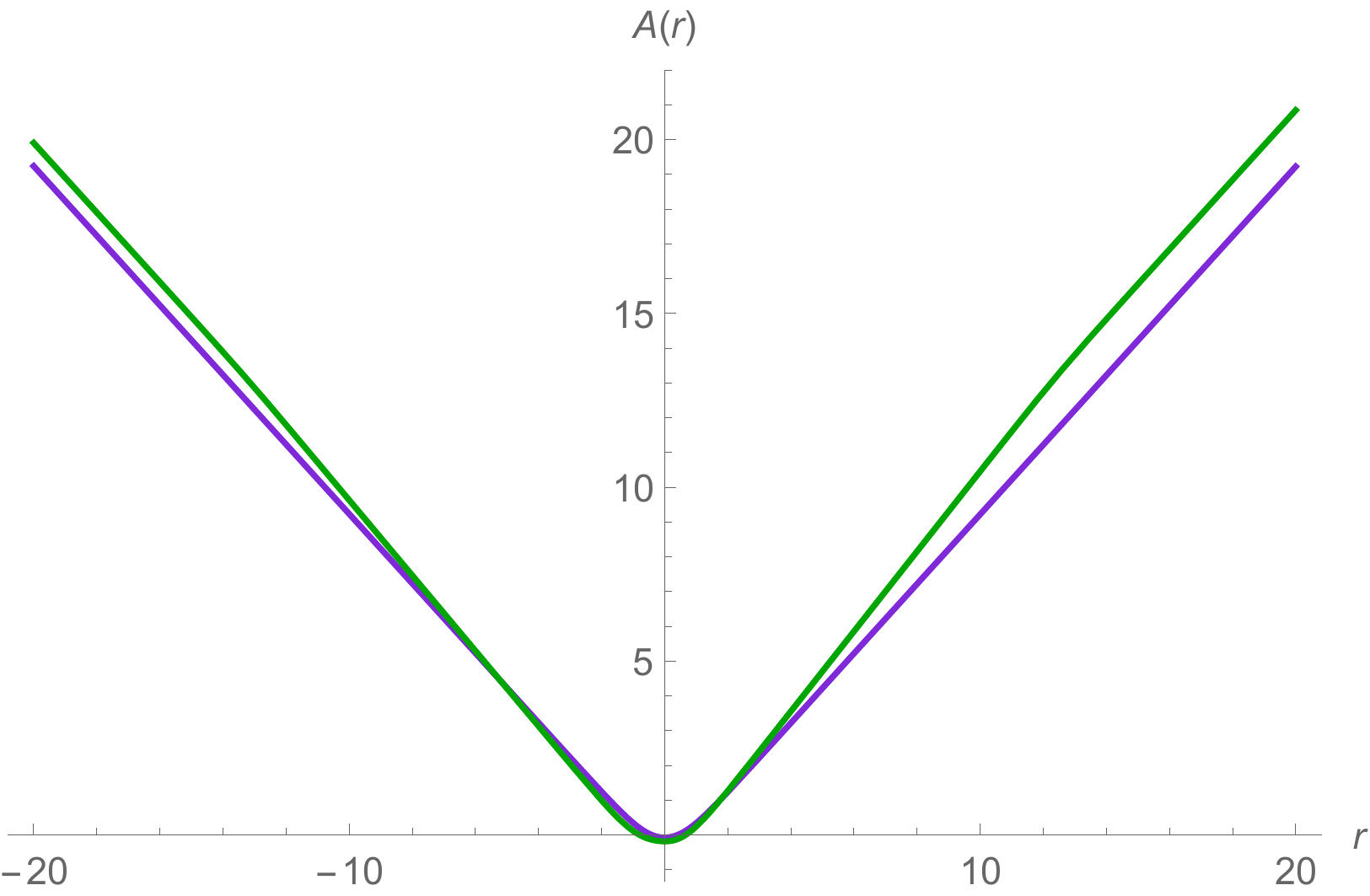}
  \caption{$A(r)$}
   \end{subfigure} 
 \begin{subfigure}[b]{0.45\linewidth}
    \includegraphics[width=\linewidth]{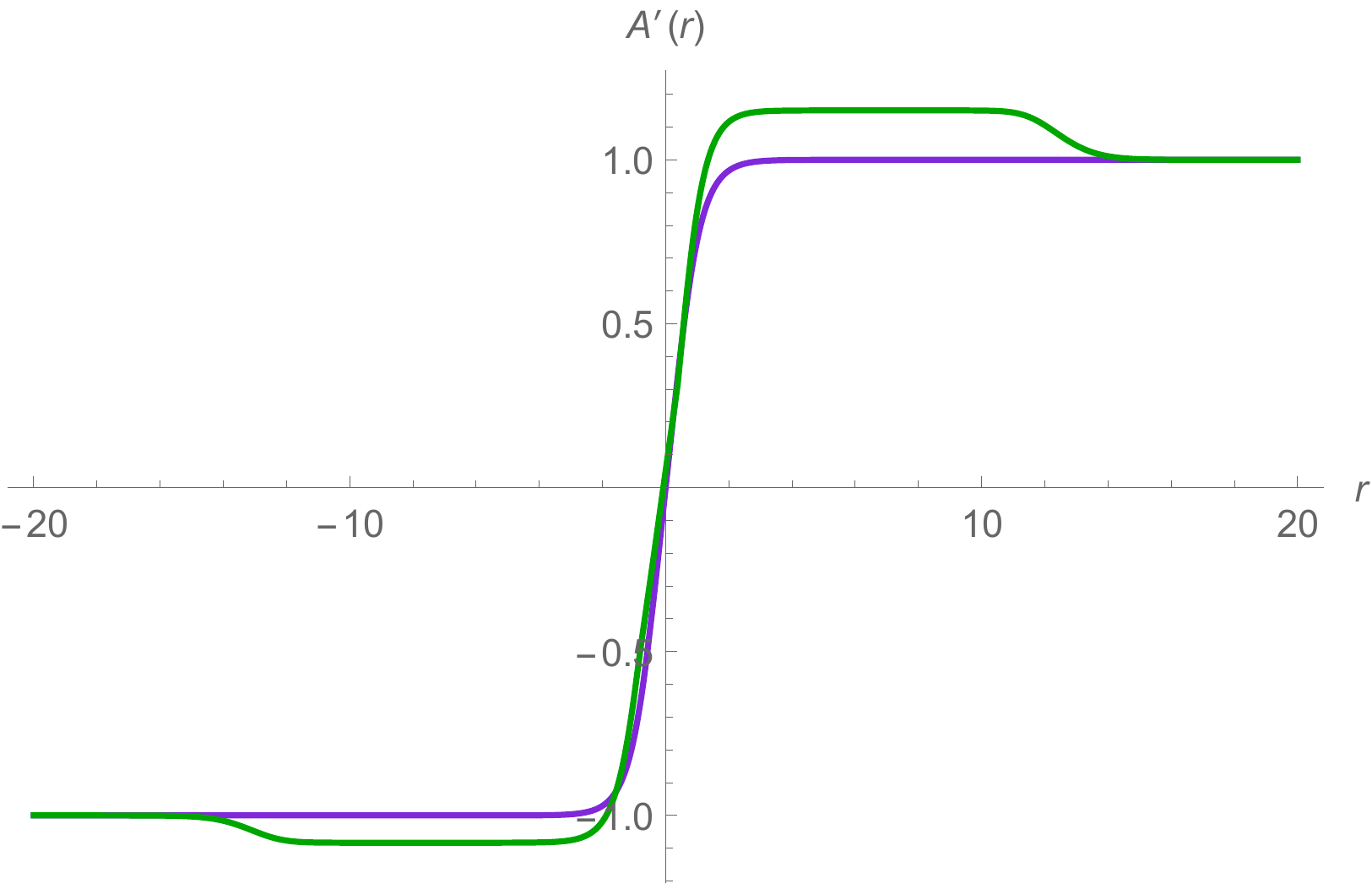}
  \caption{$A'(r)$}
   \end{subfigure} 
  \caption{Profiles of scalar fields $(\zeta,\xi,\chi,\phi)$ and the warped factor $A$ for $SO(8)/SO(8)$ and $G_2/\overline{G}_2$ Janus solutions with $\omega=\frac{\pi}{8}$ as functions of the radial coordinate $r$.}
  \label{G2G2_O_Profile}
\end{figure}

We now look at Janus solutions in the dyonic $SO(8)$ gauged supergravity with $\omega=\frac{\pi}{8}$. In this case, there are two additional $N=1$ supersymmetric $AdS_4$ critical points with $G_2$ and $SU(3)$ symmetries. The former will be denoted by $\overline{G}_2$ to avoid confusion with the previous $G_2$ critical points in the $\omega=0$ case. All the $AdS_4$ critical points of the $\omega=0$ case are also critical points of the $\omega\neq 0$ case with the positions in field space displaced from the corresponding values with $\omega=0$ except for the $N=8$ $SO(8)$ critical point that is still located at $\mc{V}=\mathbb{I}$. With more $AdS_4$ critical points, there are more possibilities for Janus solutions describing various interfaces with different conformal phases on the two sides. In the contour plot of the superpotential, the critical points with $\omega=0$ analogues will be denoted by the same color code while the $\overline{G}_2$ and $SU(3)$ points will be represented by dark green and blue dots, respectively.   
\\
\indent As in the $\omega=0$ case, there is a family of $SO(8)/SO(8)$ Janus solutions describing interfaces between $SO(8)$ symmetric conformal phases. In addition, there are $SO(8)/SO(8)$ solutions that proceed arbitrarily close to the $G_2$ and $\overline{G}_2$ points. Some examples of these solutions are shown in figures \ref{G2G2_O} and \ref{G2G2_O_Profile}. The $SO(8)/SO(8)$ solution is shown as purple line which is very similar to the solutions in \cite{warner_Janus} with $\omega=0$. Similar to the $G_2/G_2$ Janus solutions in the $\omega=0$ case, the solution represented by the green line undergoes a rapid transition between the $SO(8)$ and $G_2$ points and between the $SO(8)$ and $\overline{G}_2$ points on the two sides of the interface. We then argue that this solution describes a $G_2/\overline{G}_2$ interface with the $G_2$ and $\overline{G}_2$ conformal phases generated by the $SO(8)$ phase on the two sides.  

\begin{figure}
%[h!]
  \centering
  \begin{subfigure}[b]{0.4\linewidth}
    \includegraphics[width=\linewidth]{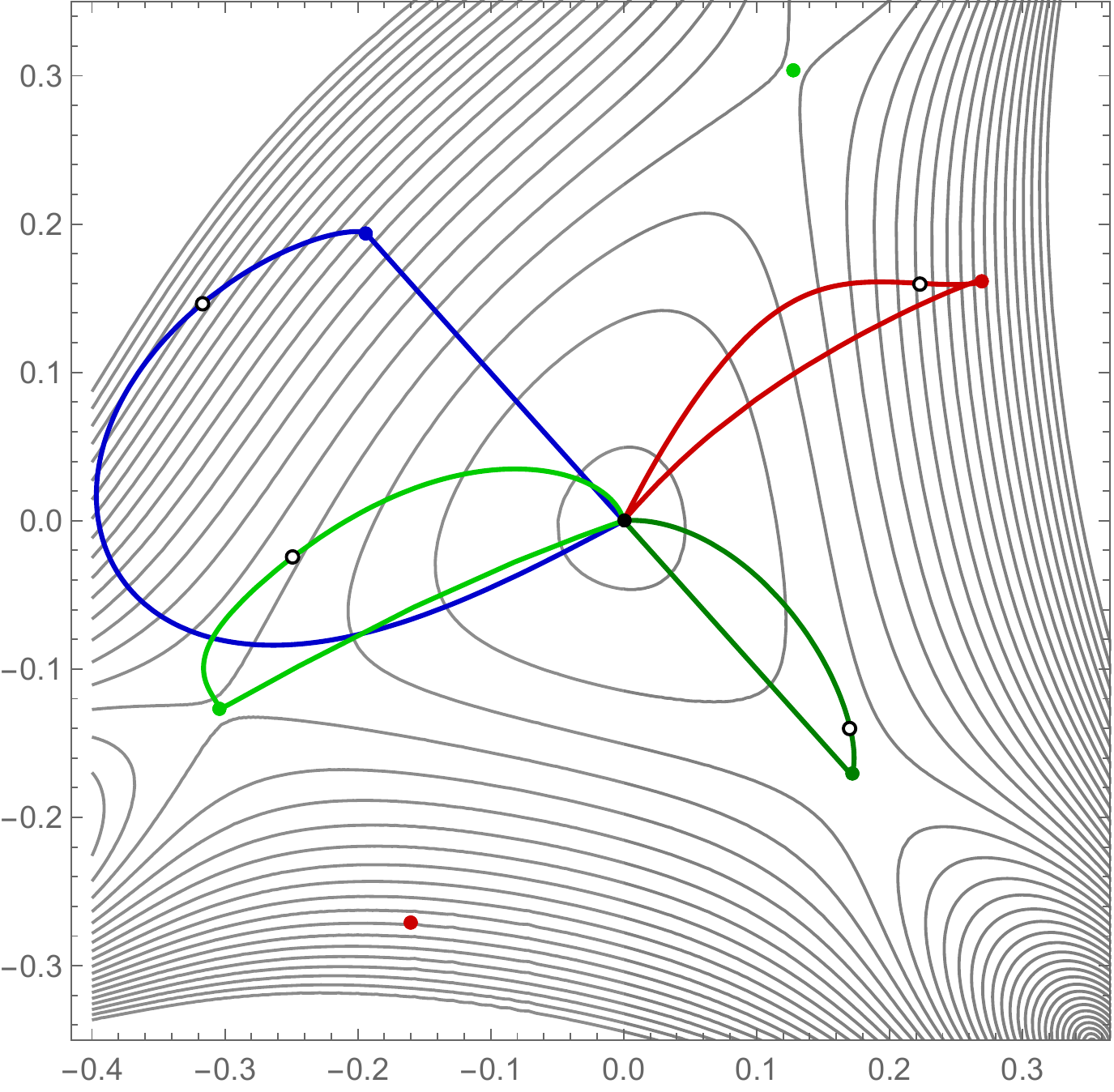}
  \caption{$(\zeta\cos\chi,\zeta\sin\chi)$ plane}
  \end{subfigure}\qquad\quad
  \begin{subfigure}[b]{0.4\linewidth}
    \includegraphics[width=\linewidth]{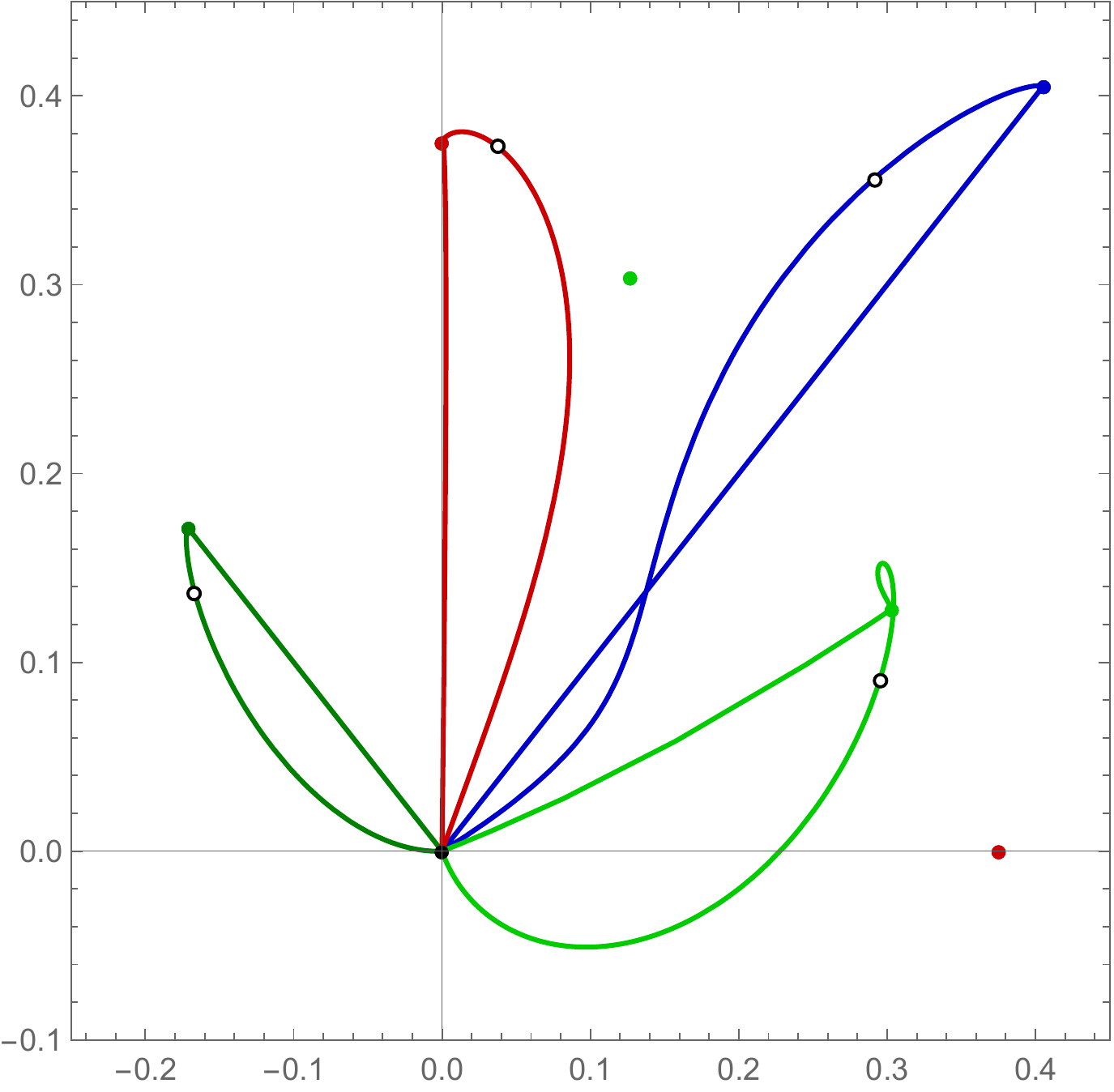}
  \caption{$(\xi\cos\phi,\xi\sin\phi)$ plane}
  \end{subfigure}
  \caption{$SO(8)/U(3)$, $SO(8)/G_2$, $SO(8)/\overline{G}_2$ and $SO(8)/SU(3)$ Janus solutions on the contour plot of the superpotential with $\omega=\frac{\pi}{8}$ are shown in red, green, dark green and blue lines, respectively.} 
  \label{SO8X_O}
\end{figure}                   
            
\begin{figure}
%[h!]
  \centering
  \begin{subfigure}[b]{0.45\linewidth}
    \includegraphics[width=\linewidth]{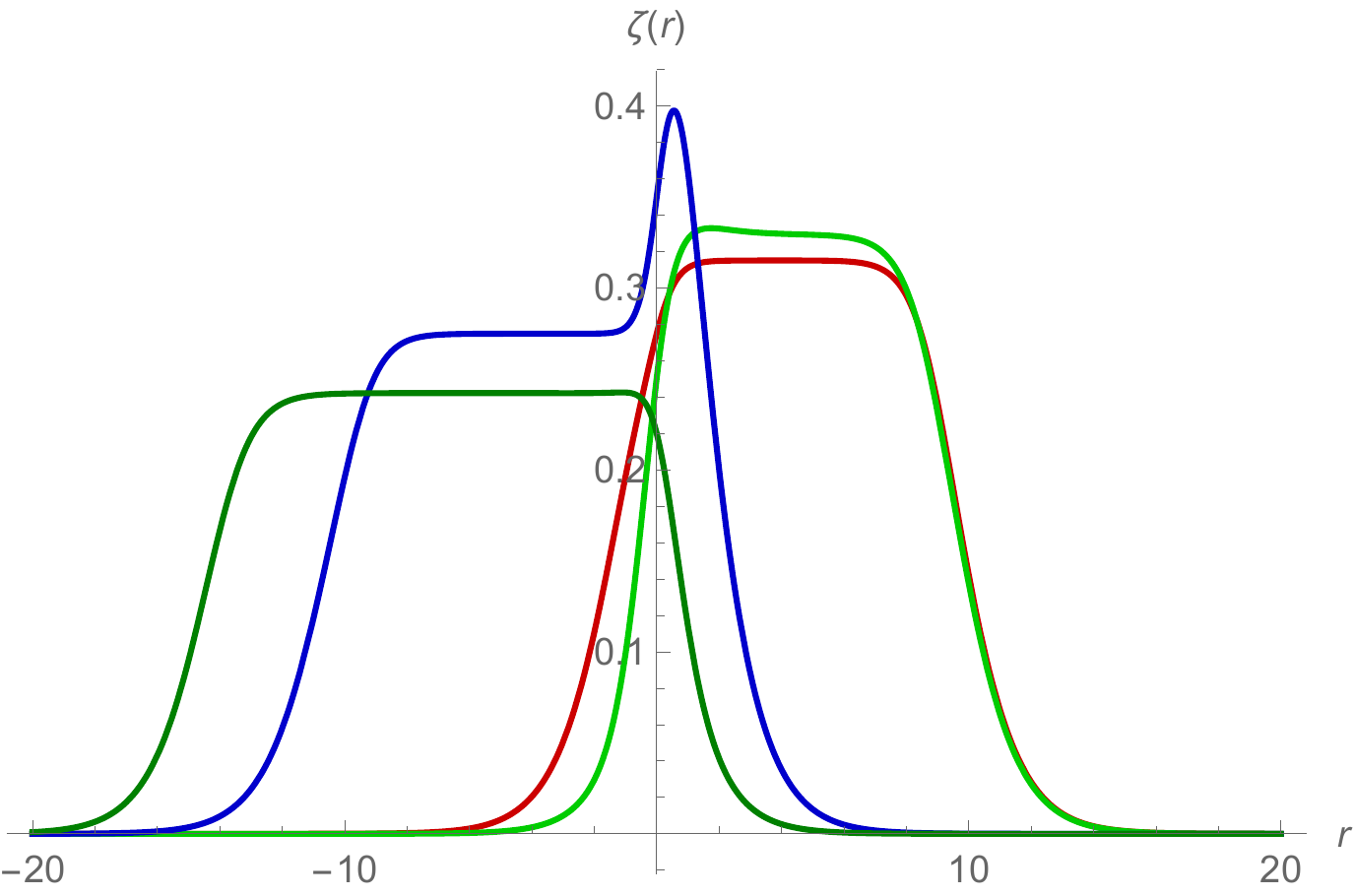}
  \caption{$\zeta(r)$}
  \end{subfigure}
  \begin{subfigure}[b]{0.45\linewidth}
    \includegraphics[width=\linewidth]{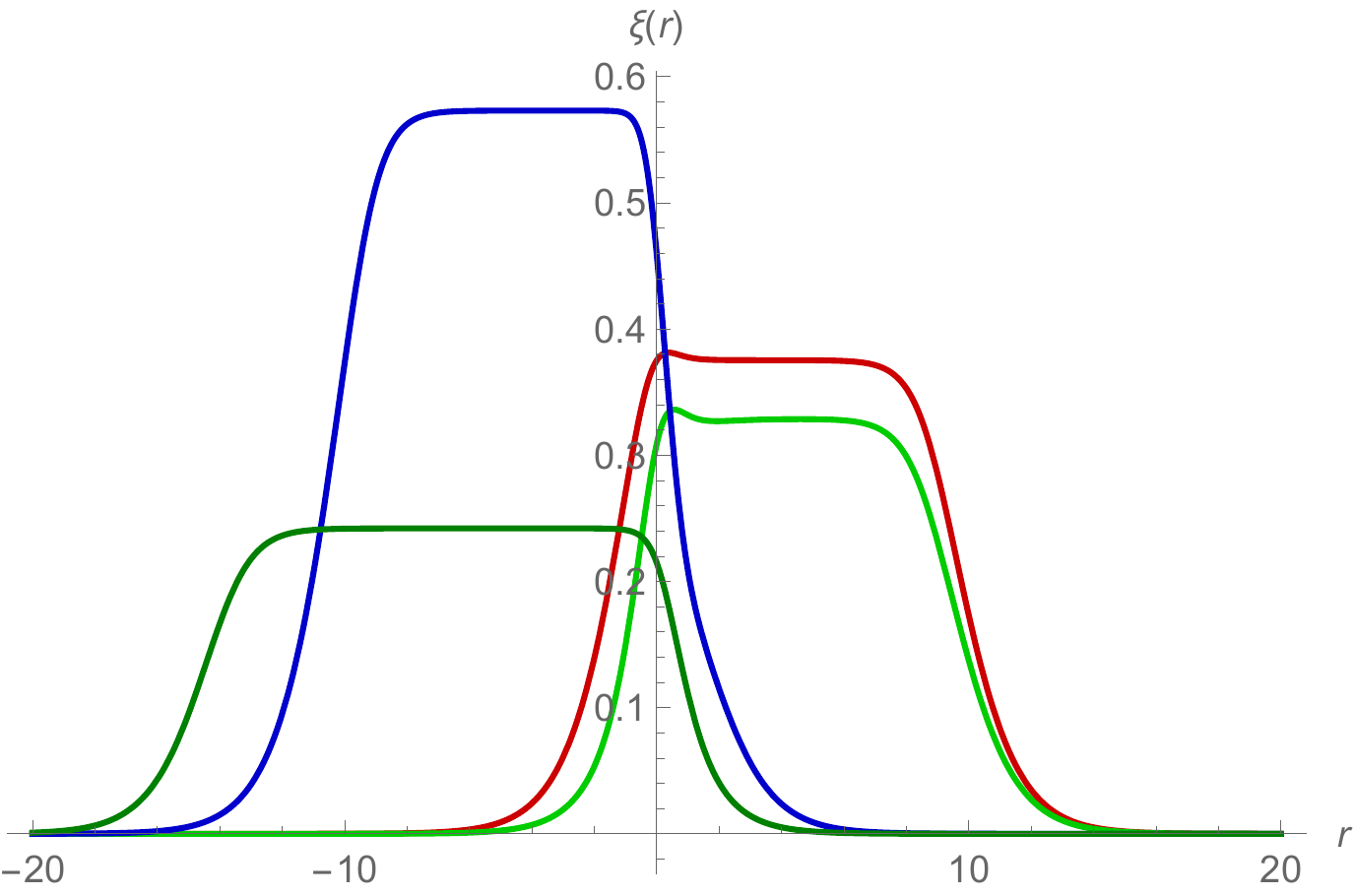}
  \caption{$\xi(r)$}
  \end{subfigure}\\
  \begin{subfigure}[b]{0.45\linewidth}
    \includegraphics[width=\linewidth]{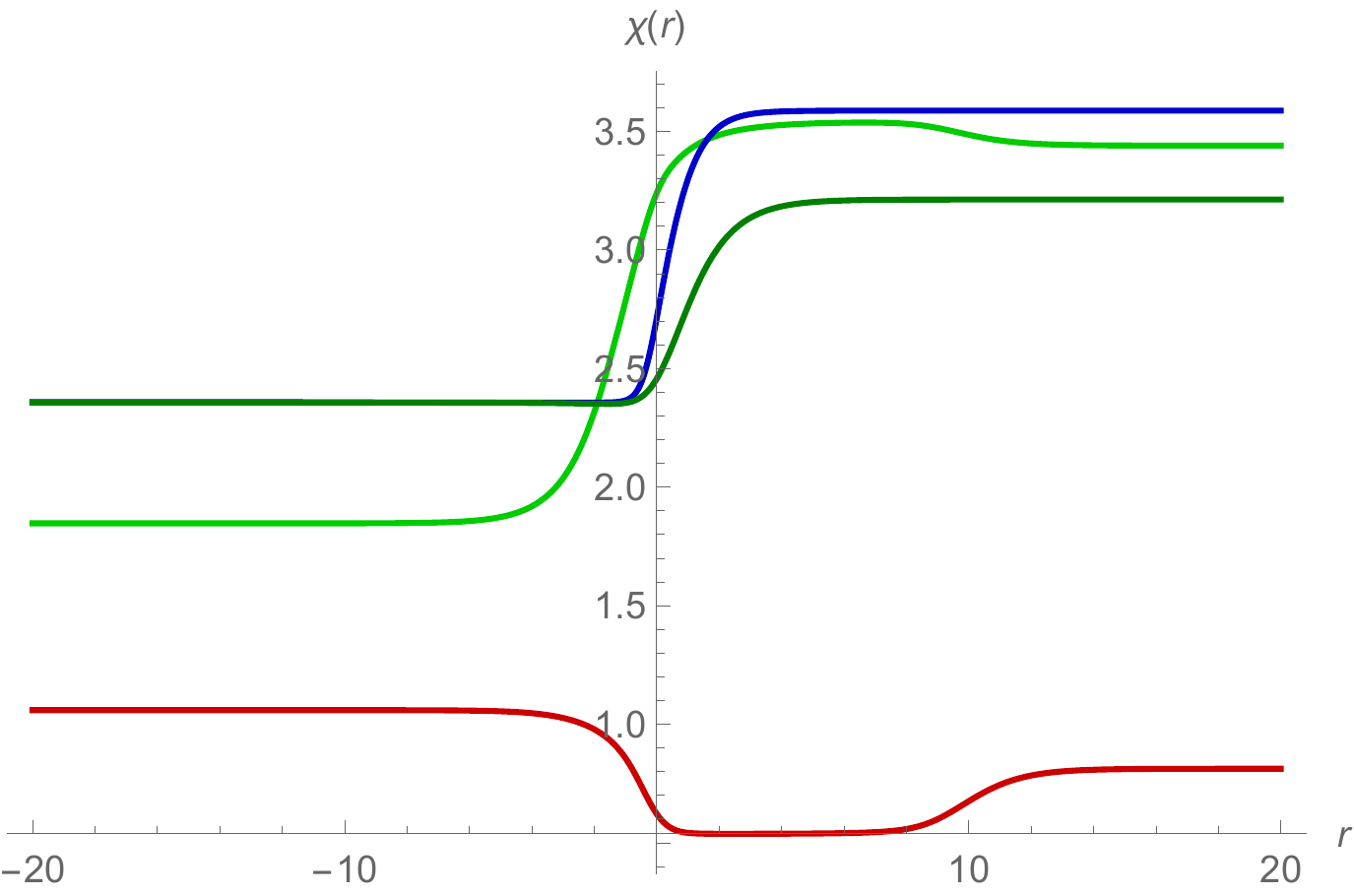}
  \caption{$\chi(r)$}
  \end{subfigure}
  \begin{subfigure}[b]{0.45\linewidth}
    \includegraphics[width=\linewidth]{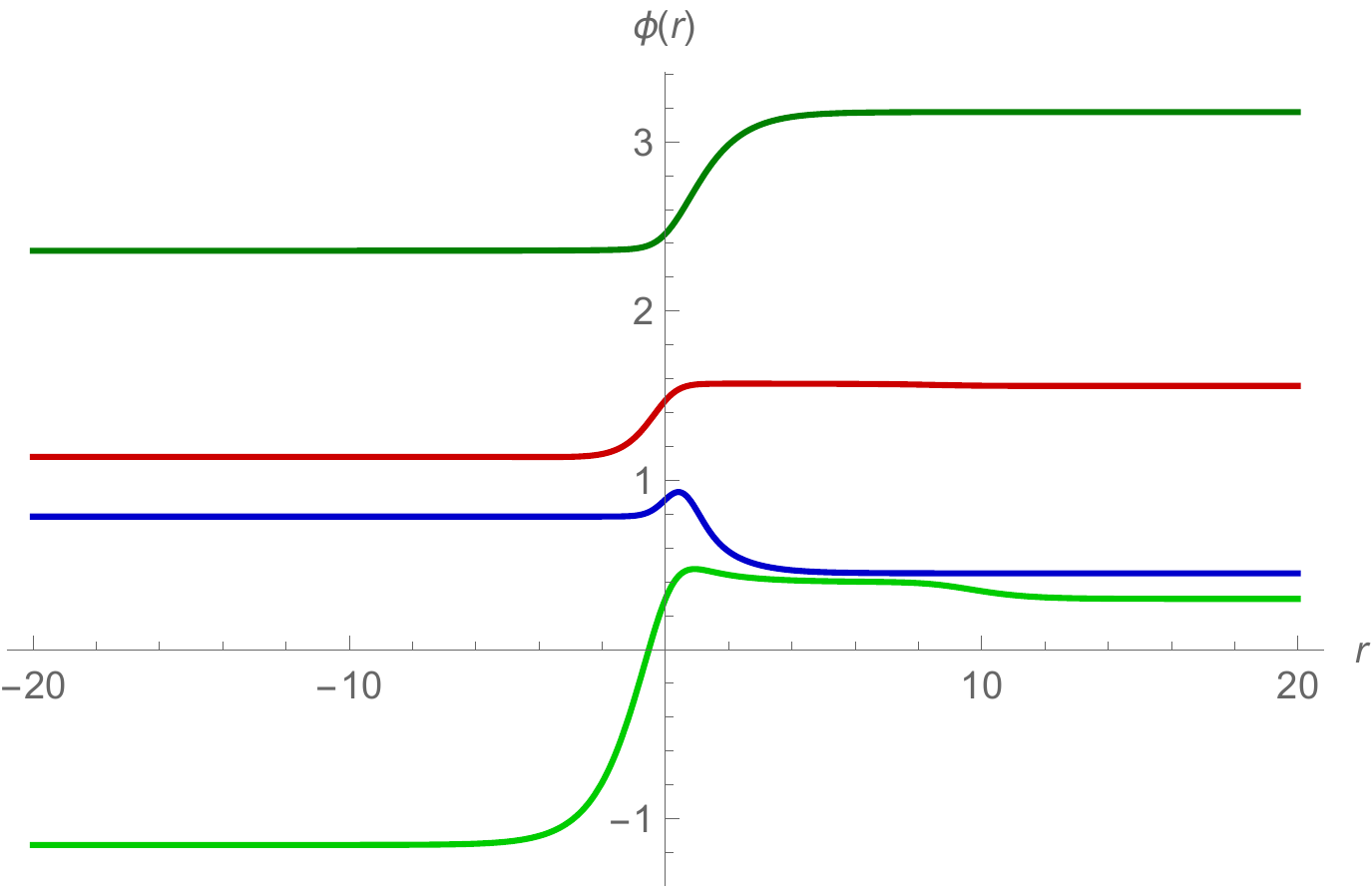}
  \caption{$\phi(r)$}
  \end{subfigure}\\
   \begin{subfigure}[b]{0.45\linewidth}
    \includegraphics[width=\linewidth]{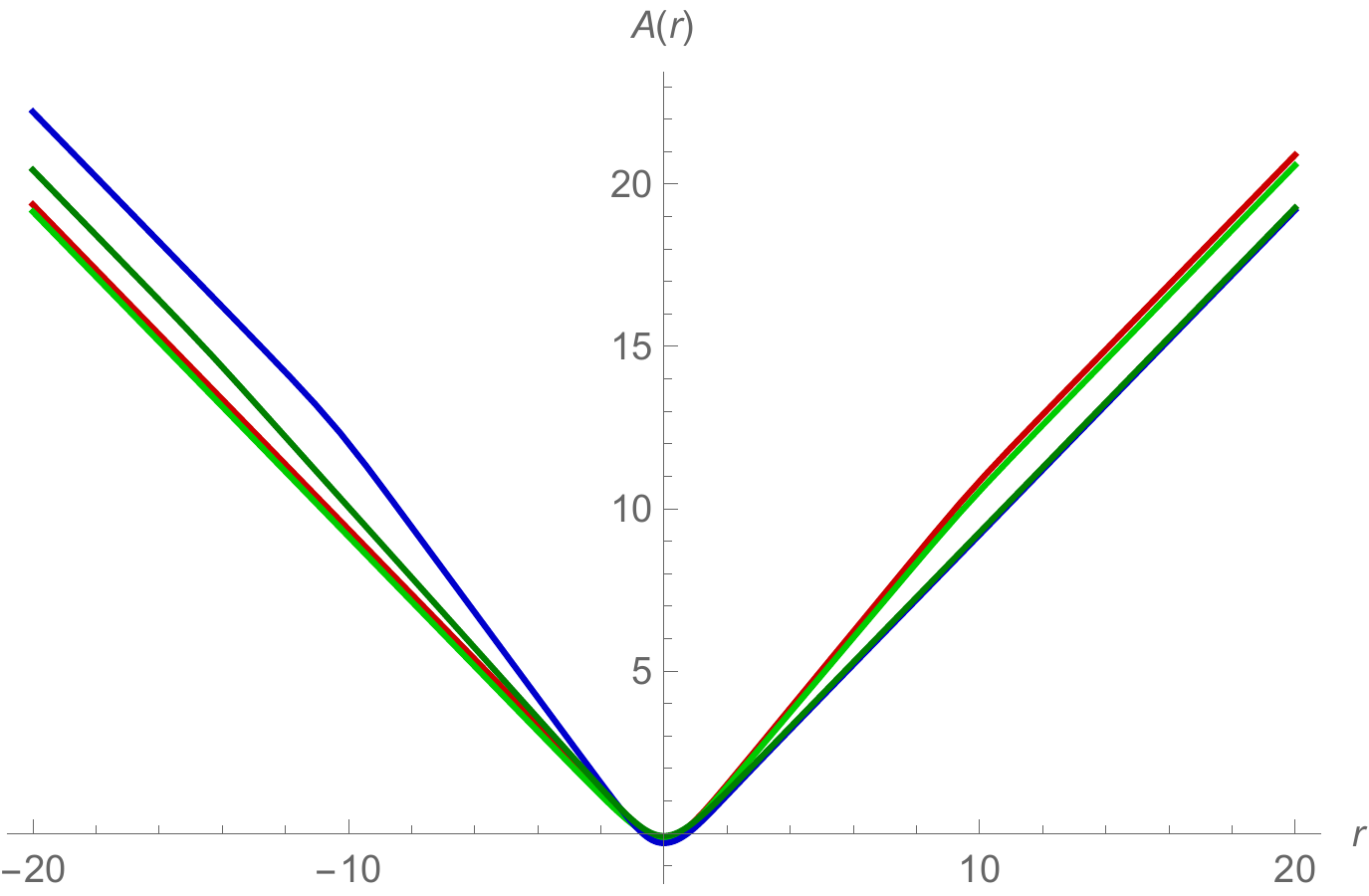}
  \caption{$A(r)$}
   \end{subfigure} 
 \begin{subfigure}[b]{0.45\linewidth}
    \includegraphics[width=\linewidth]{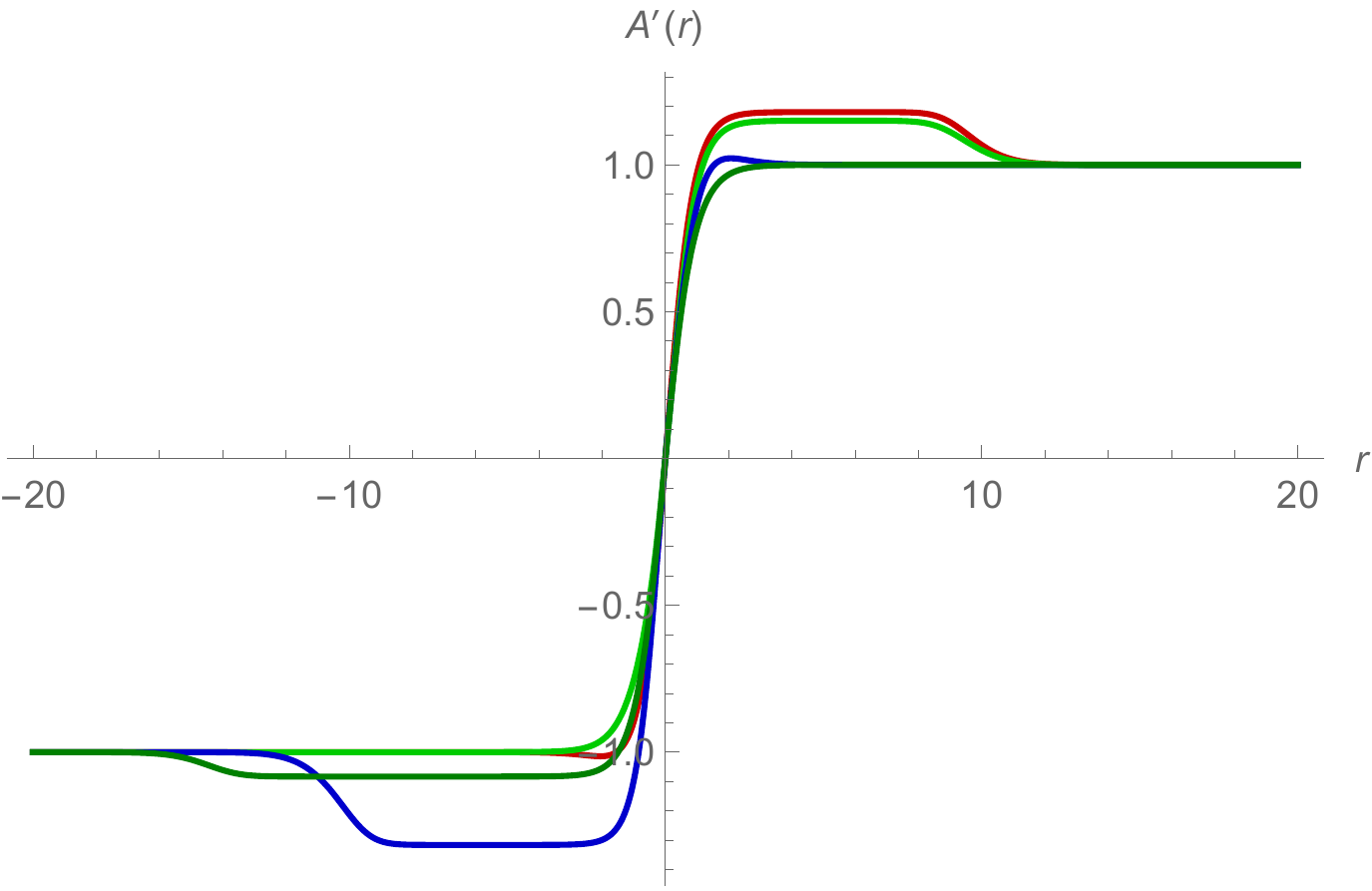}
  \caption{$A'(r)$}
   \end{subfigure} 
  \caption{Profiles of scalar fields $(\zeta,\xi,\chi,\phi)$ and the warped factor $A$ for $SO(8)/U(3)$, $SO(8)/G_2$, $SO(8)/\overline{G}_2$ and $SO(8)/SU(3)$ Janus solutions with $\omega=\frac{\pi}{8}$ as functions of the radial coordinate $r$.}
  \label{SO8X_O_Profile}
\end{figure}

\begin{figure}
%[h!]
  \centering
  \begin{subfigure}[b]{0.4\linewidth}
    \includegraphics[width=\linewidth]{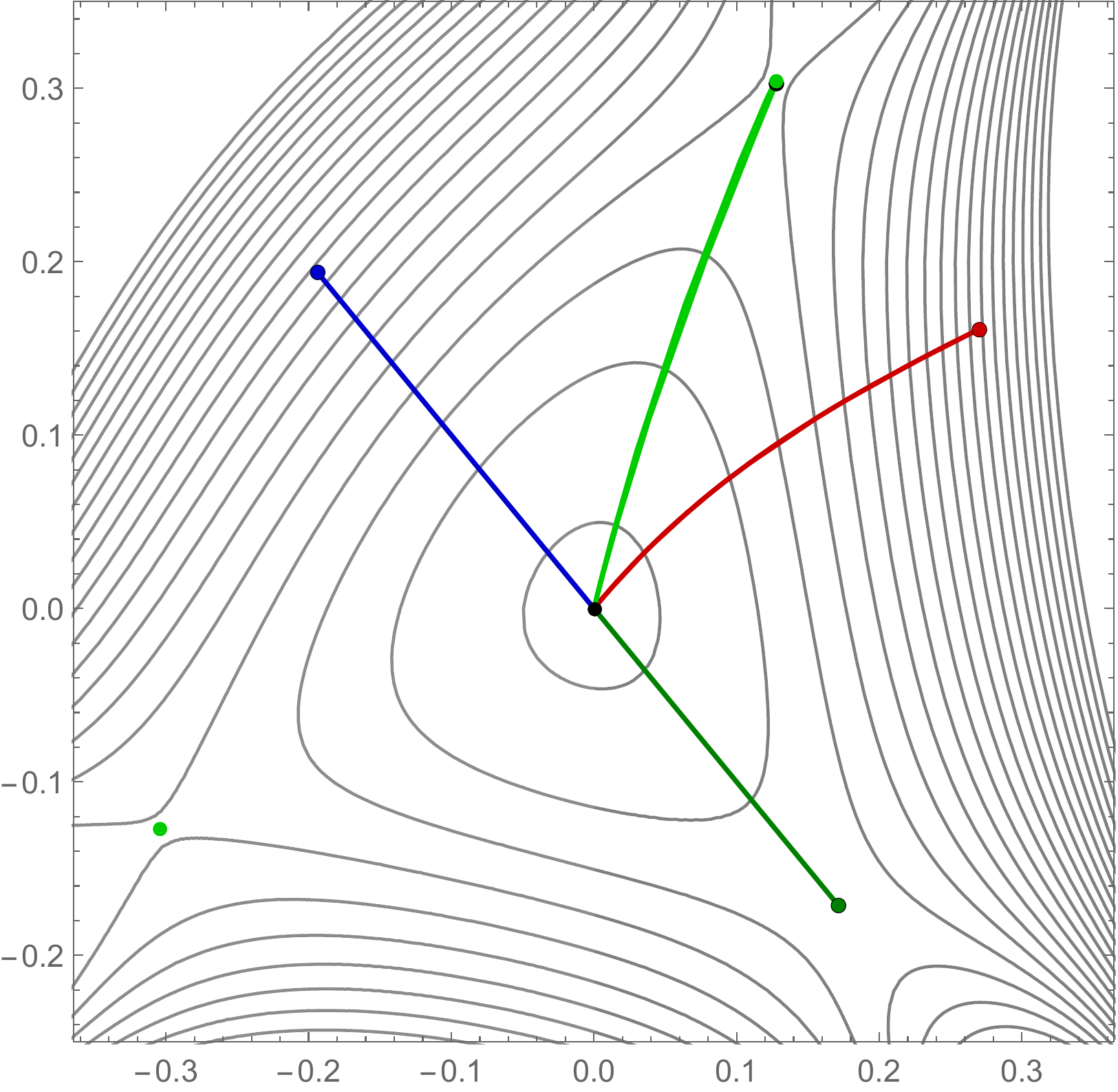}
  \caption{$(\zeta\cos\chi,\zeta\sin\chi)$ plane}
  \end{subfigure}\qquad\quad
  \begin{subfigure}[b]{0.4\linewidth}
    \includegraphics[width=\linewidth]{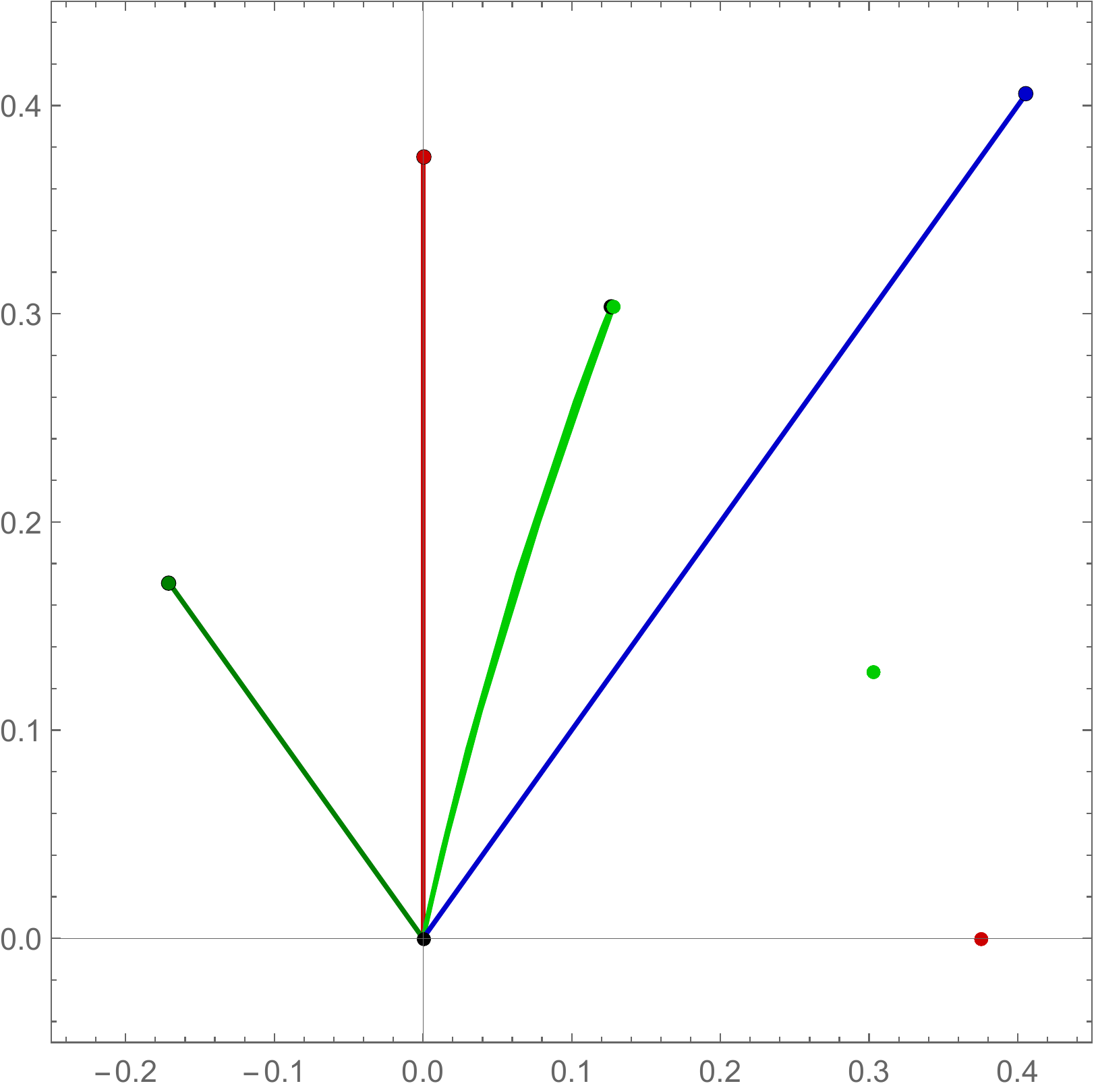}
  \caption{$(\xi\cos\phi,\xi\sin\phi)$ plane}
  \end{subfigure}
  \caption{$SO(8)/SO(8)$ Janus solutions that flow to $U(3)$, $G_2$, $\overline{G}_2$ and $SU(3)$ critical points with $\omega=\frac{\pi}{8}$ are shown on the contour plot of the superpotential by red, green, dark green and blue lines, respectively.} 
  \label{Direct_O}
\end{figure}

\begin{figure}
%[h!]
  \centering
  \begin{subfigure}[b]{0.45\linewidth}
    \includegraphics[width=\linewidth]{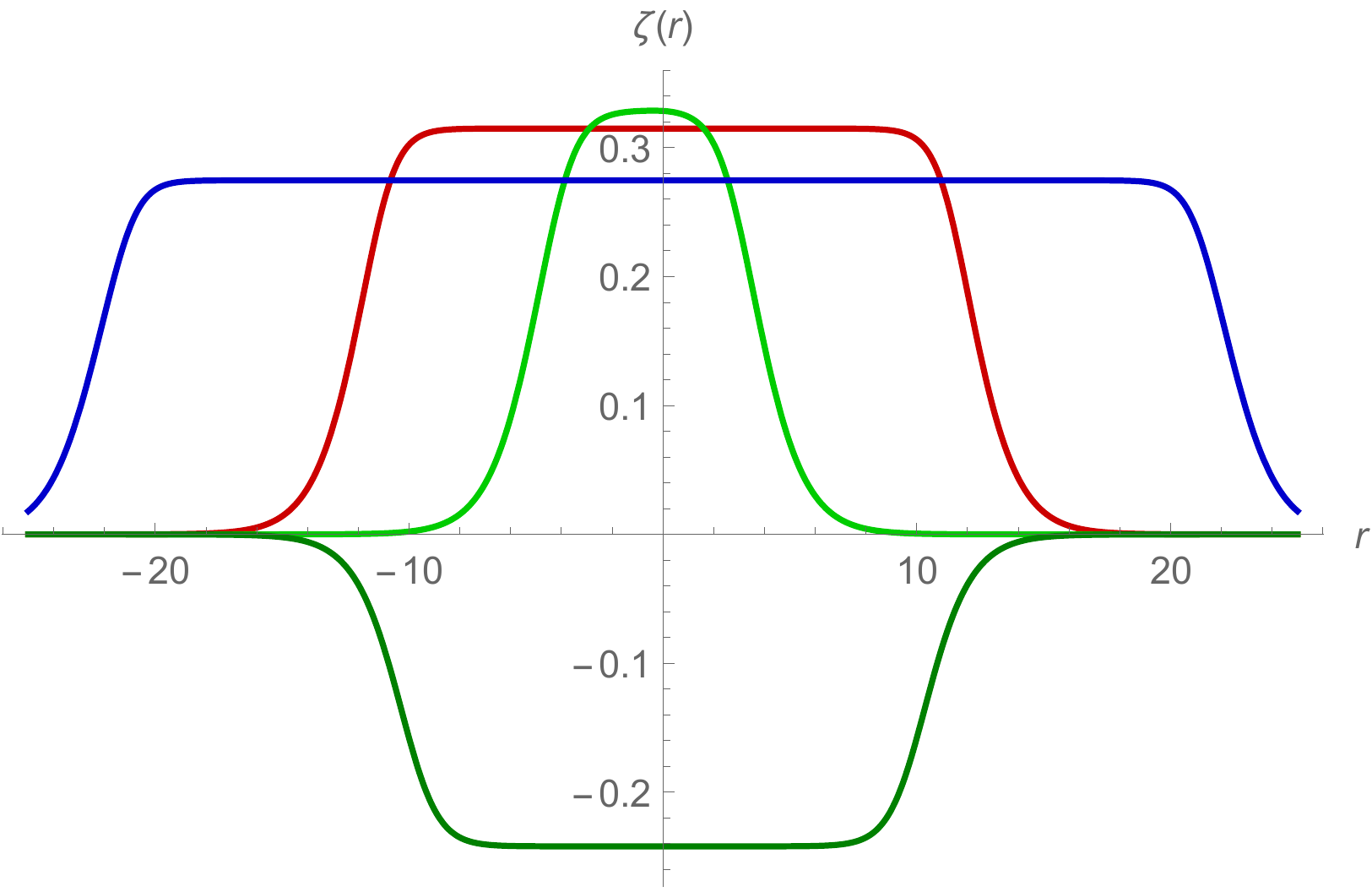}
  \caption{$\zeta(r)$}
  \end{subfigure}
  \begin{subfigure}[b]{0.45\linewidth}
    \includegraphics[width=\linewidth]{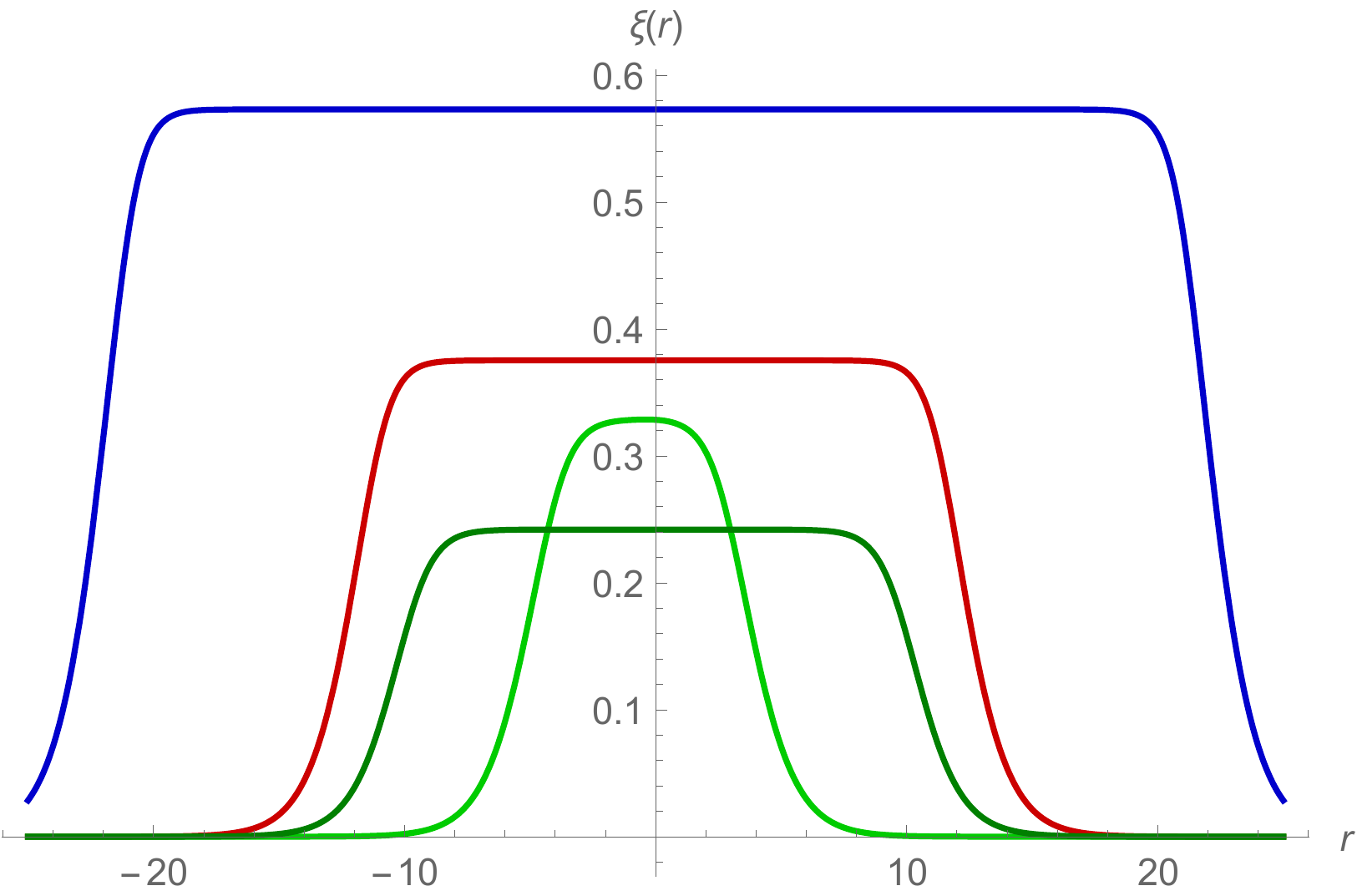}
  \caption{$\xi(r)$}
  \end{subfigure}\\
  \begin{subfigure}[b]{0.45\linewidth}
    \includegraphics[width=\linewidth]{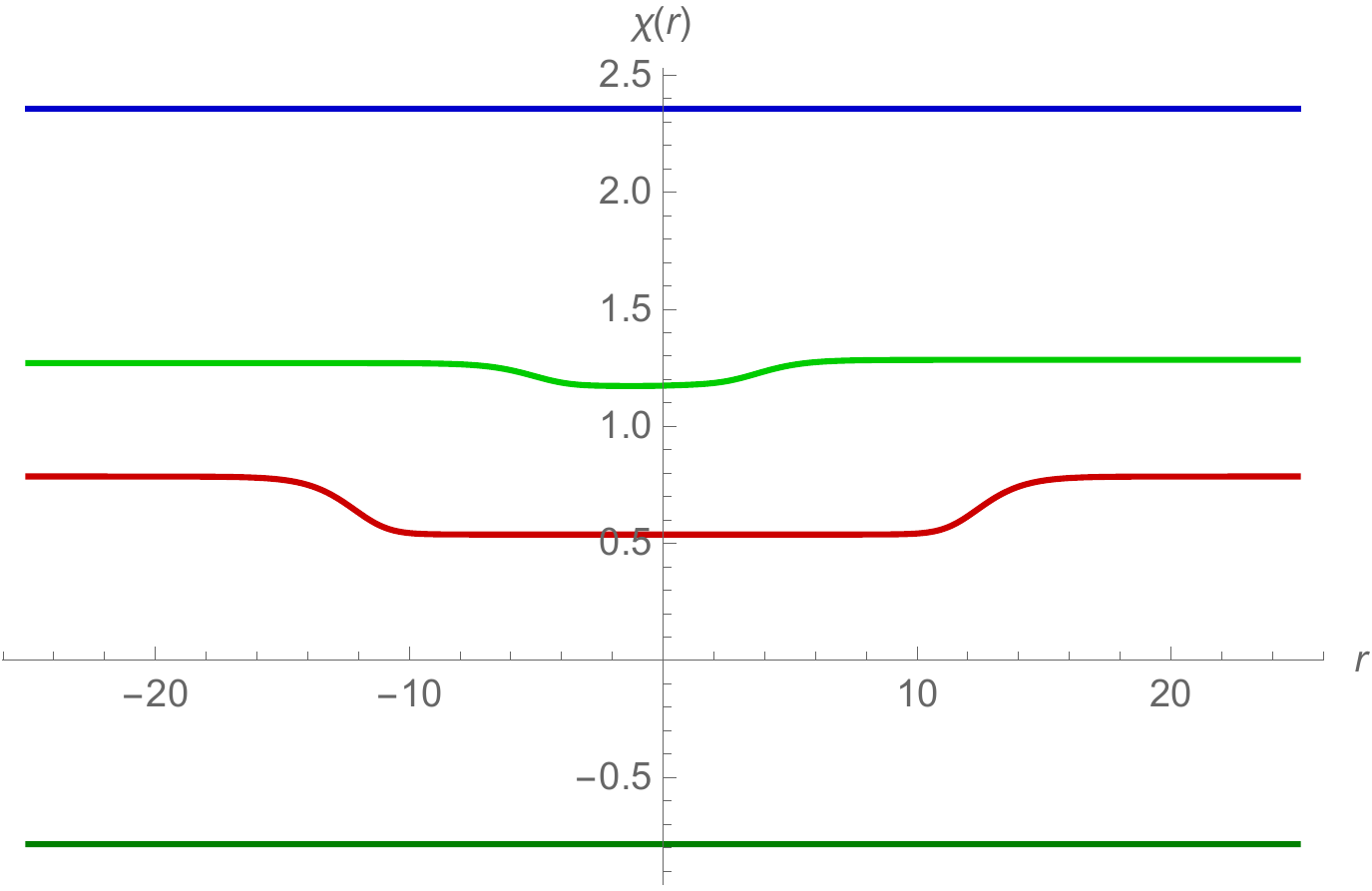}
  \caption{$\chi(r)$}
  \end{subfigure}
  \begin{subfigure}[b]{0.45\linewidth}
    \includegraphics[width=\linewidth]{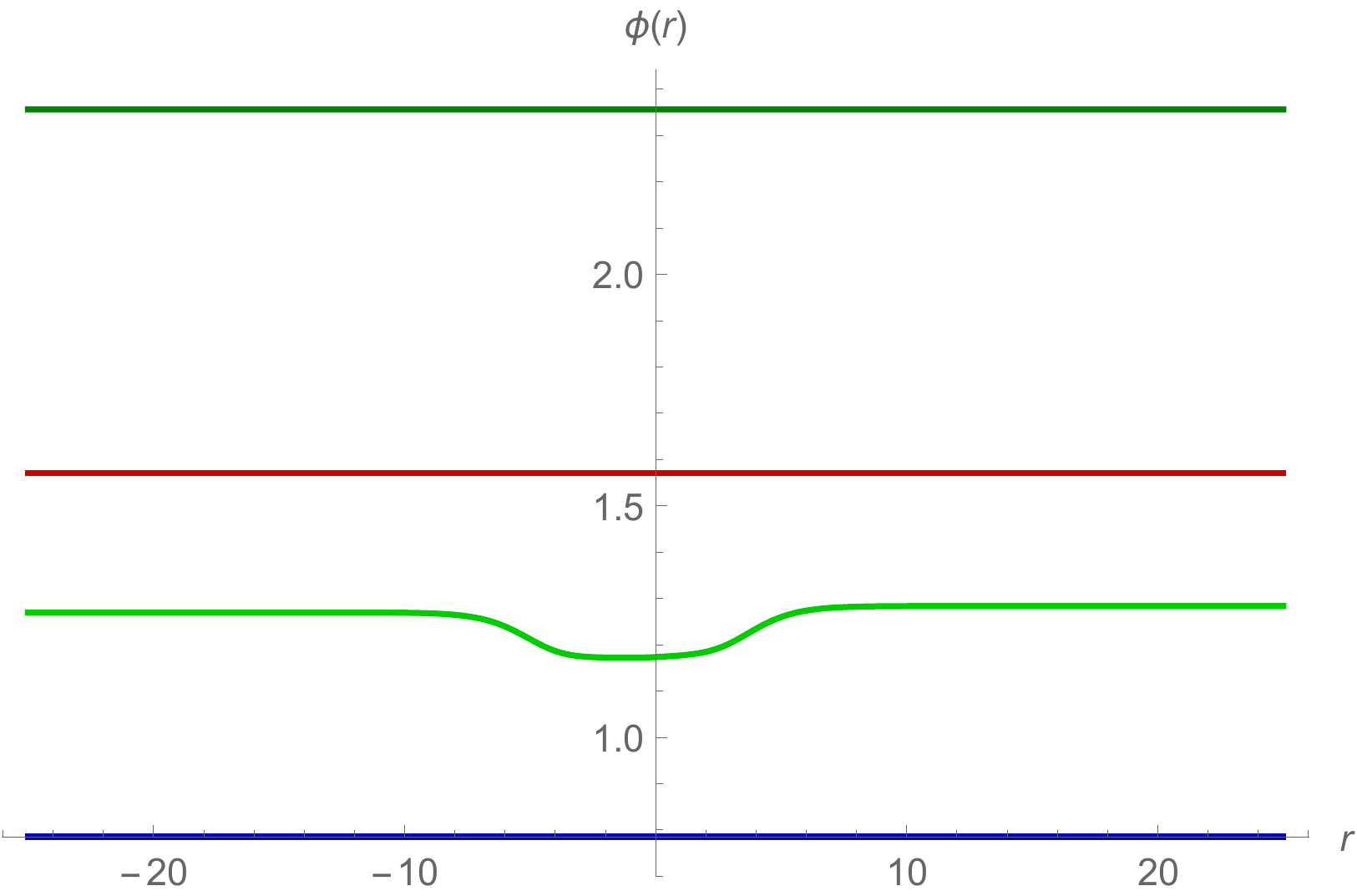}
  \caption{$\phi(r)$}
  \end{subfigure}\\
   \begin{subfigure}[b]{0.45\linewidth}
    \includegraphics[width=\linewidth]{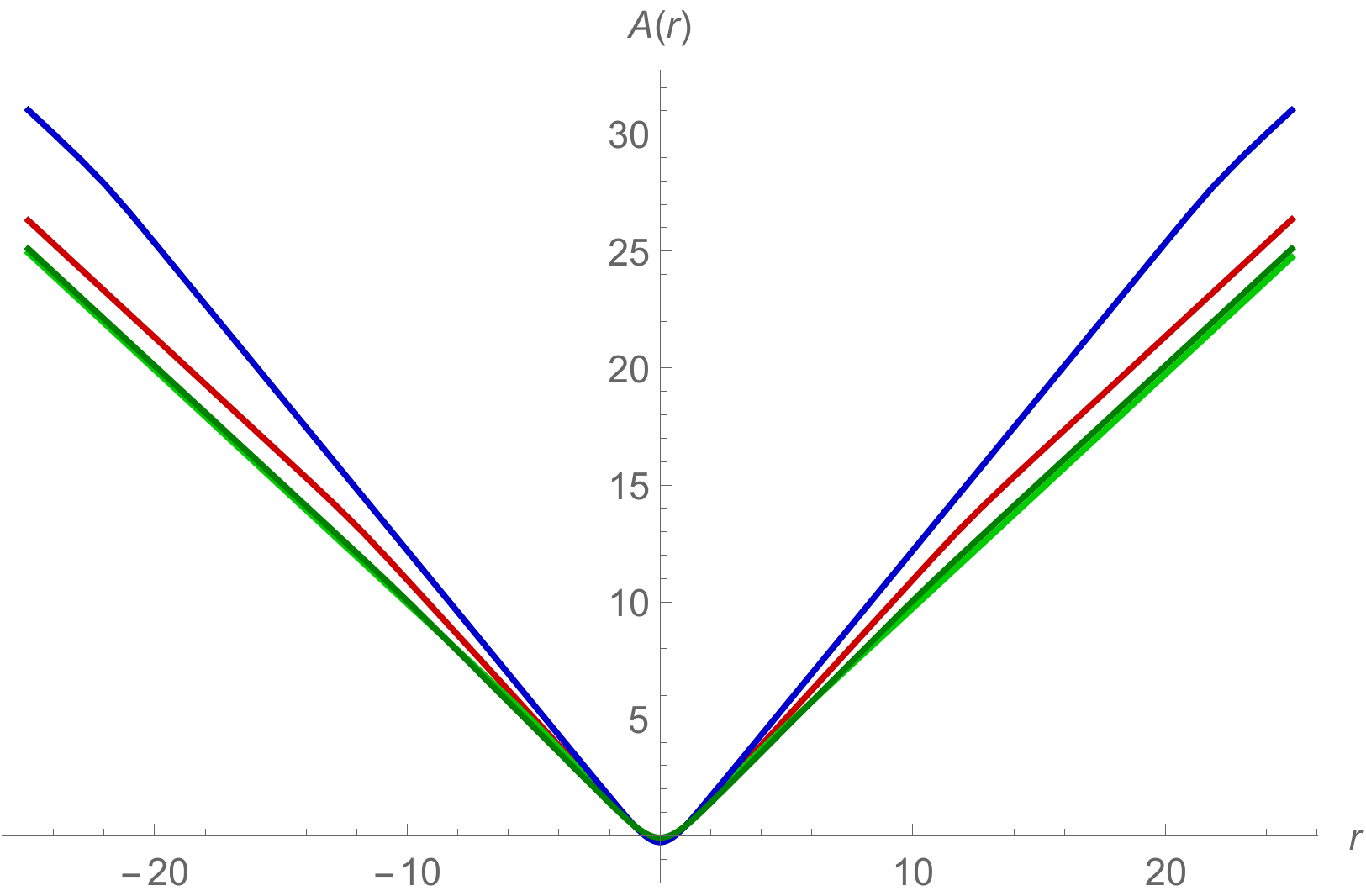}
  \caption{$A(r)$}
   \end{subfigure} 
 \begin{subfigure}[b]{0.45\linewidth}
    \includegraphics[width=\linewidth]{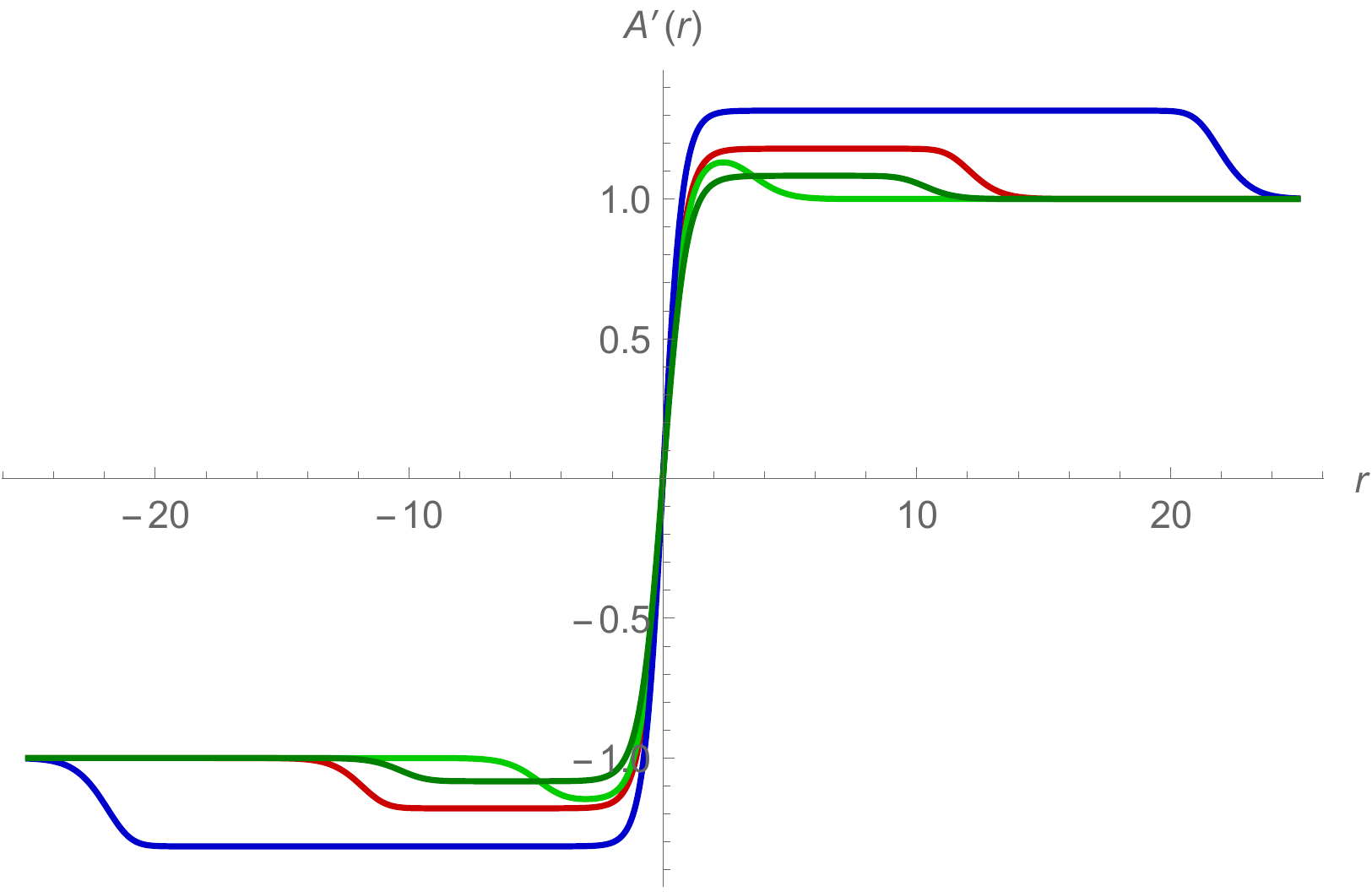}
  \caption{$A'(r)$}
   \end{subfigure} 
  \caption{Profiles of scalar fields $(\zeta,\xi,\chi,\phi)$ and of the warped factor $A$ for $SO(8)/SO(8)$ Janus solutions that flow to $U(3)$, $G_2$, $\overline{G}_2$ and $SU(3)$ critical points with $\omega=\frac{\pi}{8}$ as functions of the radial coordinate $r$.}
  \label{Direct_O_Profile}
\end{figure}

There also exist solutions describing $SO(8)/U(3)$, $SO(8)/G_2$, $SO(8)/\overline{G}_2$ and $SO(8)/SU(3)$ interfaces as shown in figures \ref{SO8X_O} and \ref{SO8X_O_Profile}. On one side of these interfaces, the $SO(8)$ phase undergoes an RG flow to the $U(3)$, $G_2$, $\overline{G}_2$ and $SU(3)$ phases. We should note here that there is an $SO(8)/G_2$ solution that flows to the $SU(3)$ critical point but we have not included this solution for readability of the figures. We also note that in $SO(8)/\overline{G}_2$ (dark green) and $SO(8)/SU(3)$ (blue) solutions, the $SO(8)$ phase is on the right while in the remaining two solutions, the $SO(8)$ phase is on the left. As in the $\omega=0$ case, there are also $SO(8)/SO(8)$ solutions that flow to $U(3)$, $G_2$, $\overline{G}_2$ and $SU(3)$ critical points as shown in figures \ref{Direct_O} and \ref{Direct_O_Profile}. 

\begin{figure}
%[h!]
  \centering
  \begin{subfigure}[b]{0.4\linewidth}
    \includegraphics[width=\linewidth]{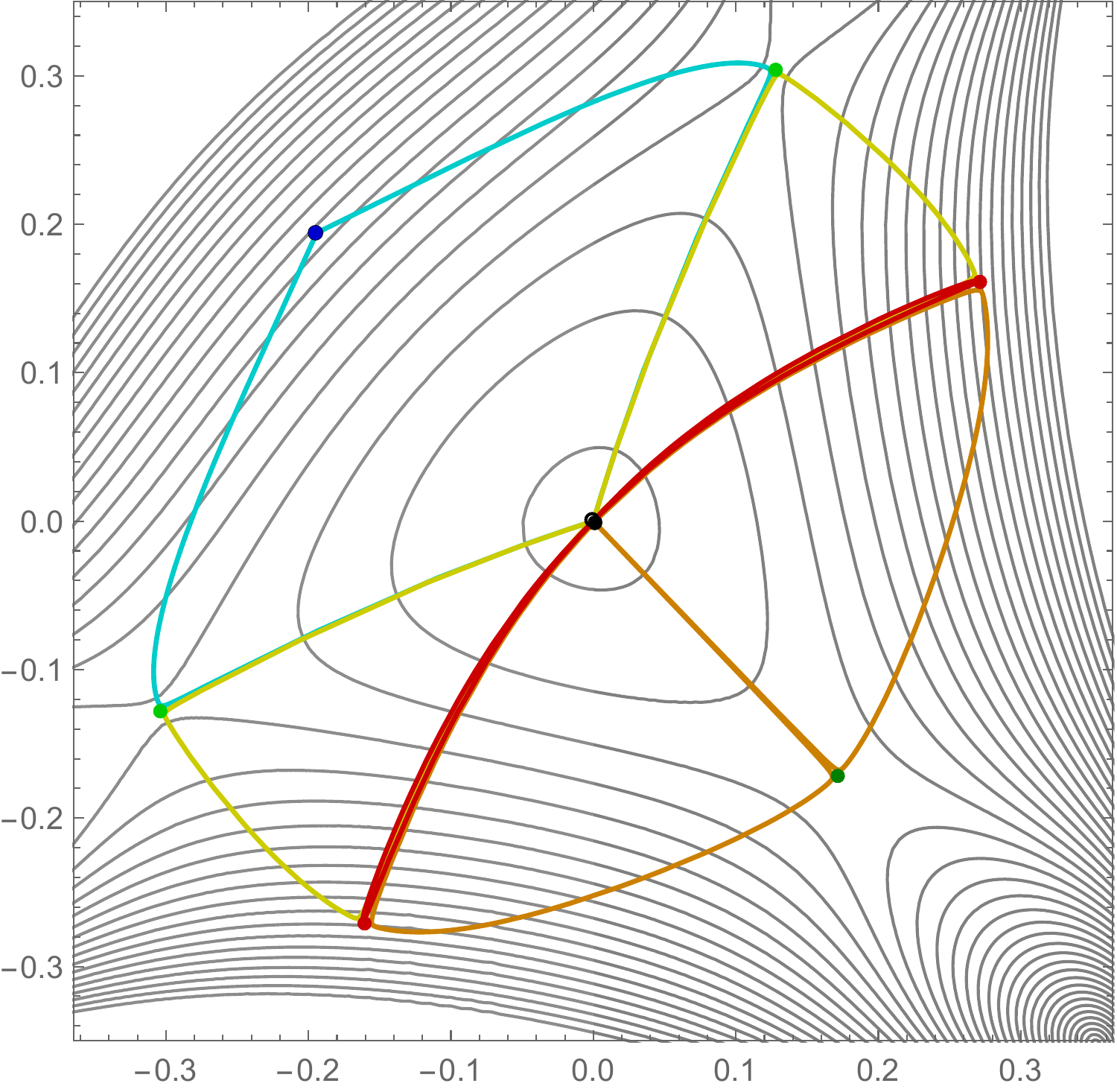}
  \caption{$(\zeta\cos\chi,\zeta\sin\chi)$ plane}
  \end{subfigure}\qquad\quad
  \begin{subfigure}[b]{0.4\linewidth}
    \includegraphics[width=\linewidth]{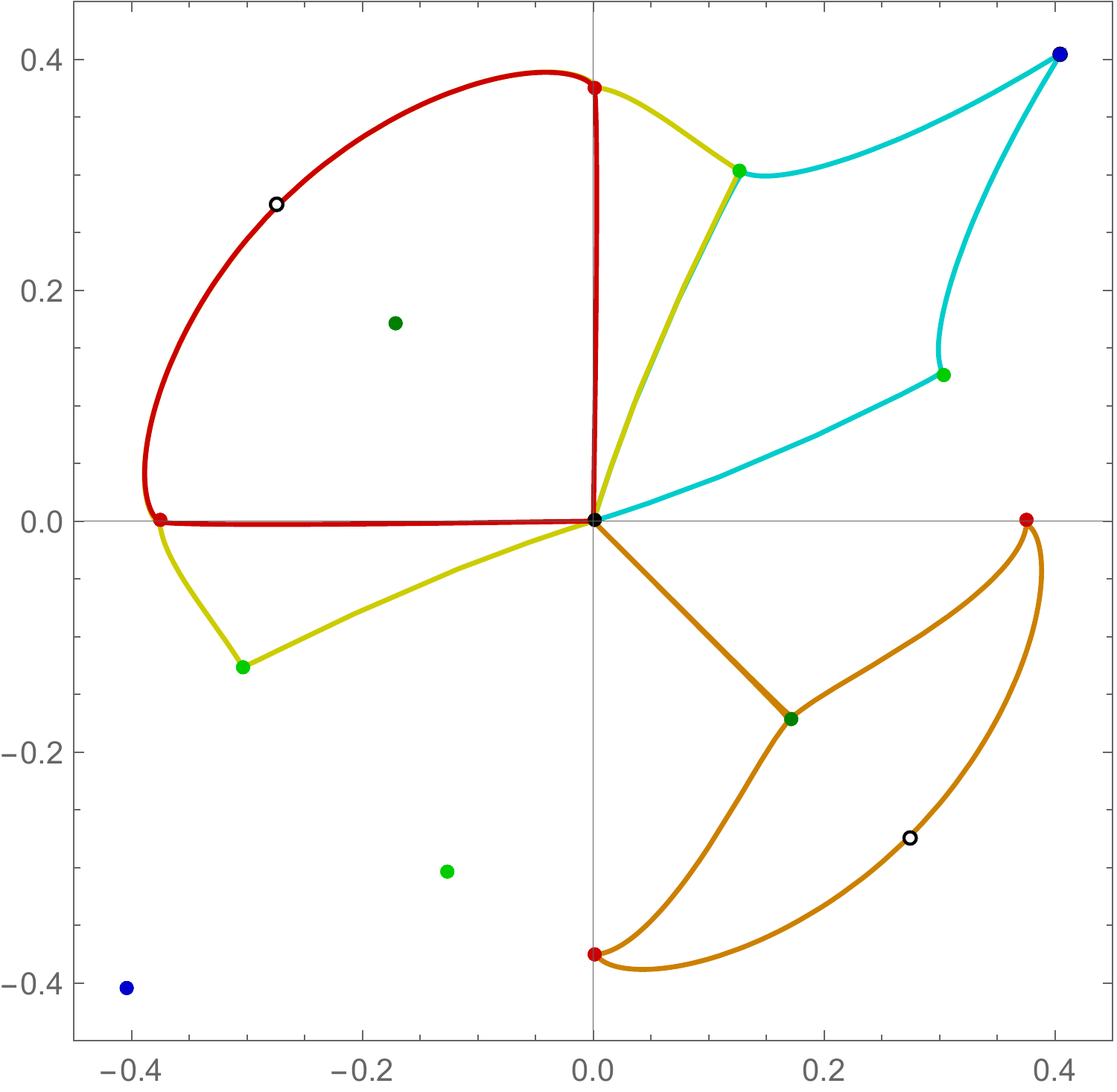}
  \caption{$(\xi\cos\phi,\xi\sin\phi)$ plane}
  \end{subfigure}
  \caption{$G_2/G_2$ Janus solutions that flow to $SU(3)$ (blue) and $U(3)$ (yellow) critical points are shown on the contour plot of the superpotential with $\omega=\frac{\pi}{8}$. The red line represents a $U(3)/U(3)$ Janus, and the orange line corresponds to a $\overline{G}_2/\overline{G}_2$ solution that flows to the $U(3)$ critical point.} 
  \label{U3U3_O}
\end{figure}                 

\begin{figure}
%[h!]
  \centering
  \begin{subfigure}[b]{0.45\linewidth}
    \includegraphics[width=\linewidth]{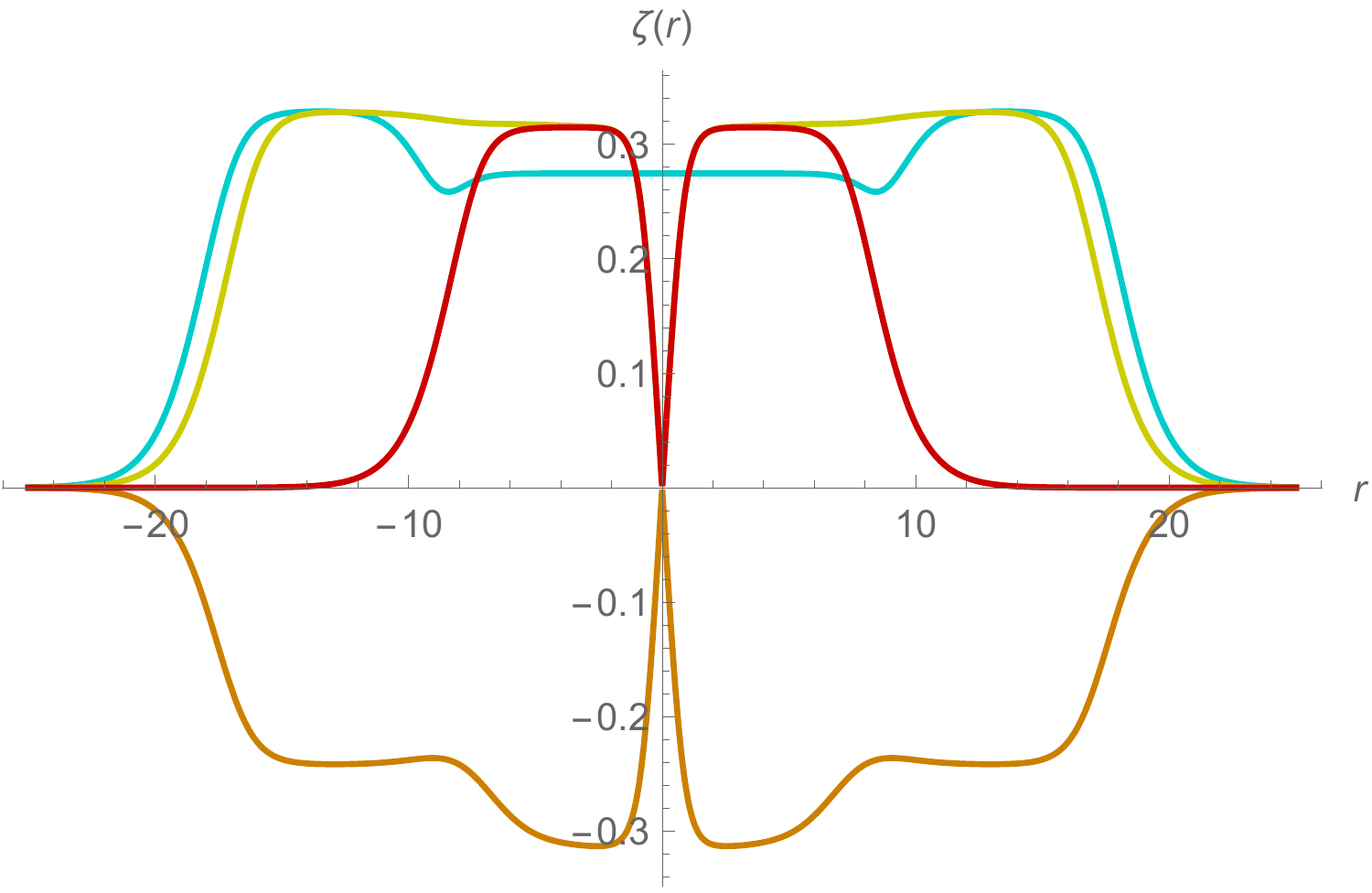}
  \caption{$\zeta(r)$}
  \end{subfigure}
  \begin{subfigure}[b]{0.45\linewidth}
    \includegraphics[width=\linewidth]{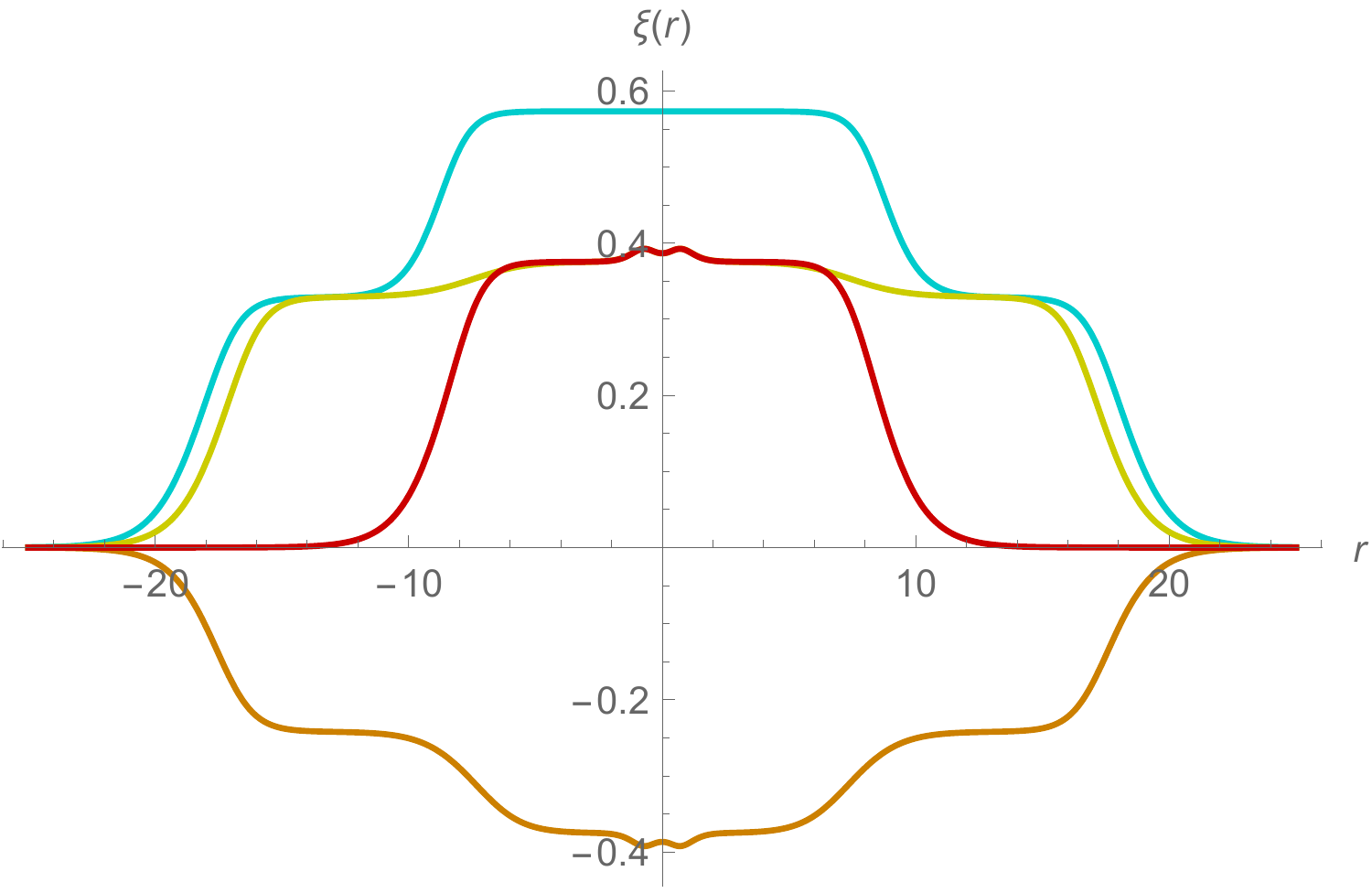}
  \caption{$\xi(r)$}
  \end{subfigure}\\
  \begin{subfigure}[b]{0.45\linewidth}
    \includegraphics[width=\linewidth]{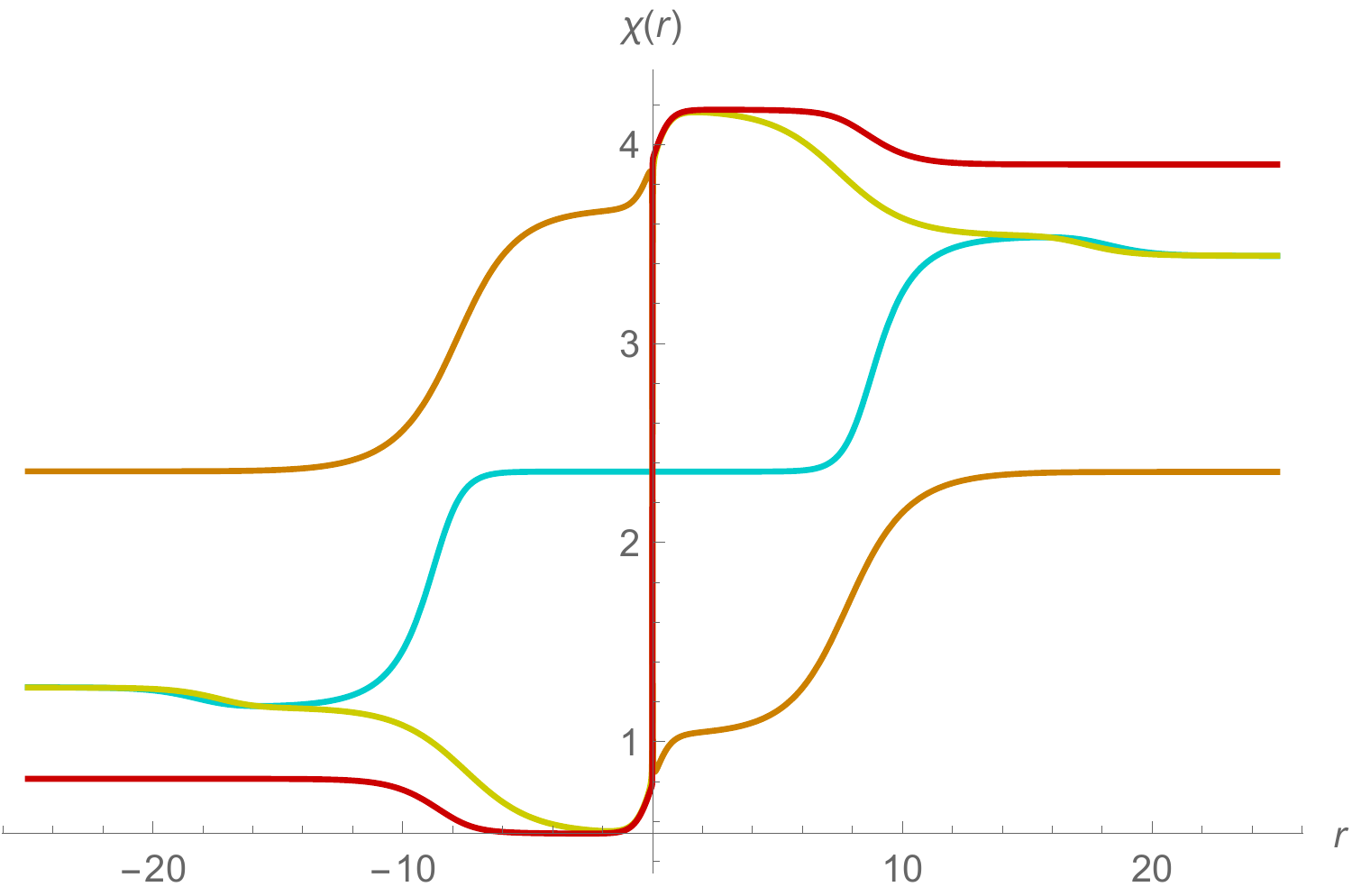}
  \caption{$\chi(r)$}
  \end{subfigure}
  \begin{subfigure}[b]{0.45\linewidth}
    \includegraphics[width=\linewidth]{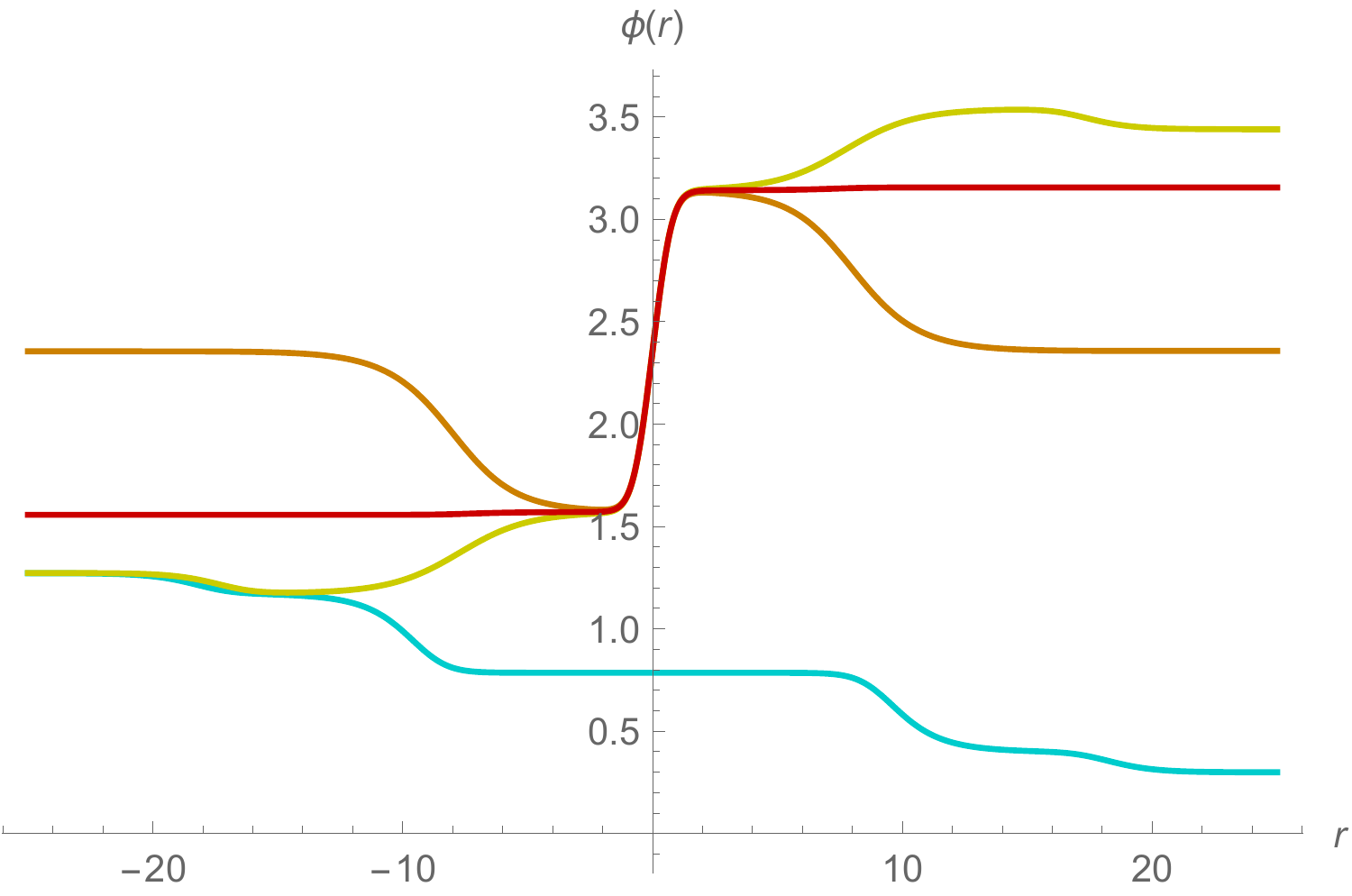}
  \caption{$\phi(r)$}
  \end{subfigure}\\
   \begin{subfigure}[b]{0.45\linewidth}
    \includegraphics[width=\linewidth]{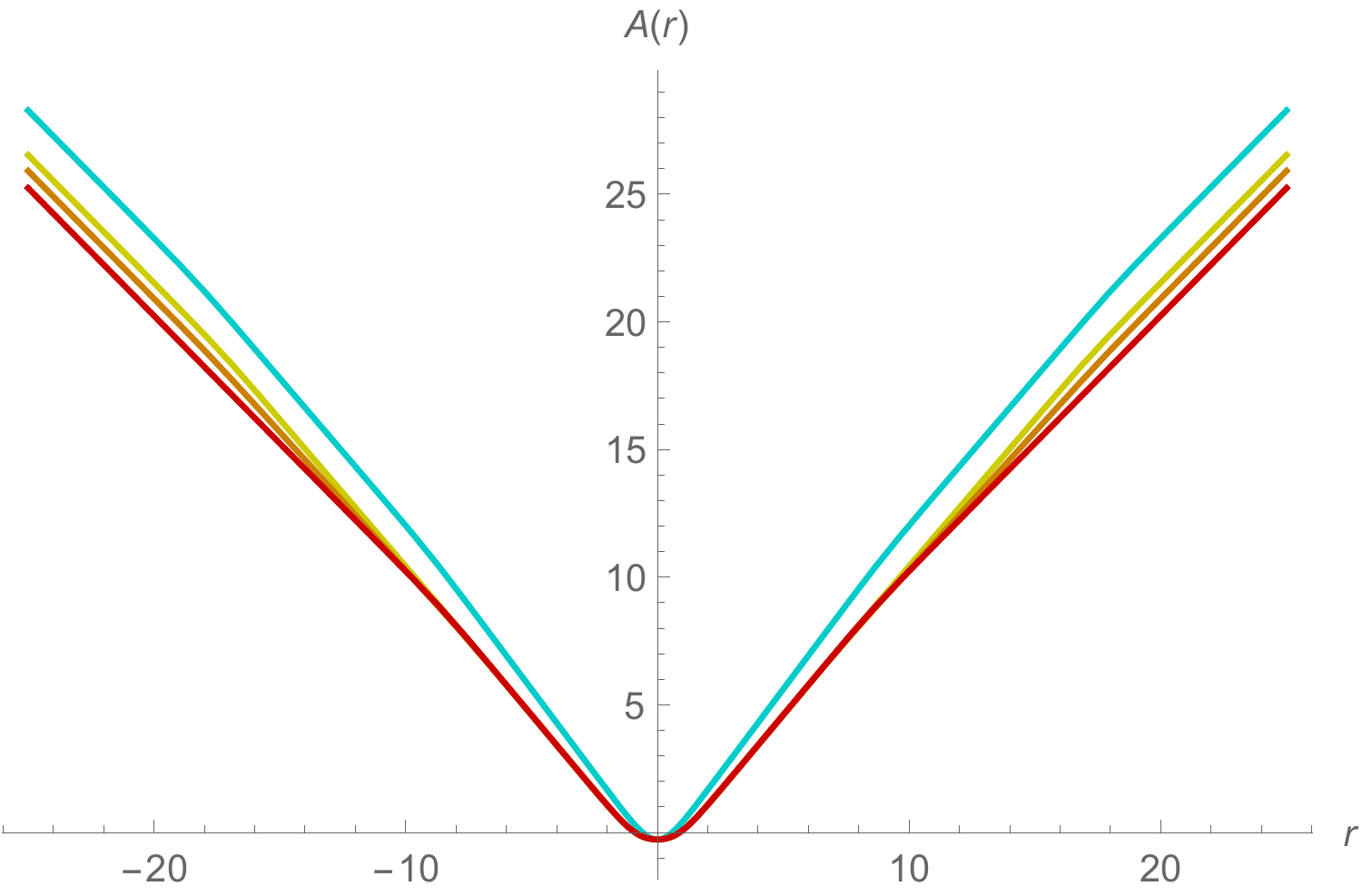}
  \caption{$A(r)$}
   \end{subfigure} 
 \begin{subfigure}[b]{0.45\linewidth}
    \includegraphics[width=\linewidth]{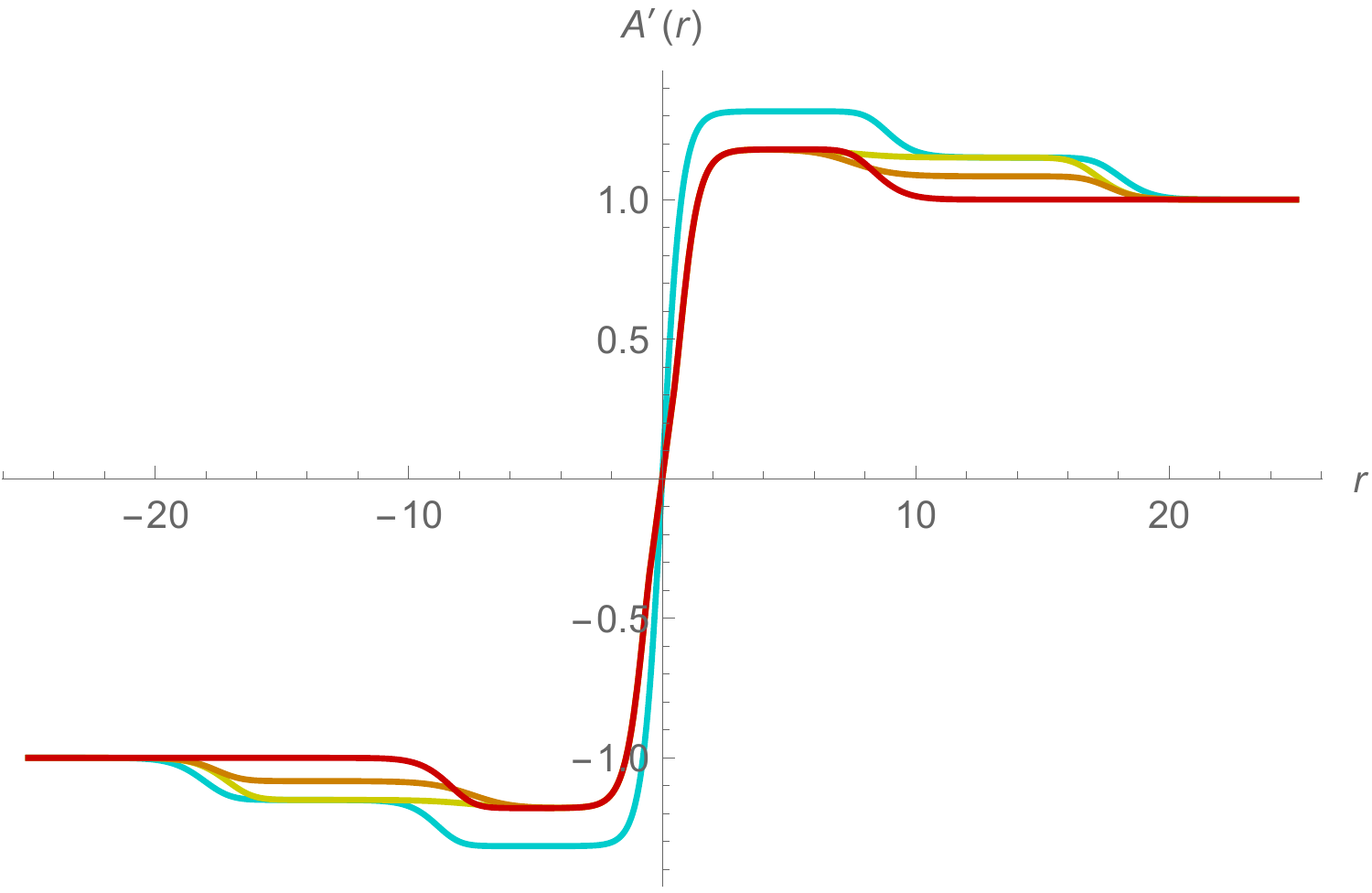}
  \caption{$A'(r)$}
   \end{subfigure} 
  \caption{Profiles of scalar fields $(\zeta,\xi,\chi,\phi)$ and the warped factor $A$ as functions of the radial coordinate $r$ for $U(3)/U(3)$ Janus together with $G_2/G_2$ and $\overline{G}_2/\overline{G}_2$ solutions that flow to $U(3)$ and $SU(3)$ critical points with $\omega=\frac{\pi}{8}$.}
  \label{U3U3_O_Profile}
\end{figure}

We end the discussion on various types of possible solutions by considering a number of solutions shown in fugures \ref{U3U3_O} and \ref{U3U3_O_Profile}. We first look at the blue line which describes a $G_2/G_2$ interface that flows to the $SU(3)$ critical point. As in the $\omega=0$ case, the initial $SO(8)$ phases on both sides undergo a rapid transition to $G_2$ phases. The red line corresponds to a $U(3)/U(3)$ interface with the $U(3)$ phases generated by the $SO(8)$ phases on each side.  Finally, the solutions represented by yellow and orange lines describe respectively $G_2/G_2$ and $\overline{G}_2/\overline{G}_2$ interfaces that flow to the $U(3)$ critical point. 
               
%%%%%%%%%%%%%%%%%%%%%%%%%%%%%%%%%%%%%%%%%%%%%%%%%%%%%%%%%%%%%%%%%%%%%%%%%%%%%%%%%%%%%%%%%%%%%%%%%%%%%%%%%%%%%%%%%%%%%%%%%%%%%%%%%%%%%%%%%
\section{Conclusions and discussions}\label{conclusion}
In this paper, we have studied supersymmetric Janus solutions of four-dimensional $N=8$ gauged supergravity with dyonic $SO(8)$ gauge group in $SU(3)$ invariant sector. For $\omega=0$, we have found Janus solutions involving the $N=2$ $U(3)$ critical point and $SO(8)/SO(8)$ solutions that flow to $G_2$ and $U(3)$ critical points similar to solutions between super Yang-Mills phases given in \cite{Minwoo_4DN8_Janus}. In addition to the $G_2/G_2$ Janus found in \cite{warner_Janus}, we have found an $SO(8)/U(3)$ Janus together with $SO(8)/G_2$ and $G_2/G_2$ solutions that flow to the $U(3)$ critical point. All these solutions can be uplifted to eleven dimensions by the $S^7$ truncation and describe conformal interfaces between $SO(8)$, $U(3)$ and $G_2$ conformal phases of the ABJM theory. These solutions extend the known solutions of \cite{warner_Janus} in which only $SO(8)/G_2$ and $G_2/G_2$ Janus solutions have been given. We have also found a family of $G_2/G_2$ Janus solutions in addition to the solution given in \cite{warner_Janus}.  
\\
\indent For $\omega=\frac{\pi}{8}$, there exist two additional supersymmetric $N=1$ $AdS_4$ critical points with $G_2$ and $SU(3)$ symmetries. We have found a number of Janus solutions describing conformal interfaces with various possible conformal phases on each side including solutions that flow to a critical point. With more different phases on each side of the interfaces and more critical points to which the solutions can flow, supersymmetric Janus solutions of dyonic $SO(8)$ gauged supergravity show a highly rich structure as expected from the analogous structure of $AdS_4$ vacua and domain walls interpolating between them. Unlike the $\omega=0$ case, the higher dimensional origin of the $SO(8)$ gauged supergravity with $\omega\neq 0$ is presently unknown. Therefore, the uplifted solutions in string/M-theory are currently not possible. The full holographic interpretation of the resulting solutions is still unclear as in the case of RG flows studied in \cite{Guariano} and \cite{Varella_N8_flow}. However, by the AdS/CFT correspondence, we expect these solutions to describe conformal interfaces between different conformal phases of the dual $N=8$ three-dimensional SCFTs.   
\\
\indent It would be interesting to uplift the $\omega=0$ solutions involving the $U(3)$ critical point to M-theory and determine the corresponding field theory deformations leading to these interfaces. The field theory description of Janus solutions that flow to a critical point also deserves further study. Embedding the $\omega$-deformed $SO(8)$ gauged supergravity in higher dimensions is clearly desirable both in the holographic context and in other applications of the dyonic $SO(8)$ gauged supergravity in string/M-theory. In particular, this would allow uplifting both the RG flows of \cite{Guariano} and \cite{Varella_N8_flow} and the Janus solutions found in this paper to string/M-theory similar to the study in \cite{Warner_N8_uplift}. It is also interesting to look for similar solutions with $0<\omega<\frac{\pi}{8}$. In this case, the two equivalent $G_2$ and $U(3)$ critical points will have different cosmological constants and become physically inequivalent critical points with the same mass spectrum and (super) symmetry. With additional two critical points, there would be many more possible solutions. Moreover, Janus solutions with $\omega\neq0$ that describe interfaces between conformal and Coulomb phases or between Coulomb phases dual to boundary SCFTs \cite{ICFT_BCFT} are also worth considering. In four-dimensional gauged supergravities, this type of solutions has been first considered in \cite{tri-sasakian-flow} and later in \cite{Minwoo_4DN8_Janus} in $N=4$ and $N=8$ gauged supergravities, respectively. We hope to come back to these issues in future work.           
\vspace{0.5cm}\\
%%%%%%%%%%%%%%%%%%%%%%%%%%%%%%%%%%%%%%%%%%%%%%%%%%%%%%%%%%%%%%%%%%%%%%%%%%%%%%%%%%%%%%%%%%%%%%%%%%%%%%%%%%%%%%%%%%%%%%%%%%%%%%%%%%%%%%%%%
{\large{\textbf{Acknowledgement}}} \\
This work is supported by The Thailand Research Fund (TRF) under grant RSA6280022.
%%%%%%%%%%%%%%%%%%%%%%%%%%%%%%%%%%%%%%%%%%%%%%%%%%%%%%%%%%%%%%%%%%%%%%%%%%%%%%%%%%%%%%%%%%%%%%%%%%%%%%%%%%%%%%%%%%%%%%%%%%%%%%%%%%%%%%%%%%%%%%%%%%%%%%%%%%%%%%%%


\begin{thebibliography}{99}
\bibitem{maldacena} J. M. Maldacena, ``The large $N$ limit of
superconformal field theories and supergravity'', Adv. Theor. Math.
Phys. \textbf{2} (1998) 231-252, arXiv: hep-th/9711200.
\bibitem{Gubser_AdS_CFT} S. S. Gubser, I. R. Klebanov and A. M. Polyakov, ``Gauge Theory Correlators from Non-Critical String Theory'', Phys. Lett. \textbf{B428} (1998) 105-114, arXiv: hep-th/9802109.
\bibitem{Witten_AdS_CFT} E. Witten, ``Anti De Sitter Space and holography'', Adv. Theor. Math. Phys. \textbf{2} (1998) 253-291, arXiv: 9802150.
\bibitem{Bak_Janus} D. Bak, M. Gutperle and S. Hirano, ``A Dilatonic Deformation of $AdS_5$ and its Field Theory
Dual'', JHEP 05 (2003) \textbf{072}, arXiv: hep-th/0304129.
\bibitem{Freedman_Janus} A. B. Clark, D. Z. Freedman, A. Karch and M. Schnabl, ``Dual of the Janus solution: An
interface conformal field theory'', Phys. Rev. \textbf{D71} (2005)
066003, arXiv: hep-th/0407073.
\bibitem{DHoker_Janus} E. D' Hoker, J. Estes and M. Gutperle, ``Interface Yang-Mills, supersymmetry, and Janus'', Nucl. Phys. \textbf{B753} (2006) 16, arXiv: hep-th/0603013.
\bibitem{Witten_Janus} D. Gaiotto and E. Witten, ``Janus Configurations, Chern-Simons Couplings, And The thetaAngle in N=4 Super Yang-Mills Theory'', JHEP 1006 (2010) 097, arXiv: 0804.2907.
\bibitem{Freedman_Holographic_dCFT} O. DeWolfe, D. Z. Freedman and H. Ooguri, ``Holography and Defect Conformal Field Theories'', Phys. Rev. \textbf{D66} (2002) 025009, arXiv: hep-th/0111135.
\bibitem{5D_Janus_CK} A. Clark and A. Karch, ``Super Janus'', JHEP 10 (2005) \textbf{094}, arXiv: hep-th/0506265.
\bibitem{5D_Janus_DHoker1} E. D'Hoker, J. Estes and M. Gutperle, ``Ten-dimensional supersymmetric Janus solutions'', Nucl. Phys. \textbf{B757} (2006) 79, arXiv: hep-th/0603012.
\bibitem{5D_Janus_DHoker2} E. D'Hoker, J. Estes and M. Gutperle, ``Exact half-BPS Type IIB interface solutions. I. Local solution and supersymmetric Janus'', JHEP 06 (2007) \textbf{021}, arXiv: 0705.0022.
\bibitem{5D_Janus_Suh} M. W. Suh, ``Supersymmetric Janus solutions in five and ten dimensions'', JHEP 09 (2011)
\textbf{064}, arXiv: 1107.2796.
\bibitem{Bobev_5D_Janus1} N. Bobev, F. F. Gautason, K. Pilch, M. Suh, J. van Muiden, ``Janus and J-fold Solutions from Sasaki-Einstein Manifolds'', Phys. Rev. \textbf{D100} (2019) 081901, arXiv: 1907.11132.
\bibitem{Bobev_5D_Janus2} N. Bobev, F. F. Gautason, K. Pilch, M. Suh, J. van Muiden, ``Holographic Interfaces in $N=4$ SYM: Janus and J-folds'', JHEP 05 (2020) \textbf{134}, arXiv: 2003.09154.
\bibitem{omega_N8_1}  G. Dall’Agata, G. Inverso and M. Trigiante, ``Evidence for a family of $SO(8)$ gauged
supergravity theories'', Phys. Rev. Lett. \textbf{109} (2012) 201301, arXiv:1209.0760.
\bibitem{omega_N8_2} G. Dall’Agata, G. Inverso and A. Marrani, ``Symplectic Deformations of Gauged Maximal Supergravity'', JHEP 07 (2014) \textbf{133}, arXiv:1405.2437.
\bibitem{SO8_deWit} B. de Wit and H. Nicolai, ``$N=8$ Supergravity'', Nucl. Phys. \textbf{B208} (1982) 323.
\bibitem{warner_Janus} N. Bobev, K. Pilchand N. P. Warner, ``Supersymmetric Janus Solutions in Four Dimensions'', JHEP 1406 (2014) 058, arXiv: 1311.4883.
\bibitem{N3_Janus} P. Karndumri, ``Supersymmetric Janus solutions in four-dimensional $N=3$ gauged
supergravity'', Phys. Rev. \textbf{D93} (2016) 125012, arXiv: 1604.06007.
\bibitem{tri-sasakian-flow} P. Karndumri, ``Supersymmetric deformations of 3D SCFTs from tri-sasakian truncation'', Eur. Phys. J. C (2017) \textbf{77}, 130, arXiv: 1610.07983.
\bibitem{orbifold_flow} P. Karndumri and K. Upathambhakul, ``Supersymmetric RG flows and Janus from type II orbifold compactification'', Eur. Phys. J. C (2017) \textbf{77}, 455, arXiv: 1704.00538.
\bibitem{Minwoo_4DN8_Janus} M. Suh, ``Supersymmetric Janus solutions of dyonic $ISO(7)$-gauged $N=8$ supergravity'', JHEP 04 (2018) \textbf{109}, arXiv: 1803.00041.
\bibitem{Kim_Janus} N. Kim and S. J. Kim, ``Re-visiting Supersymmetric Janus Solutions: A Perturbative Construction'', Chin. Phys. \textbf{C44} (2020) 7, 073104, arXiv: 2001.06789.
\bibitem{N5_flow} P. Karndumri and C. Maneerat, ``Supersymmetric solutions from $N=5$ gauged supergravity'', Phys. Rev. \textbf{D101} (2020) 126015, arXiv: 2003.05889.
\bibitem{N6_flow} P. Karndumri and J. Seeyangnok, ``Supersymmetric solutions from $N=6$ gauged supergravity'', arXiv: 2012.10978.
\bibitem{3D_Janus_de_Boer} C. Bachas, J. de Boer, R. Dijkgraaf, and H. Ooguri, ``Permeable conformal walls and
holography'', JHEP 06 (2002) \textbf{027}, arXiv:hep-th/0111210.
\bibitem{3D_Janus_Bachas} C. Bachas and M. Petropoulos, ``Anti-de-Sitter D-branes'', JHEP 02 (2001) \textbf{025}, arXiv:hep-th/0012234.
\bibitem{3D_Janus_Bak} D. Bak, M. Gutperle and S. Hirano, ``Three dimensional Janus and time-dependent black
holes'', JHEP 02 (2007) \textbf{068}, arXiv: hep-th/0701108.
\bibitem{half_BPS_AdS3_S3_ICFT} M. Chiodaroli, M. Gutperle and D.
Krym, ``Half-BPS Solutions locally asymptotic to $AdS_3\times S^3$
and interface conformal field theories'', JHEP 02 (2010)
\textbf{066}, arXiv: 0910.0466.
\bibitem{exact_half_BPS_string} M. Chiodaroli, E. D'Hoker, Y, Guo
and M. Gutperle, ``Exact half-BPS string-junction solutions in
six-dimensional supergravity'', JHEP 12 (2011) \textbf{086}, arXiv:
1107.1722.
\bibitem{multi_face_Janus} D. Bak and H. Min, ``Multi-faced Black Janus and Entanglement'', JHEP 03 (2014) \textbf{046}, arXiv: 1311.5259.
\bibitem{4D_Janus_from_11D} E. D'Hoker, J. Estes, M. Gutperle and D. Krym, ``Janus solutions in M-theory'', JHEP
06 (2009) \textbf{018}, arXiv: 0904.3313.
\bibitem{6D_Janus} M. Gutperle, J. Kaidi and H. Raj, ``Janus solutions in six-dimensional gauged supergravity'', JHEP 12 (2017) \textbf{018}, arXiv: 1709.09204.
\bibitem{3D_Janus} K. Chen and M. Gutperle ``Janus solutions in three-dimensional N=8 gauged supergravity'', arXiv: 2011.10154.
\bibitem{omega_vacua}  A. Borghese, G. Dibitetto, A. Guarino, D. Roest, and O. Varela, ``The $SU(3)$-invariant sector of new maximal supergravity,” JHEP 03 (2013) \textbf{082}, arXiv:1211.5335.
\bibitem{Guariano} A. Guarino, ``On new maximal supergravity and its BPS domain-walls'',  JHEP 02 (2014) \textbf{026}, arXiv: 1311.0785.
\bibitem{Varella_N8_flow} J. Tarrio and O. Varela, ``Electric/magnetic duality and RG flows in $AdS_4/CFT_3$'', JHEP 01 (2014) \textbf{071}, arXiv: 1311.2933.
\bibitem{Yi_4D_flow} Y. Pang, C. N. Pope and J. Rong, ``Holographic RG Flow in a New $SO(3)\times SO(3)$ Sector of $\omega$-Deformed $SO(8)$ Gauged $N=8$ Supergravity'', JHEP 08 (2015) \textbf{122}, arXiv: 1506.04270.
\bibitem{omega_Range1} A. Borghese, A. Guarino, and D. Roest, ``Triality, Periodicity and Stability of $SO(8)$ Gauged Supergravity'', JHEP 05 (2013) \textbf{107}, arXiv: 1302.6057. 
\bibitem{deWit_omega} B. de Wit and H. Nicolai, ``Deformations of gauged $SO(8)$ supergravity and supergravity in eleven dimensions'', JHEP 05 (2013) \textbf{077}, arXiv: 1302.6219.
\bibitem{warner_new_extrema} N. Warner, ``Some new extrema of the scalar potential of gauged $N=8$ supergravity'', Phys. Lett. \textbf{B128} (1983) 169.
%\bibitem{warner_N1_flow} N. Bobev, N. Halmagyi, K. Pilch and N. P. Warner, ``Holographic, $N=1$ Supersymmetric RG Flows %on M2 Branes'', 	JHEP 09 (2009) \textbf{043}, arXiv: 0901.2736.
\bibitem{Warner_N8_uplift} K. Pilch, A. Tyukov and N. P. Warner,
``$N=2$ Supersymmetric Janus Solutions and Flows: From Gauged
Supergravity to M Theory'', JHEP 05 (2016) \textbf{005}, arXiv: 1510.08090.
\bibitem{ICFT_BCFT} M. Gutperle and J. Samani, ``Holographic RG-flows and Boundary
CFTs'', Phys. Rev. \textbf{D86} (2012) 106007, arXiv: 1207.7325.
\end{thebibliography}
\end{document}